\documentclass[titlepage]{article} 
\usepackage{standalone} % load only in the main file
\pdfoutput=1
\usepackage{amsmath}
\usepackage{amsfonts}
\usepackage{fullpage}   
\usepackage{dialogue}
\usepackage{enumerate}
\usepackage{pdfpages}
\usepackage[pdfpagelabels,implicit=false]{hyperref}
\usepackage{tikz}
\usepackage{graphicx}
\usetikzlibrary{decorations.pathmorphing}
\tikzset{snake it/.style={decorate, decoration=snake}}

\newcommand{ \stod }{ \speak{Stodolsky} }
\newcommand{ \rov }{ \speak{Rovelli} }

\newcommand{ \whit }{ \speak{Whiting} }
\newcommand{ \mot }{ \speak{Mottola} }
\newcommand{ \ford }{ \speak{Ford} }
\newcommand{ \spin }{ \speak{Spindel} }
\newcommand{ \vid }{ \speak{Vidotto} }
\newcommand{ \dow }{ \speak{Dowker} }
\newcommand{ \kief }{ \speak{Kiefer} }
\newcommand{ \thft }{ \speak{'t Hooft} }
\newcommand{ \dav }{ \speak{Davies} }
\newcommand{ \mers }{ \speak{Mersini-Houghton} }
\newcommand{ \park }{ \speak{Parker} }
\newcommand{ \bard }{ \speak{Bardeen} }
\newcommand{ \misn }{ \speak{Misner} }
\newcommand{ \per }{ \speak{Perry} }
\newcommand{ \duf }{ \speak{Duff} }
\newcommand{ \louk }{ \speak{Louko} }
\newcommand{ \free }{ \speak{Freese} }
\newcommand{ \full }{ \speak{Fulling} }
\newcommand{ \stell }{ \speak{Stelle} }

\begin{document}

\begin{titlepage}

		%auto-ignore

\centering

	\vfill
	
	\vfill
	
	{\Large ``Hawking Radiation" Conference \\ Manuscript }
	
	\vspace{6pt}
	
	{\large 2015 } %Stockholm, Sweden }
	
	\vspace{12pt}
	
Laura Mersini-Houghton\(^*\)

{and\small}

Dillon N. Morse\(^*\)

\vspace{12pt}

\({}^*\)\emph{Department of Physics and Astronomy, UNC-Chapel Hill, NC 27599, USA}

\vfill

\end{titlepage}

\pagebreak

\section*{Foreword}

		   %auto-ignore

The origin of Hawking radiation and the perplexing problem of information loss in black holes has been at the center of decades of discussions. Recently the subject has seen renewed interest as new ideas, approaches and paradoxes have been suggested. Progress in our understanding of this mystery has far reaching consequences on all aspects of theoretical physics, from cosmology to quantum gravity. These are what motivated my decision to bring together the leading experts and the founding fathers of this field, for a week-long working session of concentrated effort focused on stimulating progress in the topic. Everyone I contacted enthusiastically agreed with this initiative. That is how the ‘Hawking Radiation’ conference came about.

\vspace{12pt}

Due to Stephen Hawking’s difficulty with travelling to the USA, we had to hold the conference in Europe instead of in Chapel Hill, USA. I am especially indebted to Nordita, that together with KTH Royal Swedish Institute and Stockholm University, graciously offered to co-host us and co-sponsor the conference with UNC-Chapel Hill.

\vspace{12pt}

UNC-Chapel Hill leadership, from the Chancellor’s office, the Provost’s office and the Dean’s office for the College of Arts and Science, fully encouraged and supported this initiative from the beginning. Despite the difficulties that prevented us from holding the conference in Chapel Hill, they decided to bring Chapel Hill to the conference, wherever it was held. The Chancellor Carol Folt, Vice Chancellor for University Development David Routh, Executive Vice Provost and Chief International Officer Ronald P. Strauss, the Dean of the College of Arts and Science Kevin Guskiewicz, and a full team of support from UNC-Chapel Hill, flew to Stockholm to help with all aspects of the organization, to extend Chapel Hill’s hospitality, and, to welcome the conference participants, through a series of events that included a public lecture by S. Hawking. I am very grateful to them for their support and commitment to the advancement of science. I would like to thank the UNC Global director and her staff, and the UNC Department of Communications, who offered continuous help both, during the organizing of the conference, and in the work needed afterwards in compressing, archiving and posting the video recording.

\vspace{12pt}

I would like to thank the Rektor of KTH, Peter Gudmundson, and vice Rektor of  Stockholm University, Astrid Soderbergh Widding, for their warm welcome and hospitality. The conference was sponsored by UNC-Chapel Hill, Nordita, DAMTP Cambridge University, and the Julian Schwinger Foundation. 

\vspace{12pt}

I am grateful to Yen Chin Ong at Nordita and Malcolm J. Perry at DAMTP Cambridge University from the organizing committee, the director of Nordita Katie Freese, and to Katrin Morck and Philip von Segebaden at  KTH, for their help with all aspects of the conference, and especially to my student, Dillon Morse at UNC-Chapel Hill, for transcribing the video recording. I would also like to thank my two colleagues at UNC, Jack Ng and Steve Christensen for offering their support, including help with editing the transcript.

\vspace{12pt}

The ‘Hawking Radiation’ conference, which took place August 24-30 2015, was a special and historic event that brought together founding fathers of modern physics. We are making the video recording of the whole week of talks and discussions and the proceedings here, publicly available. We have here recorded the week long talks and discussions in the hope of sharing the results, discussions, debates, enthusiasm, and ideas that came up during the conference, with those interested in the field, especially the current and future young researchers. We hope that reading this material will make every researcher feel they are participating in the conference. The video recording of this proceeding can be accessed here:

\vspace{12pt}

\url{https://www.youtube.com/playlist?list=PLJC1iUpAADPKTZuyk0lwWWQ9LRJpiRgUv}

\vspace{12pt}

We have tried to stay as close as possible to the recorded words of each speaker. Due to technical difficulties and financial limitations that allowed for only one camera recording, there may be a few minutes or sentences missing here and there, for example if a participant did not use the microphones provided.

\vspace{12pt}

\emph{\textbf{Any statements in this transcript are exclusively the property of the individual speaker. If used elsewhere, they should be quoted accordingly and have the explicit permission of the speaker.}}

\vspace{12pt}

You will notice that the voice of one special person, Jacob Bekenstein, the man who helped lay the foundations of the field, a friend and a scientist I greatly admired, is missing. He passed away a few days before the conference. I would like to dedicate this report to his memory.

\vspace{12pt}

Laura Mersini-Houghton

Department of Physics and Astronomy

UNC-Chapel Hill

\pagebreak

\includepdf[pages=-]{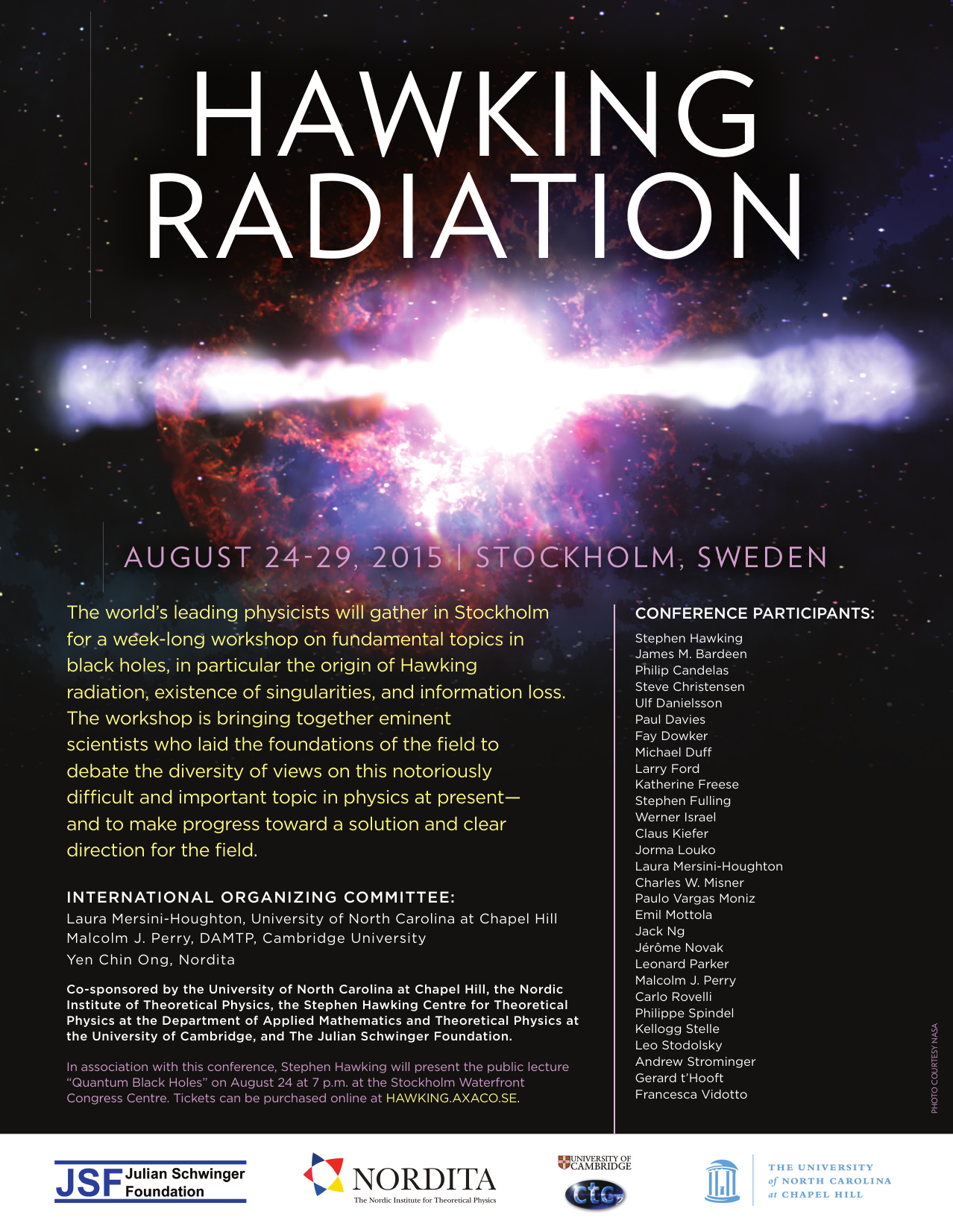}

\pagebreak

\tableofcontents

\pagebreak

\section*{Group Photos}

		%auto-ignore

\begin{figure}[!ht]
  \centering 
      \includegraphics[width=\textwidth]{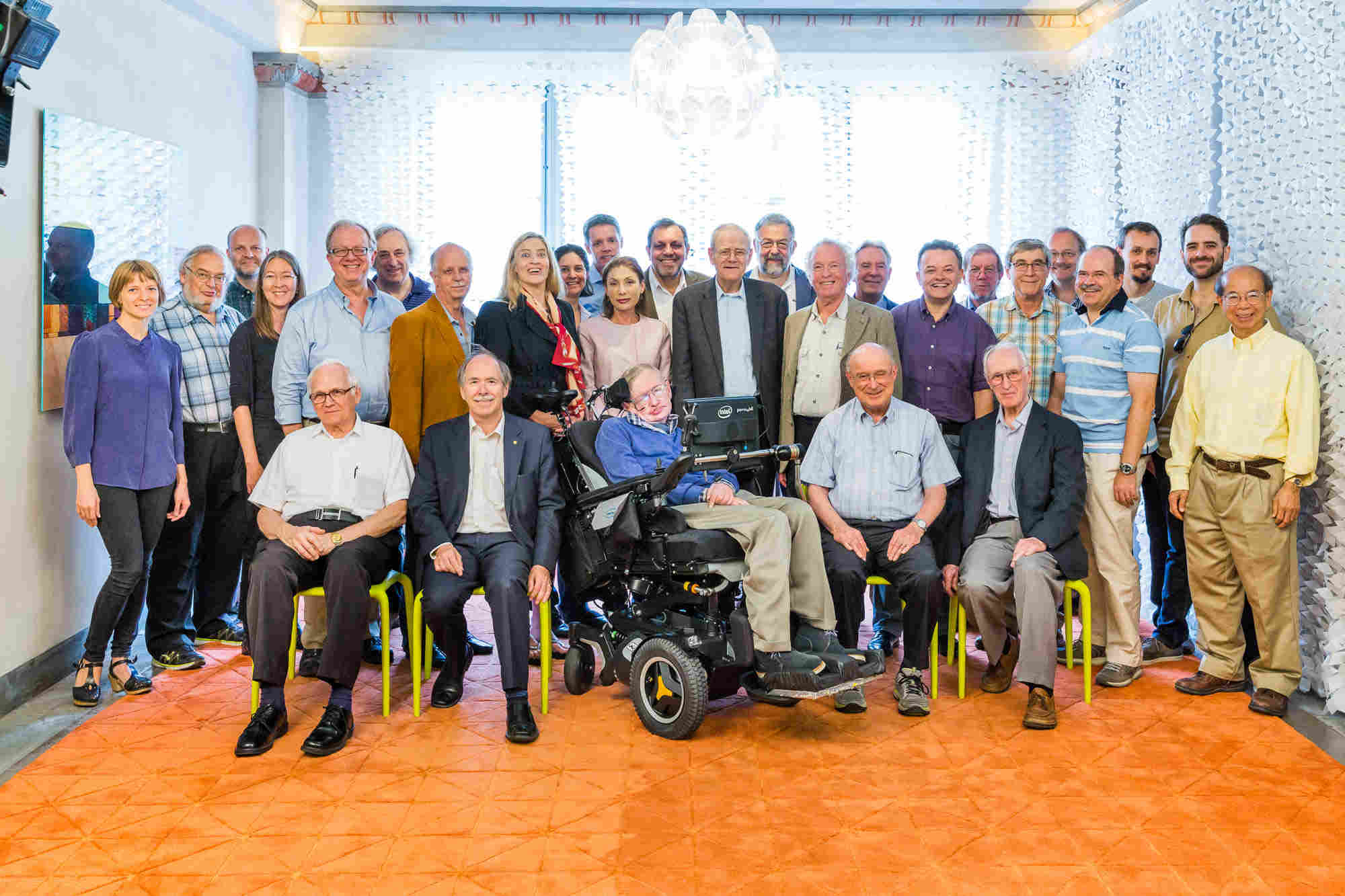}
			{\center Listed left to right.} \\ {\center \textbf{Back Row:} Jorma Louko, Malcolm Perry, Ulf Danielsson, Emil Mottola, Philippe Spindel, Michael Duff, Bernard Whiting, J\'{e}r\^{o}me Novak } \\ {\center \textbf{Middle Row:} Francesca Vidotto, Fay Dowker, Steve Christensen, Kellogg Stelle, Katherine Freese, Laura Mersini-Houghton, Leonard Parker, Leo Stodolsky, Claus Kiefer, Paul Davies, Paulo Vargas Moniz, Jack Ng } {\center \textbf{Front Row:} Stephen Fulling, Gerard 't Hooft, Stephen Hawking, James Bardeen, Charles Misner}
\end{figure}

\pagebreak

\begin{figure}[!ht]
  \centering 
      \includegraphics[width=\textwidth]{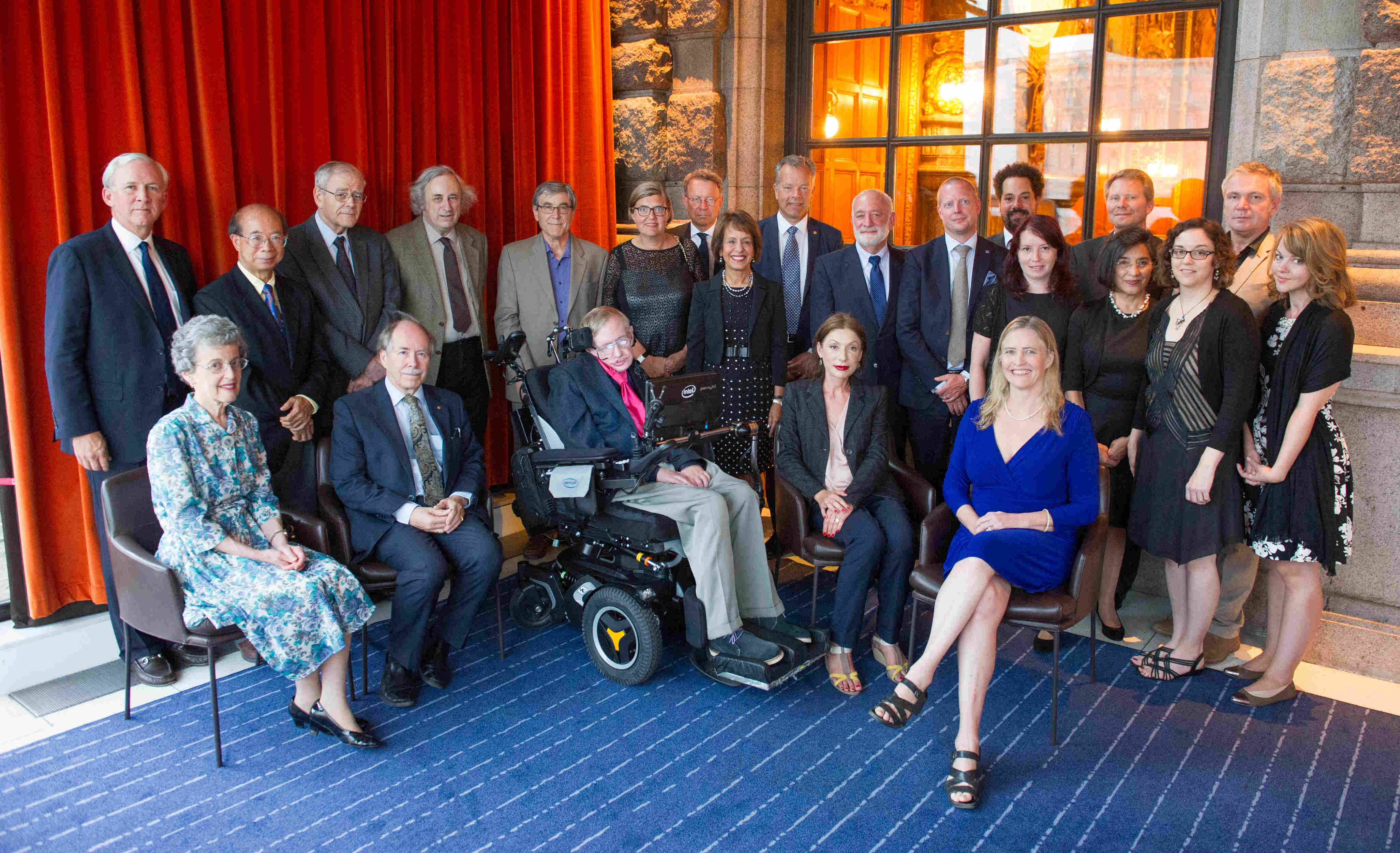}
			{\center Listed left to right.} \\ {\center \textbf{Back Row:} David Routh, Jack Ng, Leonard Parker, Malcolm Perry, Paul Davies, Astrid S\"oderbergh Widding, Anders Karlhede, Carol Folt, Peter Gudmundson, Ronald P. Strauss, Philip von Segebaden, Jonathan Wood, Radka Visanova,  Kevin Guskiewicz, Mayola Espinosa de Mee, Katie Bowler Young, Roger Couhig, Ingrid Smith } {\center \textbf{Seated:} Gloria Parker, Gerard 't Hooft, Stephen Hawking, Laura Mersini-Houghton, Katherine Freese}
\end{figure}

\pagebreak

\pagebreak

\section{Opening Remarks}
	
	   %auto-ignore

\begin{dialogue}

\speak{Peter Gudmundson} I'm extremely honored to welcome such distinguished scientists to KTH as well as to Stockholm. We are all very proud to take part in this event. KTH is a university of science and technology and, in technology or engineering we are the largest and oldest university in Sweden. I myself studied engineering physics in the beginning then became a professor, and finally ended up as president of KTH. So I have studied some physics but I don't know very much about black holes. What I know more about is black holes in economy. We don't have so many of those, we have, of course, some of them but that's another story. We have, as a university, a motto which we think is very nice: it's called science and art. So we are both science and art, mainly engineering art but also art in itself. I think this conference is for science but perhaps also a bit of art. There are beautiful theories and beautiful equations in this area so I think it's a bit of art as well. With that, I will give the word to my colleague Astrid S{\"o}derbergh Widding who is the vice-chancellor of Stockholm University.

\speak{Astrid S{\"o}derbergh Widding} Thank you. Today it's a great honor and a true pleasure for me as vice-chancellor of Stockholm University, which is also co-hosting Nordita together with KTH, to join with my colleague President Gudmundson in welcoming all of you and, in particular, of course, Professor Stephen Hawking. We are extremely pleased and proud to co-host this conference, and that so many distinguished scholars have come here to take part in this historical Hawking radiation conference. Stockholm University has a strong research tradition in the General Theory of Relativity and black holes, and therefore it is also, scientifically, very exciting for us to take part in this conference. We are extremely grateful to the University of North Carolina and it's also a particular pleasure to welcome Chancellor Folt and also her team, of course, Laura Mersini-Houghton for all efforts in arranging this together with Nordita and its director Katie Freese. I am convinced that we are all looking forward to intense, exciting, and rewarding scholarly exchanges during these days to come. Thank you.

\speak{Carol Folt} Thank you. I too extend my greetings. I have a number of colleagues with me from UNC Chapel Hill and it's such a privilege and honor to be here with you. I want to say how wonderful and welcoming we've been treated, and to come in to this beautiful room, which was designed perfectly to facilitate the amazing conversations that are going to take place in the coming week, it's such a fantastic tradition and is the perfect setting for your discussions. 

UNC Chapel Hill, which is in North Carolina, is a comprehensive university that some of you may have visited. It has 14 schools. It is not old by European standards, but it is the first public university in America, graduating its first students in the 1700's. Major research goes on across the campus, and we are very proud of our long history of research in theoretical physics. In particular, our department of Physics and Astronomy was the first faculty home of 
Professor John Wheeler, who, of course, popularized the term “black hole.” 

As a research scientist in the field of Biology, another exciting moment for me was the opportunity to visit the Nobel museum yesterday. While I was reading some of the citations and quotes, I was really taken by a Eugene O'Neill quote that said, ``Happiness is for the
timid, scientists are not timid.'' So this is a place not for the timid, where you have come to debate and think about a forty-year-old puzzle, as Professor Hawking called it. 

I want to say a special thanks to my colleague, Dr. Laura Mersini-Houghton, who is one of the people who spent so much time organizing this special gathering. She told me that she was having telephone conversations with many people about these great subjects of debate. And in a world of e-communication and social media everybody just said: Why are we doing this online? Let's get together, face-to-face, and have a conversation. So I believe this is fantastic that you are moving from Facebook to in-person, face-to-face conversations to move this discussion process forward. We're so pleased and honored to be here. Thank you.

\speak{Katie Freese} Hi, I'm Katie Freese, director of Nordita. Well welcome everybody, I wanted to say a few words about Nordita. I came from Michigan to Stockholm because Nordita is one of the most wonderful, prestigious international institutes for theoretical physics in the world. It was founded in 1957 in a rather interesting situation. CERN, the center for research nucleaire, Copenhagen was competing for it. Well, they lost to Geneva so Niels Bohr decided to create an institute in Copenhagen in 1957 and it was illustrious from the beginning. Very many famous people there, many Nobel prize winners working there for a long long time. This is now a sixty year old place which, in 2007, moved here to Stockholm hosted by the two universities and you've heard from the leaders of these two universities. We have, just so you know, in-house four faculty members, up to twenty Nordita fellows who all get jobs afterward as post-docs and then go on to faculty positions elsewhere, so it's a great place. We're very happy to have everybody here for this wonderful conference and very grateful to Laura for getting this going. So welcome and I'll pass it on to you.

\speak{Laura Mersini-Houghton} You are all here! Thank you for coming! I am extremely pleased, honored, and humbled that we have all of you here, three Chancellors, from KTH, Stockholm University and UNC-Chapel Hill, and the Provost and the Dean of UNC-Chapel Hill. Some of you have flown halfway across the world to be here today. Recently, especially because of the economic crises that the world has been going through, support has been gradually shifting away from science and the universities which are, first and foremost, temples of knowledge. So, the fact that we have these Chancellors here to honor us and welcome us to this conference shows that these universities really stress their goal and attention on the mission of remaining temples of knowledge and discovery. So, we have to come up with the solution to the black hole problem this week. Otherwise.., no pressure!

\vspace{12pt}

There is a story, maybe many of you have heard, about Faraday. The British Chancellor of The Exchequer, equivalent to a Minister of Finance, visited the lab where Faraday was demonstrating electricity. At the end he said: `this is all very impressive, and it sounds fascinating. Most of it is above my head, but what is it good for?' Faraday's answer was: `well I don't know Sir but I'm sure you'll tax it one day.' So, that's another aspect of knowledge, that although the discoveries and breakthroughs are motivated by knowledge for the sake of knowledge, their implications are not immediately obvious. But of course, this is the century of technology and science, thus everything around us is based on these discoveries and developments. 

\vspace{12pt}

I'd like to say a word about a friend and a colleague, Jacob Bekenstein. He was invited here and he would have come. But he didn't... because he died last week. It was sudden. It was a heart attack. He wrote to me only a month ago. He was one of the founding father of black hole thermodynamics, and it's very sad for us that he's not here. 

\vspace{12pt}

We'll start the conference now. Thank you for coming. Thank you.

\end{dialogue}

\pagebreak

\section{Monday, August 24 2015 \\ \textit{\small Convener: J. Ng} }

		\pagebreak
		\subsection{Backreaction and Conformal Symmetry \\ \textit{\small Gerard 't Hooft}  }
				{\small (Slides from the talk can be found at \url{physics.unc.edu/~dnmorse/Hawking_Conference_Slides}.)}			
				
%				\includepdf[pages={1,2,3,5,8,9,11,12,14,16-22,24-}]{./Slides/THooft.pdf}

%		\pagebreak
		\subsection*{Discussion}
				%auto-ignore

\begin{dialogue}

\speak{Ford} Can you say a little bit about whether your idea on conformal symmetry will lead to a viable classical limit. It is well known that if you add quadratic terms, \(R^2\) terms, in to the classical Lagrangian you run in to problems with the stability of flat space and agreement with various tests of general relativity. Can you comment on that?

\speak{'t Hooft} Well I haven't made those studies as extensively as I wanted to as I was involved with another project as well. In principle of course the whole Lagrangian has a classical limit. The anomalies have to cancel and in the classical limit they might not. That's not not totally obvious. In principle there is a conformally invariant theory and it means that scales of things are not what you think they are and you always have to keep that in mind when you consider this theory. My proposal is to put the negative metric gravitello field, which is the cause of the problems that you mentioned because of the negative metric is unstable. But I put it to the other side of the equation. I have the energy momentum tensor of matter, that generates a gravitational field which you get by solving such-and-such differential equations and then that tells you that there is a gravitational field. The \(1/k^4\) propagator suggests that gravity will be screened at very small distances, so much so that it never becomes too strong at the Planck scale and that would be its internal reason for being renormalizable: that gravity would very inevitably shrink at very small distances. That's what this theory would suggest.

\speak{Stelle} I'm going to disagree with you about spin 2 fields. You get rid of the apparent ghost for the \(\omega\) field simply because you have conformal symmetry, that actually removes it from the spectrum so its apparent wrong sign is of no consequence. A spin 2 massive field, however, it's really there and you can't gauge it away; so moving it to the right hand side is just shuffling the cards, you're not going to get rid of the problem.

\speak{'t Hooft} This would have been my reaction too. So you're quite right. Maybe there is a way out along these lines, I haven't figured out all the details. I sympathize very much with what you say.

\speak{Stelle} There are two things I could say about that. One is that the asymptotic safety idea is one that has been pursued, in particular by Niedermaier who has written a living reviews long article with Reuter. The idea there is to use the renormalizable \(R\) plus \(R^2\) plus Weyl squared theory and the asymptotic freedom of the two quadratic curvature terms such that the negative metric states are, in principle, present within the Hilbert space. The coupling to those states can go to zero at large momentum so that you never wake them up basically. That's essentially the idea. The question in any asymptotic safety discussion is: does it really happen? There are violent arguments back and forth about scheme dependence and so forth and for that you would need another whole conference. 

\speak{'t Hooft} The only thing I can say to my defense, I understand what you say is that, in conformal symmetry, familiar notions such as energy and time and distance all disappear so you have to realize that a conformal transformation is quite a big departure from the situation where you can have intuitive ideas about stability, about positivity and so on. All these go down the drain quite a bit, so you have to think twice before you say this can't be right. So I said okay, normally speaking I would say exactly the things that you said, I agree with that, but this is not an ordinary theory. This involves space and time itself, you have to gauge transform to get the right coordinates and then you have to realize that the coordinates I use are not representing ordinary distances and time scales correctly and mass and energy don't mean anything. Now you have to think again what do you really mean when you talk about stability of a solution. That then becomes the nasty question and I don't have a ready answer to it. I'll only say maybe the answer is more complicated than just a straight rejection as one would be inclined to do. Only a few months ago I would have exactly the same answer, say no, this doesn't make any sense because of the wrong sign. But maybe the answer is not that simple.

\speak{Stelle} I can add one other thing which is classical, and it may be a rather dangerous point of view, I understand from Toby Wisman that the classic proof of stability of flat space in general relativity extended to general relativity plus a scalar field still works even if you flip the sign of the scalar field so that you make it a ghost. The largely dispersive nature of gravity causes flat space still to be stable within perturbation theory. If you consider non perturbative theory it's a bad solution. So it's possible that the spin 2 modes here behave in the same way. Whether you want to trust yourself to live in a theory that's only perturbatively stable I think I leave to a philosophy conference. 

\speak{'t Hooft} Anyway, I won't be claiming that I have \textit{the} correct theory of gravity before I have the answer to such questions, and I can't answer your question today. So that must be noted, there is still a lot of work to be done and maybe this is a false alley like so many other alleys.

\speak{Stelle} Also, about the conformal anomalies, if you really insist that all conformal anomalies cancel then I don't have any objection to your having only the Weyl squared term as the counterterm, but you do have to take into account that gravity has its own conformal anomalies because of the Gaus-Bonnet invariant. You have to make sure that that coefficient, as well as everything else in the standard model, that all the other conformal anomalies cancel out. So that's a rather tight constraint.

\speak{'t Hooft} Yes it is very tight, so probably the bottom line is that we have to find a better formalism to build this in so that these things happen automatically and not by brute force. It's going to be a tremendous amount of work and you fail in the end because you forgot something. So yes, but the thing I was very much intrigued about is that if I look at the standard model with this \(\omega\) field added and the other field added and then demand that it has to be on a fixed point in this renormalization group, that's a very interesting demand. It's not totally silly and it's not hopeless, there may be a model to be found that happens to sit at that fixed point with interesting values of these couplings, which are all calculable now by the demand that you have to be at that fixed point. So that makes it, to my mind, an exciting possibility which it still remains to be seen if it makes any sense or not. 

\speak{Parker} The particle, you could call it something like the gravitino, it has a mass term in it and does it also interact with other forms of matter, you know where it has more degrees of freedom because of the spins and I was wondering if it could affect the classical limit of gravitation?

\speak{'t Hooft} I didn't quite understand the question.

\speak{Parker} You know, there is the paradox where if you have a massive graviton that you have this effect on the Mercury orbit and the path of light because of the extra degrees of freedom.

\speak{'t Hooft} I have a massive graviton, but I also have the original graviton. The massive graviton has a mass which could be anywhere between a TeV and the Planck scale, it could be a wide range of possibilities which does not affect the motion of planets or classical general relativistic experiments, of course I would insist that general relativity as you know it today still makes sense in the large scale domain of the universe. Only at very tiny scales something has to go, and of course the black holes we are most worried about are the microscopic black holes. The divergences that we are worried about are divergences at the Planck scale, so there you want to cure the theory, but I don't want to modify the theory at very large scales.

\speak{Parker} I see. Thank you.

\speak{Stelle} It generates Yukawa terms. It generates Yukawa terms so depending on the scale of the mass you either worry about those or you don't.

\speak{'t Hooft} In my case the Yukawa terms will kill gravity at distances much shorter, it would make it \(1/k^4\) instead of \(1/k^2\). So that means that the gravitational force will dwindle beyond that scale and it has been tested until a millimeter or so or a few microns. There this apparently doesn't happen, but who knows what happens beyond that.

\speak{Stelle} I'll talk about some of those things in my talk.

\speak{Mottola} As I recall the conformal anomaly terms, the terms that contribute to the anomaly coefficients, tend to enter with the same sign for both the Weyl squared and the Euler, Gaus-Bonnet. Even in super Yang-Mills, \(N=4\) I believe, there is a still a non zero anomaly. Do we know of any theory which has an exactly zero conformal anomaly? 

\speak{'t Hooft} I don't know. I think that if there is such a theory it should be constructed by methods totally different than what we're used to now to build that. But, as you say, some anomalies have very obnoxious constant signs. I thought there was something about the gravitino 

\speak{Stelle} Pure Weyl squared supergravity has vanishing conformal anomaly if it is coupled to a matter sector with a dimension four gauge group. Funny constraint which doesn't fit the standard model.

\speak{'t Hooft} Thank you for that remark. There is a way to construct such theories but I personally think that is not the theory we really want because it has too much symmetry. The super Yang-Mills, supergravity theories, to my taste, have too much symmetry to be good candidates for the real world. But maybe I'm wrong there and maybe one can turn those theories in to useful theories.

\speak{Mottola} We'll then have to work to break that symmetry spontaneously to reach our world. 

\speak{'t Hooft} To find how to make such a theory will be a big question, yes.

\speak{Ford} I'd like to go back to the question of semi-classical phenomenology a little bit more. It's well known that in Einstein's theory if we write down an equation for gravity waves in an expanding universe that equation is not conformally invariant. In fact it looks, in the transverse trace-free gauge, like a massless minimally coupled scalar field and that fact is used, for example, in trying to predict the gravity wave modes we might expect to see in inflationary models. If gravity is exactly conformally invariant, would you expect to see deviations on cosmological scales?

\speak{'t Hooft} The words I've been using might have been a bit confusing because I'm not claiming a big modification of physics at the scales we are familiar with. The ordinary Einstein gravity should still hold exactly, I don't want to touch that, and that includes many of the cosmological conservations one has in cosmological models. That is still in the domain of physics where I would not expect a major modification of gravity. But when you talk about scale in that case you talk about, again, the relevant scales, the \(\phi\) field divided by the field I called \(\omega\). Those ratios are now the things that you normally call scales. I have no constraint on that. But I have the constraint that, if you disentangle those fields, then if one of those fields goes to zero nothing should happen. And that is a big deviation from what we normally think. But that always means it's information about the infinitely small distance limit of nature. So this dilaton field I introduced, there is no \(1/\omega\) term means that if this \(\omega\) field goes to zero nothing happens. What it really means is that if you go to distance scales much shorter than the Planck scale you will see nothing happening. And that is the piece of physics that we have very bad intuitions about, we don't understand it very well. So this is why your remark is somewhat unusual, but it should not affect physics at very mundane scales. So you can continue doing cosmology the way you are used to. 

\speak{Dowker} Is one of the ambitions for your candidate theory that you should be able to calculate the low energy cosmological constant, the one that we actually measure and observe today, from it?

\speak{'t Hooft} The cosmological constant, I'm not sure I understood your question correctly, but the cosmological constant is one of those parameters which should be completely calculable. The thing that I miss completely and I don't understand is the hierarchy problem, what is it that makes the cosmological constant as small as it is today? That question has not been answered by my theory unfortunately, I wish it had but I have to confess that no, I do not have a good answer as to why this number is so tremendously small. This is a big mystery today, but the remarkable thing about the theory is: small or big or whatever, it's calculable. It should be a number you can work out with a fancy Mathematica program and you will find the first hundred and twenty two decimal places to be zero. Why that is so, there should be an answer to that but I don't have it.

\speak{Dowker} There is no indication about how that might emerge?

\speak{'t Hooft} It's calculable but we don't understand why it is so small and that's because there are too many details of this theory which are totally mysterious.

\speak{ Dowker } One might say that one would expect it to be of order one.

\speak{'t Hooft} Of course, as has been remarked by umpteen people, if you just plug in the numbers you expect the cosmological constant to be equal to one. That's completely wrong, for many reasons it can't be right. So nature chose another way, I'm sure there will be people doing physics as long as we don't have the answer to the question why this is so. We want to have the answer to that question.

\speak{Fulling} I'm trying to understand the second half of the talk and so I have several related questions. First, did I understand correctly that the split of the standard Einstein tensor into \(\omega\) times \(\hat{g}\), is something that is somewhat arbitrary? So there is something like a gauge transformation in a generalized sense and you said that observers outside the black hole should use a different one than observers falling in. So the first question is, why is that so? Why shouldn't any observer be free to use any frame or any gauge?

\speak{'t Hooft} The way anybody would set up a theory of gravitation and other forces is by defining a field as compared to the value it has in the vacuum. So we say that the vacuum expectation value of that field is one, then you do your perturbations. Unfortunately the different observers around black holes disagree about the vacuum state. So then they disagree about the vacuum state when one says `my \(\omega\) field is a field, is 1 or \(1/G\), in vacuum and this is the field I'll use.' The other says `no I'm going to use a different \(\omega\) field because I want my field to be exactly that value but in \textit{my} vacuum not in yours.' So they disagree about which number to take and that disagreement will linger on in the other disagreements that they have. One observer sees that if you throw a pebble into a black hole it will shrink in size because the black hole slowly shrinks in size as well. The other observer says `no, no. I go with the pebble, I don't see it shrink at all.' So they disagree about sizes. Because they do that they disagree about everything else as well.

\speak{Fulling} So that's what I'm trying to understand. I know what a Boulware vacuum is for a scalar field or an electromagnetic field, when you say the Boulware vacuum are you talking about the Boulware vacuum for the \(\omega\) field? Or for every field?

\speak{'t Hooft} For all the fields, as you know, all the particle species that you can make up in your standard model will all be radiated away by the black hole. So they all contribute to whatever vacuum you have, whether its the Hartle-Hawking vacuum or the Boulware vacuum. All these particles contribute so you have to take them all together. Maybe I haven't answered the question sufficiently.

\speak{Fulling} I've always thought that the big contribution that Bill Unruh made was to understand that what is interesting about the Rindler quantization was not the Rindler vacuum, which would be the Boulware vacuum in a black hole theory, but the Minkowski vacuum or Hartle-Hawing vacuum looked at from a Rindler point of view. If you are in an accelerated state of motion you perceive thermal radiation. But there is still a unified physical picture of what's going on. The two pictures are not really in contradiction. The physical state is either a Hartle-Hawking type vacuum or an Unruh type vacuum depending on initial conditions. Why do we even need to talk about Boulware vacuum? 

\speak{'t Hooft} The situation is simple and transparent as long as you don't take the backreaction into account. The fact is that all these Hawking particles actually do carry the gravitational field whether you like it or not, because they're particles they are the source of gravitational fields. In the standard description of Rindler space and, of course, the black hole horizon you don't take that into account properly, you are very much tempted to leave it out. But leaving it out is actually a big mistake. Primarily so because Hawking radiation has basically an infinite spectrum. If you do it naively you just follow the rules and you forget about backreaction you find that the Hawking spectrum, in fact the Boulware vacuum, cannot exist because it is an infinite departure from the states of the Hilbert space because it diverges at the horizon. This divergence means that Hawking particles would carry away an infinite amount of energy and mass and then would generate an unlimited gravitational field. Something is completely wrong if you do that, it's obvious that you have to somehow handle the backreaction more carefully. That's why this meeting is very important. This is about the backreaction of Hawking particles. We have to understand the backreaction. It's there where you discover that particles carry an enormous amount of gravity with them so that spacetime behind these particles looks totally different for that observer. But the ingoing observer doesn't see those Hawking particles so they disagree, they don't disagree about the matter tensor so much at our side of the horizon but they disagree what it is at the other side. If you go beyond that curtain of Hawking particles, what is the matter tensor there? Now they both disagree because one says `no, you have to take those particles in to account, they are the source of gravitational fields. The matter that will obey Einstein's equations will be different behind than it was before.' But the other observer says `no, no, Hawking particles are just mathematical artifacts of you sitting in the wrong frame. I'm sitting in the right frame, I see no Hawking particles, so I also see no change in the metric tensor.' So now you see where the clash comes, the clash is not just general relativity and coordinate transformations and as such the clash is in a more basic question, what does the metric tensor mean? What does it stand for? And how come the different observers must see a different number there? That's not just a general coordinate transformation, it's something more, and that's what I'm interested in. 

\speak{Fulling} But a covariantly renormalized stress tensor, the expectation value of the stress tensor in a particular state, is absolutely defined. It doesn't depend on what Fock space you are using. Now I agree I do not know how to take account of backreaction in detail.

\speak{'t Hooft} If you do that, people have tried to do that and usually find that energy momentum tensor, the vacuum expectation value, is very small or zero for Hawking particles. But you also know that can't be right because Hawking particles can eventually carry away all the mass and energy of the black hole. 

\speak{Fulling} It depends on where you're looking. 

\speak{'t Hooft} They must carry lots of energy and momentum eventually. So that number cannot be as small as some people claim it is. They claim it is small because there are these particles which go in to the black hole which balance out in a sense so that it looks very small. But you also know very far away from the black hole that the particles did come out, eventually carrying away all the mass. So you can't have it that cheaply. There is something that needs to be further explained and worked out in a proper theory, how do we take into account this backreaction of the matter? 

\speak{Rovelli} I have a very naive question, it's probably related to what Stephen was saying just now, I realize there is something I got confused about. At the very beginning you sort of build the theory in two steps. First you just replace \(g\) with the two fields \(\omega\) and \(\hat{g}\), then you said that \(\hat{g}\) knows about conformal structure, the light cone structure, which of course is obvious and \(\omega\) knows about the scale. I would say, at least at this stage, that \(\omega\) does not know about the scale, only the product \(\omega \hat{g}\) knows about the scale, because of the previous interpretation and because what we do in GR is not to just define the scale but read out the scale from the behavior of matter, how it couples to the gravitational field. I guess the question is, is what I said right or not and does it change when you actually modify the theory by changing the Lagrangian? So are you giving to \(\omega\) a physical interpretation related to scale which only appears at the second point, or are you making a hypothesis that \(\omega\) knows about the scale just by itself and not together with \(\hat{g}\) as I would suspect at the first step?

\speak{'t Hooft} All this is very much a question of gauge fixing because now we have added a new local gauge symmetry, called conformal symmetry, which has to be fixed somehow by choosing your gauge. When I looked at this the first time I really thought about gauge fixing by just gauge fixing \(\hat{g}\). Say, for instance, the determinant of \(\hat{g}\) is one. That's a nice way to fix the gauge. Once you say that then all the rest is done by the \(\omega\) field and then \(\omega\) fixes the scale, so that remark was a remnant of that insight. Now I no longer am so much interested in fixing the gauge by saying the determinant of \(\hat{g}\) has to be one, the gauge fixing can be done at any later stage any way you like, it doesn't really matter much how you fix the gauge.

\speak{Rovelli} Right, if you gauge fix \(\hat{g}\), determinant of \(\hat{g}\) equal to one for instance, then of course the scale is coded into \(\omega\). If you don't it's not, is that right?

\speak{'t Hooft} Right right. So basically you have a theory where the coordinates mean even less than they did before. So the coordinates themselves are no longer insisting what the scales are. You need these coordinates to renormalize the theory, you have to go to very small distances and then make a subtraction. But if you make that subtraction, matter should be free of anomalies so there should not be a \(\beta\) function associated with that because then you're in trouble. This is now the main ingredient of this theory.

\speak{Whiting} You suggested during your talk that if you submitted a paper with this \(\omega\) theory in it it might well not get published. What do you think it would take to have something that was physically calculable from the things you have described today so that we could say, well, this doesn't measure up or this does. 

\speak{'t Hooft} I think the second half of the talk contains the things where I deviate from standard physics. One of the big deviations is what happens when this \(\omega\) field which I introduced, when it goes to zero. I claim nothing should happen because in the Higgs case also nothing happens when the Higgs field goes to zero and that’s a big deviation from standard physics. Now I'm proposing something physically nontrivial. I think at least whether it's right or it's wrong at least its something one can discuss. But I'm proposing a physical change, a change in our view of what's happening physically. What predictions it will eventually give, nobody knows. I made one prediction which is that everything is calculable, next question is to calculate it and I've not succeeded in doing that, that's much more difficult. 

\speak{ Whiting } Does that mean that there must be some other fundamental constants other than things like the speed of light and the gravitational constant?

\speak{'t Hooft} No, just algebra. I think the constants will come by postulating the algebra of the standard model particles and gravity interacting with them and so on and so forth. All this is algebra. So you give me the algebra and I'll calculate everything else. 

\speak{Mottolla} As one more followup to these questions, your example of the electroweak theory was very nice but we know the scale there was set by the Fermi constant and the scale at which new physics comes in is therefore set by that. So I think this was asked already but I wasn't clear on the answer, are you suggesting this conformal symmetry, the new symmetry, will come in at the Planck scale, as the Newtonian constant sets the scale just like the Fermi scale, or are you suggesting a different scale?

\speak{'t Hooft} No, well, you could ask a similar question for the electroweak theory, you could ask is the electroweak theory locally gauge invariant at the Fermi scale or at what scales? Of course the only correct answer is that it is locally gauge invariant on any scale. So conformal invariance happens at all scales. The fact is, of course, that we don't recognize it today in ordinary physics, everything has definite size. That is because this field I called \(\omega\) is a dominant and important field, its value determines everything and so the sizes of atoms and sizes of people and sizes of planets all compare to this \(\omega\) field. If I vary the \(\omega\) field people look smaller or larger but in ordinary physics it's irrelevant, you just say well this is just a dummy nonsensical field, I put it equal to one and then everything is fine again and then you have a definite scale. But it becomes particularly relevant at the Planck scale where you want microscopic black holes and also we want to understand these divergences of gravity and you want to understand where the standard model comes from. All these questions have not been answered today and so we are searching for answers there and we all believe that the answers are somewhere near the Planck domain. How near is again a question of discussion, some people think the Planck domain could be just across the LHC, then that would be marvelous if true but I don't believe it. Anyway, the question becomes completely irrelevant at the Planck scale just like in the electroweak theory the question of local symmetry became irrelevant at the Fermi scale.

\speak{Mottolla} I agree but a question I guess that we'll discuss much of the rest of the week is how something at the Planck scale, new physics, could help us with large black holes?

\speak{'t Hooft} I don't believe it helps us much with large black holes. I think large black holes don't need quantum mechanics for their existence and understanding their physics. But for large black holes in principle, you know large black holes emit Hawking particles but emission is so tremendously weak that nobody can ever see it. It would be much much weaker than the cosmic background radiation that they are submerged in anyway. So it would be completely manageable in all reasonableness, so for that reason the black holes at the centers of galaxies or black holes which arise from collapsing stars, for all those the Hawking radiation is highly irrelevant. It's there in a formal sense, mathematically it's there, but it's not important.

\speak{Mottolla} So in that regime do you expect we can use semi-classical gravity?

\speak{'t Hooft} Yes. 

\end{dialogue}  

		\pagebreak
		\subsection{Backreaction of Hawking Radiation and Singularities \\ \textit{\small Laura Mersini-Houghton} }
				{\small (Slides from the talk can be found at \url{physics.unc.edu/~dnmorse/Hawking_Conference_Slides}.)}			
%				\includepdf[pages={1-3,5-}]{./Slides/Mersini.pdf}
	
%		\pagebreak
		\subsection*{Discussion}
				%auto-ignore

\begin{dialogue}

\speak{Davies} I'm sure its coming up again and again this week, but people talk about the information paradox. Going back to [classical general relativity] there was never a paradox because there was a singularity and information could disappear in to the singularity, there was no reason to suppose that it would come out and so everything was consistent. Of course you could argue that if you have a fully quantized theory that singularity would be replaced by something else, it might be replaced by another spacetime region, an asymptotically flat spacetime region. The problem about unitarity is it's got to be unitary on the whole superspace or whatever you want to call it, there is no reason it needs to be unitary in our cosmic patch. So I've never understood why this is fundamentally a problem. I can see that that may be the wrong picture.

\speak{Mersini-Houghton} In that approach you have just offered one way of getting around the information paradox.

\speak{Davies} Right, you never used to worry about information because you defined it at the singularity. Everyone knew it wasn't really lost. Stephen wrote a paper in, what, 1978, and it seemed that issue was un-contentious. I can understand that some people might want to replace the singularity with some more complex thing.

\speak{Mersini-Houghton} For me the hard part is how can anyone take the existence of singularities seriously? We had this problem in cosmology, that the universe was supposed to start with a singularity. That was a mathematical finding, nobody really thought that physically there was a singularity there.

\speak{Mersini-Houghton} [Responding to a question about the horizon finder experiment] I haven't paid attention to the details. I heard about it this May actually, from Ferreira, but it's called the Horizon Finder. I should have looked it up in more detail, we can check it later. 

\speak{Bardeen} Its a microwave interferometer, long baseline interferometer, sub-millimeter experiment which will resolve the black hole in the center of our galaxy. The hope is to be able to see the shadow of an event horizon.

\speak{Mersini-Houghton} Do you know when it will be launched?

\speak{Bardeen} It involves coordinating very short microwave radio telescope interferometry on long baselines between Hawaii, the South Pole, Chile, a couple of places in Europe I think, in California, and so on. A lot of different baselines to be able to say something about the image of the black hole in the center of our galaxy. When implemented it will have the resolution to resolve things on the scale of the event horizon, in particular one can try to see if there is a shadow. Of course it is complicated by the fact that you have accreting gas falling in the the black hole, swirling around in an accretion disk perhaps, which is radiating. It involves some modeling, but for a large range of models one should see some sort of shadow along lines of sight that would be coming from the inside of the black hole.

\speak{Davies} Laura, I would just like to make what I think will be a helpful comment about your presentation which is that this discussion of temperature versus stress-energy-momentum tensor. Out at infinity there is of course a very simple relationship between them, it's a thermal flux and we know what \(T_{\mu\nu}\) is for that. In the region of high spacetime curvature there is a complete breakdown between the particle language and the \(T_{\mu\nu}\) language. For Einstein's theory it's only \(T_{\mu\nu}\) that counts. Particles are defined globally on the whole manifold, \(T_{\mu\nu}\) is defined locally and that is the quantity that is going to determine what happens to the collapsing star, which is what I think you were saying. It's worth emphasizing that. It's only out at infinity that you get a simple relationship between the two.

\speak{Mersini-Houghton} I completely agree! Also we need to emphasize that you can speak of temperature after you have thermalized.

\speak{Davies} That's right.

\speak{Mersini-Houghton} Only then it makes sense to have this parameter to describe the spectrum. 

\speak{Whiting} Laura, can you go back to your last transparency again, the one where you had your conclusions? Here it is saying that the collapse reverses before the horizon or the singularity forms, and yet the diagram you have there shows both the singularity and an horizon. I would be much more interested in knowing what comes out of your calculations in showing what's not on the right hand side of your diagram. You say that you can't complete the evolution but you must have up to a certain point.

\speak{Mersini-Houghton} It's really very very close to this region here where the bounce occurs, but the numerical program breaks down very close.

\speak{Whiting} But you must at least have that sketched in curve.

\speak{Mersini-Houghton} Definitely you can see the turn around the bounce you can see that it bounces. But what you can't see is does it explode after that? Does it have shell crossing? 

\speak{Davies} So if I take a point, say where \(r = 2M\), which is just where the turnaround is occurring, if I take a null line in from there it hits the singularity. Now in your evolution, can you construct that null line?

\speak{Mersini-Houghton} Yes, through the parameter \(\theta\), through \(\theta\) the expansion parameter. We did check that the singularity will never be there because \(\theta\) never becomes zero. It never crosses zero.

\speak{Davies} So that means something is already different in your calculation from what you show in that diagram, up to the part that you can.

\speak{Mersini-Houghton} Yes, that was meant as a sketch to show the bounce of the star.

\speak{Whiting} I understand, but I really want to know what you do see.

\speak{Mersini-Houghton} We check the expansion parameter and it never crosses zero so I know there will never even be a trapped surface, not even a temporary trapped surface. 

\speak{Whiting} So if we take the the domain computed and look at the Penrose diagram for that.

\speak{Mersini-Houghton} I should have stopped this diagram, this hand drawn sketch really here.

\speak{Whiting} But you do see the null curves that go from your last surface, you do see the null curves that go right into \(r=0\)?

\speak{Mersini-Houghton} Yes, that's where the boundary conditions are, of course, we can follow the evolution inside all the way to \(r=0\). Yes. 

\speak{Whiting} I'd like to see the Penrose diagram for that. 

\speak{Mersini-Houghton} I wish someone would do it. I can't do the numerics. I want to know does this star explode or does it just hang around forever like that? In reality I really don't know what happens after the bounce. Look at how soon after the bounce has occurred the program breaks down. We know for sure that the star bounces, it turns around, the collapse first slows down and then it stops, but then Harold's program breaks down. So, I really don't know how the star evolves after that.

\speak{Bardeen} Laura, you say you look at the conservation laws, energy and momentum conservation, in the interior of the star right? But I believe that there is a serious problem with global conservation in that you have a sudden transition from outgoing positive energy just outside the star, a very large energy flux to overcome the gravitational redshift, and then a very large negative energy flux going inward. Well the energy flux is positive, it's positive energy going out outside and negative energy going in inside. Now that means that there is a very large discontinuity in the radial stress at the surface of the star or wherever that transition occurs. That violates momentum conservation violently. At the surface of the star the radial stress has to be continuous, you can't have a step function. That means then that since you have positive pressure outside, in the outgoing positive energy, outgoing radiation, negative pressure inside, there's a huge force acting on the star inwards due to the imbalance. Which I don't think you've taken into account.

\speak{Mersini-Houghton} Just so that I understand, you are talking about the fuzzy region outside, near the surface of the star, where the pair creation is taking place. If we were to solve for energy conservation there in this picture of a flux of positive energy particles going away from the star and the ingoing flux of `negative energy` going in to the star.

\speak{Bardeen} Yes, so energy conservation is okay, it's momentum conservation that I believe is not okay.

\speak{Mersini-Houghton} If we were to solve for stress energy conservation outside the star, which I haven't, since I'm only concerned about the interior dynamics of the star for. You think we have a problem? For all practical purposes in this approach all I worried about is modeling radiation crossing the surface of the star going in. Inside the star everything is conserved because I explicitly solved for energy conservation. But outside, in this picture, where these particles, this flux, is produced you are saying that if I were to solve in this fuzzy region where the pair creation is occurring then I would find out that the radial component of the stress-energy tensor would not be conserved?

\speak{Bardeen} You'll find that your assumption about what you're assuming about the energy-momentum tensor going from the interior to the exterior.

\speak{Mersini-Houghton} No, the other way around. We assumed the ingoing flux crosses the surface from the exterior towards the center of the star.

\speak{Bardeen} Well, either way. It's inconsistent with momentum conservation. 

\speak{Mersini-Houghton} So the picture of pair creation, where a positive energy particle flies off to future infinity and the `negative energy' part loosely speaking, an ingoing negative energy flux crossing the surface of the star, you're saying that's incorrect because it breaks the radial component?

\speak{Bardeen} That's inconsistent if you insist that it happens over an infinitesimal distance just outside the star, and if you insist that the magnitude of the energy momentum flux be much much larger than what I think any reasonable estimate would be which is of order \(M_{\text{Planck}}^2/M^4 \) which is what you'd expect for this sort of Hawking flux if it's not produced right at the surface of the star. What I'm saying, your assuming it's produced just outside the surface of the star is inconsistent with momentum conservation.

\speak{Mersini-Houghton} There are two things there: one is if I assume that this pair creation occurs very close to the surface of the star, which I think is what you would argue in favor of, is that correct? And the second question was how big is the magnitude of the luminosity at the surface. If I understood this is what you are trying to ask me\ldots

\speak{Bardeen} So you are extrapolating back from future null infinity, the only way you can get a large backreaction if you say that that energy flux is coming, you know positive energy going out, is coming from very close to where \(r = 2M\). Because of the gravitational redshift the energy flux has to be huge and it can be much larger than, it can cancel, the very small \(M_{\text{Planck}}^2/M^4\) by correcting it by an enormous redshift factor, going backwards in time it's a blue shift factor. Now the problem with that is that if you have then a sudden transition from positive energy density outside the star and negative energy inside, for your model of the radiation that also implies a positive radial pressure, radial stress, outside and a negative radial stress inside and there's a delta function in the gradient of the pressure which is not balanced by anything.

\speak{Mersini-Houghton} Okay I'll break this in three bite sizes. Is this pair creation occurring very close to the surface of the star? Not necessarily! We have had this discussion before, the region where particle creation is occurring is that fuzzy region which can be anywhere outside the surface of the star all the way up to \(3M\), which I think is what you believe is happening and what Don Page believes. It's a fuzzy region, it doesn't make sense to talk of particles or where they're created as Paul mentioned, just of the stress-energy tensor.

\speak{Bardeen} The magnitude of the negative energy flux into the star depends critically on how close.

\speak{Mersini-Houghton} That was your second question, what is the luminosity at the surface?

\speak{Bardeen} At what radius do you assume, the assumption about the radius at which there's a transition from ingoing negative energy to positive outgoing energy. If that occurs at \(r = 3M\) or something then there's not a large redshift correction and the Hawking flux, which is very low, means that the negative energy flux going  inward is very small, order \(M_{\text{Planck}}^2/M^4\), but that has a negligible affect on the dynamics of the star. The only way you get a significant affect on the dynamics of the star is if you assume that transition occurs infinitesimally close to the surface of the star when the surface of the star is infinitesimally close to \(r=2M\).

\speak{Mersini-Houghton} No, no, I understand the rest of how, what the implications would be, but the three problems you are talking about is, you need the particle creation to be, which really you should be speaking in terms of stress energy because only that makes sense, you need that flux to be really close to the surface of the star. Your second claim is that you need the luminosity of that radiation to be really large, like infinite, to obtain any significant effect, correct? 

\speak{Bardeen} You need a huge energy flux, negative energy flux which is very large compared to what you would expect dimensionally for a quantum process.

\speak{Mersini-Houghton} [Pointing at the slide] So this is the boundary condition for the flux luminosity. The redshift, \(F(s)\), is evaluated at the surface and \(U\) is the stellar material velocity evaluated at the surface. 

\speak{Bardeen} I think what you basically assume is that the radiation is produced right at the surface of the star, right?

\speak{Mersini-Houghton} No. \(L_S\) is just the luminosity at the star's surface which is just the one you'd obtain, the redshifted version of what you would obtain for Hawking luminosity at infinity, the one which gives the surface temperature. So that is an input. There is the consistency condition that this surface luminosity is the one that would give you the correct expression for Hawking luminosity at future infinity. 

\speak{Bardeen} The point, then, as the surface of the star approaches there is a very large redshift of the radiation going outward, which means that, compared to the Hawking luminosity, you need to have a luminosity just outside the surface of the star which is larger to compensate for the redshift. Now when the surface of the star gets very very close to \(r=2M\), because the redshift is infinite at \(r=2M\), you can have an arbitrarily large energy flux and pressure and energy density just outside the surface of the star. But what I'm saying is that in order to get significant backreaction on the star from the negative energy going in that has to be so large that you have an unacceptable discontinuity in the radial stress because outgoing radiation streaming outward with positive energy has a positive radial stress, equal to the energy density and the energy flux, while the negative energy going inward has a negative radial stress and energy density. That means then that there is a delta function gradient in the radial stress which implies a huge force on something.

\speak{Mersini-Houghton} I understand the radial stress concern, but don't you have the exact same problem in the traditional picture of vacuum pair creation outside the star? In this problem I'm not concerned at all what is happening outside, assuming the typical picture to be correct. All I care about is that there is this flux coming inside the star and we are only studying the interior of the star. But don't you have that exact same problem in the usual picture of Hawking radiation in terms of pair creation?

\speak{Bardeen} No. Because it's not coming from the surface of the black hole. What I'm going to argue in my talk is that it's coming from \(r = 3M\) where the redshift is modest, it's a factor of 2 or so, and therefore the energy flux that you have when you make the transition from positive energy going out to negative energy flux going in, for one thing its not concentrated over a small distance it's over a range of order \(M\) in radius and it's also not very large.

\speak{Mersini-Houghton} These are different questions. The flux in my approach is not arbitrarily large either, and it is not produced at the surface. The magnitude of the energy flux or the radial pressure is one question, conserving it is a different question.  So whether particle pair creation happens at \(r=2M\) or \(r=3M\) or \(r=5M\) doesn't matter: if you have got a problem with the conservation of pressure you have it at \(2M\), \(3M\), and so on.

\speak{Bardeen} So energy conservation is one issue and that requires that the energy flux behave in a certain way which takes into account the redshift. You also have to have momentum conservation.

\speak{Mersini-Houghton} Yes absolutely. As I said, we didn't worry about the exterior of the star because we took the standard picture, the pictorial way of it being vacuum pair creation.

\speak{Bardeen} You have to be able to argue that nothing is going drastically wrong elsewhere when you're looking at just a certain part of a problem.

\speak{Mersini-Houghton} No no, I'm not saying it's going right or wrong, I'm saying that we didn't even bother with the exterior because we have no reason to think the standard picture in the exterior is drastically wrong. We haven't solved for the exterior. We just assumed the typical picture of vacuum pair creation where you have one particle going in, the other flying away. I understand what you are saying, in that picture you would have a radial pressure build up because you wouldn't satisfy the total stress energy conservation. In that case all you have outside is vacuum, the only thing you have is radiation. That is the only thing you have to conserve. I haven't studied the exterior of the star, because I know that it has already been studied, why is that a problem? So placing this radiation at radius \(2M\) or radius \(3M\) does not make a difference when it comes to the conservation law. It makes a difference on the magnitude pf radial pressure, how big it is or how little their difference is, but it does not change the violation of the conservation law. It's two separate questions. 

\speak{Bardeen} No, you can construct perfectly smooth solutions satisfying momentum conservation.

\speak{Mersini-Houghton} So the one that satisfies conservation: are you saying that the stres-energy tensor of radiation should not have this form for the partner that goes inside the star? 

\speak{Bardeen} Well that's another issue but I'm giving you that.

\speak{Fulling} Could I try to clarify this a little bit. I think that the problem comes from, if you'll go back to your picture of the negative radiation coming from the surface of the star. You have a region, if I understand correctly, you assume or you assert that pair creation occurs on a surface which is the surface of the star.

\speak{Mersini-Houghton} No no, that was a confusion that we really fought hard for six months. I absolutely do not assume that pair creation occurs on the surface. It's \textit{vacuum} pair creation. It occurs outside on empty space.

\speak{Fulling} If that is the case then there will be a large region, a four dimensional spacetime region in which this pair creation is taking place. There will be positive flux going outward and negative flux going inward in the same place, they aren't separated into two sides.

\speak{Mersini-Houghton} Yes, that's exactly the picture.

\speak{Fulling} If you already accept that then a lot of the disagreement I think disappears. That isn't what some of us understood you were saying.

\speak{Bardeen} The thing is, if that doesn't occur right at the surface of the star\ldots

\speak{Fulling} It occurs just outside the horizon and all the way out.

\speak{Bardeen} Then the negative energy coming in is not enough to have any real effect on the dynamics of the star, that's the issue. Either the pair creation is concentrated in a thin layer at the surface of the star, so there is a huge gravitational redshift when the radiation propagates out, or it is spread out and in that case there isn't a huge gravitational redshift and in that case the negative energy coming in is much less than she assumes and has no effect on the dynamics of the star. 

\speak{Mersini-Houghton} So there are two issues here, stress-energy conservation and the magnitude. How big it is. That is I think what Jim is asking.

\speak{Fulling} Jim, isn't the argument that goes back to Hawking in 1975 by energy conservation the total energy flux over the horizon must balance what comes out and therefore the apparent horizon must shrink?

\speak{Bardeen} Because of the gravitational redshift the magnitude of that flux is very small, of order \(M_{\text{Planck}}^2/M^4\), at some radius which is not close to \(r=2M\).

\speak{Fulling} I think we're arguing the same side of the question. I agree with you, that is the correct picture. What I wanted to recall was that in the 70's we had detailed calculations of the stress tensor for the two dimensional model of the Schwarzschild solution and it didn't take into account back reaction but it gives a very nice consistent picture in which the conservation law for the stress tensor reduces to ODE's in which in each case there is a source which is the curvature scalar. It has nothing to do with the matter, it comes from the curvature scalar. So something is happening all the way out, not just inside the star or at the surface of the star, and nothing much is happening at the early stages where the curvature is not yet big.

\speak{Mersini-Houghton} I made that point too, that most of it occurs as you are getting close to the horizon. 

\speak{Davies} The mischief comes because people insist on talking about particles when \(T_{\mu\nu}\) is the only relevant thing. Forget particles. They're not well defined. 

\speak{Mottolla} Can I just ask a question Laura, do you or do you not have a conserved stress tensor?

\speak{Mersini-Houghton} I do in the interior.

\speak{Mottolla} But not across the surface?

\speak{Mersini-Houghton} No, I didn't bother because I took the standard picture for the outside. All I care is that from the standard picture there is a flux coming in.

\speak{Mottolla} I don't want a picture. I just want to know, is it conserved everywhere?

\speak{Mersini-Houghton} No, I did not study the exterior.

\speak{Mottolla} You didn't study it? Or it is not conserved.

\speak{Mersini-Houghton} I did not study it. I wouldn't go out of my way to choose a picture where the stress energy is not conserved.

\speak{Mottolla} I just cannot follow the equations so I'm just asking.

\speak{Mersini-Houghton} Let me describe those equations in three sentences. Assume everything in exterior is exactly as has been calculated in the 70's, as in the `bible' of the field, Birrell and Davies. Take that calculation and worry only about the part of the flux that goes in the interior of the star. That part is new, that wasn't studied before. That's where the interesting stuff lies. You say whatever happens outside was studied before, we know the answer, now let me sit in the interior of the star and find out what happens to the star if I have the flux coming from the outside through the surface. 

\speak{Mottolla} But I think the issue is that that's supposed to be continuous, and if it's continuous then you have conservation everywhere and there isn't a problem. Jim, I don't understand where this discontinuity comes from unless you put it in.

\speak{Bardeen} No, the discontinuity comes from the fact that a radial flux of radiation doesn't just have an energy flux, it also has necessarily an energy density and radial stress.

\speak{Mottolla} Right, now does she put that in or not?

\speak{Bardeen} Its the discontinuity in the radial stress that raises the problem with momentum conservation.

\speak{Mottolla} But Jim, it depends on what her equations do. If she's put that in properly then the conservation is there. But you're saying, or tell me if I'm wrong, that she has not put that in properly.

\speak{Bardeen} Well she, I think, did not check energy momentum conservation outside.

\speak{Mersini-Houghton} No because I know its been done before and it is conserved. Those are the seminal papers. All I took is that flux, that partner in the pair creation, that's all I care about. If we are in this room and there is something coming from the door and someone has already worried about conserving stuff outside the door, why would I care about what's left outside? All I care about is what flux is coming in this room.

\speak{Bardeen} In the rest frame of the star, how big the energy flux is depends critically on where you start the negative energy flux from and in order for it to be big enough to have the absorption of that negative energy flux by the star impact the dynamics of the star appreciably, it has to come from infinitesimally close to the surface of the star.

\speak{Mersini-Houghton} How big it is and can it stop the collapse, that is a totally different issue, it's a question of magnitude not conservation. Which is the reason why I bothered to solve the same problem with the very simple analytic approach because here we can't hide behind numerics and say this is what the plot looks like. Here we can follow the story. It is very easy to see whatever the magnitude of this radiation that goes in is: you will always find a bounce solution to this Friedman equation no matter how small or big the incoming flux is. \(a\) is the scale factor of the interior metric in this very simple approach, so you have a `closed universe' inside your star and you ask `can I find a bounce point for this metric?' It's only three ingredients that define the evolution of the collapse, it's a very easy problem: curvature, matter, radiation. And the answer is that there's always a solution no matter what your radiation flux is, how big or how small. But there's always a bounce solution!

\speak{Misner} My take away from this is that, according to Jim, if you're going to get radiation going out and generate it from a region near the horizon, as the surface of the star gets near the horizon, you are going to generate radiation going out. You want to get a desired result at infinity, and because the redshift that means you're going to have huge amounts in the rest frame of the star's surface. Counteracting that, he says, there will be a negative flow of radiation into the material and that's accompanied by essentially a negative pressure. Now it's well known that with negative pressure you can avoid a singularity and get a bounce. That occurs even in cosmological solutions. If you want a bounce it doesn't help to make the equation of state harder and harder. You have to get negative pressure to allow a bounce. So it appears to be what's happening is that a moderate amount of outgoing radiation, if it's generated near the horizon as the matter surface gets near the horizon, is going to be huge there and the counterbalancing negative energy radiation pouring into the star is going to be huge and if it is big enough it will produce a bounce sooner.

\speak{Mersini-Houghton}

\speak{Bardeen} Yes, so it needs to be huge and then you're violating momentum conservation at the surface. 

\speak{Ford} I would like to clarify a little bit more the issue of the scale at which this bounce occurs. Normally we would expect if we were doing backreaction with a quantum stress tensor that you would get a large effect on the geometry only at the Planck scale. If I understand correctly you are talking about a bounce at a much larger scale, is that correct?

\speak{Mersini-Houghton} Hm\ldots Yes for the masses that we checked, but the mass parameters that we managed to program were from 4 Planck units to 80 Planck units, so it's not a very firm yes.

\speak{Ford} So in your numerical study you really are talking about close to the Planck scale, you are not really talking about large stars at all then. So it's not so clear that you would get a bounce with an astrophysical sized star, is that correct?

\speak{Mersini-Houghton} Well, 80 Planck masses, that's the closest one can get to astrophysical size with numerics.

\speak{Ford} I would still call that as pretty close to the Planck scale.

\speak{Mersini-Houghton} Sure, Steve promised me for a year that he was going to do this for me, but he hasn't. 

\speak{Davies} Can I just add to the mischief here by saying, when people talk about where do the pairs get created, where do the particles come from, I keep wanting to say that if you have an excited state of the hydrogen atom it will decay by emitting a photon and the backreaction is that the electron gets returned to its ground state. We could spend hours saying `does the photon come from the surface of the electron, or the surface of the proton, or halfway in between, or three radii out?' It's all nonsense.

\speak{Stodolsky} In particle physics we have the concept of the Formation Zone, which is the distance it takes to create a particle This is a very familiar concept and it means that, for example, a high energy photon hitting the nucleus produces a pion and also it produces the pion over a distance which is much bigger than the nucleus. This just follows from the Lorentz factor which is the inverse of the Compton wavelength of the pion. There are experimental manifestations of this. So I agree that you guys should be careful talking about where particles are coming from, especially when you have these enormous redshifts and so on. This an established experimentally proven concept. My understanding is that we are always working with locally defined energy momentum tensor quantities which are defined at a point and we have forgotten the uncertainty principle. So watch out. For example, to define a field at a point you need coherence over different energy states, or different momentum states, as I'll explain in my talk on Friday. You have some kind of coherence between the different multiplicities, there are some phases in there which are playing a role. I don't understand this stuff well enough in detail to say more about it, but you should be aware that you are making some very strong assumptions.

\speak{'t Hooft} I'm a bit confused about the arguments about the energy-momentum tensor. I heard someone say that you shouldn't talk about particles but you should talk about energy-momentum tensors, but the energy-momentum tensor is also an operator. So if you haven't got it diagonalized in the frame that you are looking at then you don't know exactly what spacetime you are looking at because your spacetime metric depends on your energy-momentum tensor. So there are all these ambiguities which confuse the whole issue, which is why I thought that we have to introduce extra symmetry notions to say that you have the option to transform this spacetime into that spacetime. You are discussing the same physics but you are using a different metric. I think one of these days we have to go in that direction to understand fully what we are doing. For instance, in one of the pictures that you showed, Laura, there was a picture where you have this pair creation happening at this point zero but the flux that goes in is negative, right? So that should balance out with the flux at the positive side so there is nothing at that zero, not even a purple line, it is just going more like that. But then that should also, possibly but not obviously, modify your Penrose diagram. I don't believe why there should be a singularity in your Penrose diagram. I think that is something from the more classical picture, that there ought to be a Penrose diagram for a black hole but I think also that thing is an operator, not just something that you can draw on a piece of paper just like that. And the altered picture will probably be that you need the trivial Penrose diagram if you want to get some categorization for all states in Hilbert space, but even that could be problematic. So we have to learn that we have to modify and extend our theory to get the proper language to describe these things. The classical language doesn't seem to work very well to me. 

\speak{Mersini-Houghton} Thank you. I completely agree. I wanted to talk about the geometry and I didn't because your discussion about the conformal field and particle creation, but absolutely, one has to find out what the geometry of the interior of the star is and solve for these functions \(\phi\), \(\lambda\), and \(R\). That depends on what extra stuff comes in inside the star.

\speak{Whiting} I wonder if I could draw something on the board since we have a board here? So I'd like to hear more about what's done in high energy physics will fit in to this picture.

\begin{figure}[h]

\centering
\begin{minipage}{0.45\textwidth}
\centering

\begin{tikzpicture}
\draw[line width=0.3mm] (-3,3) -- (0,0);
\draw[line width=0.3mm] (0,0) -- (3,3);
\draw (-1,1) -- (1,3);
\draw (1,3) -- (2,2);

\draw[line width=0.05mm] (0.2,0.2) -- (-0.8,1.2);
\draw[line width=0.05mm] (0.4,0.4) -- (-0.6,1.4);
\draw[line width=0.05mm] (0.6,0.6) -- (-0.4,1.6);
\draw[line width=0.05mm] (0.8,0.8) -- (-0.2,1.8);
\draw[line width=0.05mm] (1,1) -- (-0.0,2);
\draw[line width=0.05mm] (1.2,1.2) -- (0.2,2.2);
\draw[line width=0.05mm] (1.4,1.4) -- (0.4,2.4);
\draw[line width=0.05mm] (1.6,1.6) -- (0.6,2.6);
\draw[line width=0.05mm] (1.8,1.8) -- (0.8,2.8);

\draw[line width=0.05mm] (-0.2,0.2) -- (1.8,2.2);
\draw[line width=0.05mm] (-0.4,0.4) -- (1.6,2.4);
\draw[line width=0.05mm] (-0.6,0.6) -- (1.4,2.6);
\draw[line width=0.05mm] (-0.8,0.8) -- (1.2,2.8);

\filldraw[black] (1.4,2.6) circle (0pt) node[anchor=west] {\(s\)};

\draw [->] (2,2.5) -- (2.5,3);
\draw [->] (1.7,2.8) -- (2.2,3.3);
\draw [->] (1.4,3.1) -- (1.9,3.6);

\filldraw[black] (-0.4,2.1) circle (0pt) node[anchor=west] {\(\bar{s}\)};

\draw [->] (-1,1.5) -- (-1.5,2);
\draw [->] (-0.7,1.8) -- (-1.2,2.3);
\draw [->] (-0.4,2.1) -- (-0.9,2.6);
\draw [->] (-0.1,2.4) -- (-0.6,2.9);
\draw [->] (0.2,2.7) -- (-0.3,3.2);
\draw [->] (0.5,3) -- (0,3.5);

\draw[line width=0.8mm, dashed] (0,0) -- (1,3);

\label{DiagL}
\end{tikzpicture}

\caption{}
\end{minipage}\hfill
\begin{minipage}{0.45\textwidth}
\centering

\begin{tikzpicture}
\draw[line width=0.3mm] (-3,3) -- (0,0);
\draw[line width=0.3mm] (0,0) -- (3,3);
\draw (1,1) -- (-1,3);
\draw (-1,3) -- (-2,2);

\draw[line width=0.05mm] (0.2,0.2) -- (-1.8,2.2);
\draw[line width=0.05mm] (0.4,0.4) -- (-1.6,2.4);
\draw[line width=0.05mm] (0.6,0.6) -- (-1.4,2.6);
\draw[line width=0.05mm] (0.8,0.8) -- (-1.2,2.8);
\draw[line width=0.05mm] (1,1) -- (-1.0,3);

\draw[line width=0.05mm] (-0.2,0.2) -- (0.8,1.2);
\draw[line width=0.05mm] (-0.4,0.4) -- (0.6,1.4);
\draw[line width=0.05mm] (-0.6,0.6) -- (0.4,1.6);
\draw[line width=0.05mm] (-0.8,0.8) -- (0.2,1.8);
\draw[line width=0.05mm] (-1,1) -- (0,2);
\draw[line width=0.05mm] (-1.2,1.2) -- (-0.2,2.2);
\draw[line width=0.05mm] (-1.4,1.4) -- (-0.4,2.4);
\draw[line width=0.05mm] (-1.6,1.6) -- (-0.6,2.6);
\draw[line width=0.05mm] (-1.8,1.8) -- (-0.8,2.8);

\filldraw[black] (-1.8,2.65) circle (0pt) node[anchor=west] {\(\bar{s}'\)};

\draw [->] (1,1.5) -- (1.5,2);
\draw [->] (0.7,1.8) -- (1.2,2.3);
\draw [->] (0.4,2.1) -- (0.9,2.6);
\draw [->] (0.1,2.4) -- (0.6,2.9);
\draw [->] (-0.2,2.7) -- (0.3,3.2);
\draw [->] (-0.5,3) -- (0,3.5);

\filldraw[black] (0.0,2.1) circle (0pt) node[anchor=west] {\(s'\)};

\draw [->] (-2,2.5) -- (-2.5,3);
\draw [->] (-1.7,2.8) -- (-2.2,3.3);
\draw [->] (-1.4,3.1) -- (-1.9,3.6);

\end{tikzpicture}

\label{DiagR}
\caption{}
\end{minipage}

\end{figure}

If we consider some spacetime region [shown in Figure 1] we consider essentially the same amount of spacetime region but defined with respect to the null cone this way [shown in Figure 2]. If there is some flux going out here [right arrows of Figure 1] then basically the affine length [\(s\)] of this null line tells us something about the density of this flux and the affine length [\(\bar{s}\)] of this null line tells us something about the density of the flux [left arrows in Figure 1], and since these affine lengths don't have to be, as I've drawn them they're nowhere near equal, the flux densities are effectively very different. On the other hand the same volume of spacetime, just essentially flipped in with respect to out, would have a relatively higher flux here [left arrows of Figure 2] than here [right arrows of Figure 2]. It always turns out that we can draw a timelike line [dashed line in Figure 1] in here because this is a null cone and this is somewhere in the interior of the null cone, so if this timelike line were to represent the orbit of some particle or the orbit of the surface of the star or something like that, this would look like a high flux going out and a low flux going in, but simply by boosting it to this frame locally it would look like a low flux going out or a high flux going in. The difference between these, although I got them by flipping them, the difference between them is a boost. So unless we know in what local frame we should be constructing these local fluxes we don't really know whether one is high and one is low or the other way around. The reason I would like to hear more about the high energy physics discussion is because it affects the total volume that we should be putting in here whether we want to discuss either one of these or the other. I think the arguments that we've been having about whether the flux is high or low depend on some local frame that we don't know very much about. I claim we don't know very much about it because, if we go back to a paper more than twenty years ago, when we're trying to look at a two dimensional problem and a null shell collapsing and expanding we were able to construct, because it was very simple. We had a null shell, we had flat space on the inside and flat space on the outside, we had some kind of geometry like this so that this was flat Minkowski, this was Schwarzschild. We had some \(r = \text{constant}\) line that we could draw here [bottom of Figure \ref{D3}] and since this was flat there was some other \(r = \text{constant}\) line we could draw here [top of Figure \ref{D3}]. But it turns out that in the natural scale, the natural choice of frames that were relevant for discussing some non-boosted Schwarzschild on the outside the effective place where \(r=0\) was was being boosted away from the point of the intersection. So I don't know whether we should consider this point as being boosted away from where we're trying to emit the radiation or if it's more like this which is typically what people might be thinking of. So I claim we don't know, I sort of rather regret we didn't follow up understanding this further when we did this because this is boosted very close to null. Whereas when we typically look at how \(r=0\) goes closer and closer to null as the singularity forms it's actually curving towards that null, the null for outgoing radiation, rather than the null for ingoing radiation. So I don't really understand why this disparity happens, this was a result that we never followed up on. But I do think that this discussion about densities of ingoing and outgoing flux can only be concrete when we know what frame we want to hold that discussion in, but I don't think we have it yet.
\end{dialogue}

\begin{figure}[h]
\centering
\begin{tikzpicture}
	
	\draw (0,0) -- (2,2);
	\draw (0,0) -- (2,-2);
	\draw [dashed] (0,0) -- (-2,2);
	\draw [dashed] (0,0) -- (-1.5,-1.5);
	
	\filldraw[black] (0.3,0) circle (0pt) node[anchor=west] {\tiny Schwarzschild};
	\filldraw[black] (-0.66,-1) circle (0pt) node[anchor=west] {\tiny Minkowksi};
	
	\draw [thin] (-1.5,-1.5) -- (-1.5,-2.5);
	\filldraw[black] (-1.45,-2.0) circle (0pt) node[anchor=west] {\tiny \(r\)=const};
	
	\draw [thin] (-2,2) -- (-2,3);
	\filldraw[black] (-1.95,2.5) circle (0pt) node[anchor=west] {\tiny \(r\)=const};
	
	\draw [line width=0.08mm] (-2,2) .. controls (-2,-0.6) and (-1.5,0.6) .. (-1.5,-1.5);
	\filldraw[black] (-1.79,0.17) circle (0.5pt);
	
\end{tikzpicture}
\caption{}%
\label{D3}%
\end{figure}  

		\pagebreak
		\subsection{Physical Interpretation of the Semi-Classical Energy-Momentum Tensor \\ \textit{\small James Bardeen} }
				{\small (Slides from the talk can be found at \url{physics.unc.edu/~dnmorse/Hawking_Conference_Slides}.)}			

%				\includepdf[pages=-]{./Slides/Bardeen.pdf}

%		\pagebreak
		\subsection*{Discussion}
			
				%auto-ignore

\begin{dialogue}

\speak{Ford} I'd like to comment a little bit on the validity of the semi-classical theory. I think basically I agree with the picture that you've given in the early part of your talk, with maybe one refinement and that is that the states we are dealing with are not eigenstates of stress-energy so that means that there are going to be quantum fluctuations on the stress tensor. So of course what you're dealing with is an expectation value which is doing some sort of an average and I would suggest that probably it's at best a time average. You can see that from just considering the fact that a distant observer watching the Hawking flux as it comes out sees about one quantum per horizon crossing time come out.

\speak{Bardeen} Even less, it's actually very small.

\speak{Ford} Less than that then. Still it's clear that there are going to be large statistical fluctuations and the flux only has a meaning as a time average over many horizon crossing times. Of course as you go further inward it becomes less clear how big the fluctuations are so that at the very best you're dealing with some type of time average.

\speak{Bardeen} I agree, anything except a time average doesn't make any sense in the semi-classical treatment.

\speak{Rovelli} I want to follow up on this, the slide before that you said that the semi-classical theory, the quantum field theory on curved spacetime, should be taken very seriously which obviously is true. Everything you said is obviously totally consistent within this framework but we know that physically this framework does neglect the fluctuations of the geometry itself, of the light cone itself, which might be large. In fact it seems to me that in your own work you mention at the beginning, as well as what Giddings has been doing recently, and what we have been doing recently with Planck stars, there is a number of attempts to explore the possibility that the fluctuations of the geometry in the region outside the horizon might not be captured by quantum field theory on curved spacetime. This seems to me a direction which has not been explored enough.

\speak{Bardeen} You could say that, on the other hand, for a very large black hole those fluctuations are very small. It's true that over a long time they might\ldots but you wouldn't expect coherence over a long time. What you can show, if you accept as a long term time average this semi-classical energy-momentum tensor, you can calculate the first order back reaction on the geometry. What you find is that the geometry basically stays Schwarzschild with a decreasing mass. You can use advanced Eddington-Finklestein coordinates in terms of the advanced time, you can show that the general spherical mass function has a time derivative which is the same at all radii, it doesn't depend on radius at all. So the exterior geometry, in the semi-classical approximation, remains Schwarzschild and fluctuations, if they're very small and not coherent over long times, shouldn't produce a very large affect.

\speak{Rovelli} But this is within the approximation itself because, just exactly what we heard a moment ago, you're using the expectation value of \(T_{\mu\nu}\). In other words, you are treating the geometry classically again. If there is a possible entanglement with the geometry, it might be small of course, but I don't think, given the confusion we have about black holes so far, this can be dismissed by just saying any actual fluctuation of the geometry which is not captured by quantum field theory on curved spacetime is relevant. Your own recent suggestion, in some sense, required that.

\speak{Bardeen} I agree there could be effects of that kind. 

\speak{Bardeen} [responding to a question about a slide] It doesn't make a difference to Schwarzschild. I think there is an issue with some of the attempts to derive analytic formulas for the energy momentum tensor in that they involve considering an ultra-static conformally transformed metric where you consider a conformal transformation, where you take out the \(g_{tt}\) as a conformal factor, and there then of course the Ricci tensor is not zero in the ultra-static metric, so you have a much more complicated conformal anomaly. Furthermore they consider a local effective action in order to derive an energy momentum tensor with terms in the action which are related to the coefficients in the conformal anomaly plus a part which is associated with the conformally invariant part of the field. So for spin zero the results they got from this method seemed to work reasonably well, not perfectly by any means. For spin one there are a lot of issue where people got results which were in really violent disagreement with the numerical results and furthermore weren't even internally consistent and they had singular behavior at infinity or on the horizon. That's discussed in the Jensen and Ottewill paper where they compare some of the different formulas. There was a history where people got very widely different results. Also this whole business of using an effective action. Then they get the results, then they do a conformal transformation back to the physical spacetime to get the physical energy momentum tensor. I don't claim to understand all the subtleties of that but it seems to me, certainly in my mind anyway, that there are reasons to suspect that that whole procedure might not be valid. Gravity is intrinsically non-local, assuming you have a local effective action may not be valid. Certainly not for the gravitational perturbations of the field, maybe for the lower spins. Whether the conformal anomalies are really being treated correctly by the conformal transformation back to the physical space, I'm not entirely sure. They do sort of a fudge where they correct the conformal anomaly in the physical spacetime by adding a term but that makes me suspect the whole procedure.

\speak{'t Hooft} To return to your conclusions, one of your conclusions was that we should not consider much information on the stretched horizon, one of the last slides. You said it in the beginning and you say it in the end again, no information can be stored in a stretched horizon. I would say quite the opposite, if you look at the outgoing Hawking particles but you take their wavefunctions and you look at their wavefunctions in position space. Now if you then go back to the past, you find those being blueshifted by exponential amounts so that means that the most minute displacement of the horizon there will give you a totally different spectrum of outgoing particles as if the most minute information of the metric at the event horizon in the past contains all the information you ever wanted to have, and even more, about the outgoing particles. Indeed these positions are being jiggled around by the ingoing particles, the ingoing particles move around the horizon because of their backreaction and the outgoing particles immediately depend on that. So the stretched horizon is exactly a place where you can transfer ingoing information into outgoing information that way. That you can calculate, you get even more information than you want. That’s the basic problem there.

\speak{Bardeen} My response is that you shouldn't try to extrapolate the outgoing Hawking radiation wavefunction back to inside around \(r=3M\).

\speak{'t Hooft} No, not inside, on the horizon. In fact on the stretched horizon if you will, so it's just about visible for the outside world if you wait long enough. There is an enormous amount of information there, but it all depends on which basis you use for Hilbert space. Again the comment also has to be made, if you say things classically you don't know what Hilbert space and you think your statements are universally valid. But now we say no no, these things are all operators and if they are diagonal in one basis they are not diagonal in another basis. Since we are doing quantum mechanics I would say that we have to look at the particles in any basis we like, let's take a basis where we identify their positions on the horizon, or near the horizon, and then all this information turns out to be extremely crucial to describe the outgoing particles. 

\speak{Bardeen} My argument against that is that there is no real existence of those particles anywhere that close to the horizon. 

\speak{'t Hooft} Well you can turn time around and just extrapolate a certain state to that state which you've got

\speak{Bardeen} Take the outgoing Hawking radiation basis and extrapolate backward?

\speak{'t Hooft} Right, and that all depends on how you represent your states in Hilbert space, whether you can do that or not. I guess you are looking at the picture where the energy momentum tensor is diagonalized and for some reason you come to the opposite conclusion. But if, I claim, we do quantum mechanics we can describe things in any basis that we find convenient to us and the most convenient basis is the one where you see the information of the outgoing particles.

\speak{Bardeen} Yes but it's sort of like in Rindler space where you just have ordinary Minkowski spacetime and you consider a basis, a Rindler basis.

\speak{'t Hooft} We are talking about the vacuum at the horizon and the vacuum is full of vacuum fluctuations at all stages, these vacuum fluctuations they are transmitted into the future from the past, they just follow some sort of rule.

\speak{Bardeen} You certainly have vacuum fluctuations and the question is does something happening, say in the distant past near the horizon, have any effect on the vacuum fluctuations at a much later time at the horizon.

\speak{'t Hooft} What happens there is that you have ingoing particles making a black hole. So there are particles going in to the horizon which produce the black hole, and later perhaps some more particles, but anyway whatever particle goes in to the horizon it jiggles about the coordinates of that horizon so that outgoing particles are being dragged along, just a tiny little amount, but it's sufficient because that is expanded exponentially in time, so very soon whatever the ingoing particles did on the horizon is going to be imaged by the outgoing particles. This is my way of seeing how the information can creep back out, just with this effect. That means that there are enormous amounts of information just exactly at the stretched horizon.

\speak{Stodolsky} I think that feeds in to a question which always confuses me when I hear these discussions: what the devil is information? You're talking about it as if it's a quantity which moves around and stays here for a while then goes someplace else and so on. When I learned physics I learned about conservation of momentum, charge; I never heard of conservation of information. So what, when you talk about information, what quantity actually is it in your calculations? 

\speak{Bardeen} You can think of it as a Von Neumann entropy, though you have to be careful because if you take a sharp surface then there are vacuum correlations in just the vacuum fluctuations across that surface which are very large.

\speak{Stodolsky} Yes but for instance when you spoke about the information doing this and that, which of your quantities did you use from your previous slides?

\speak{Bardeen} I'm not sure I know how to answer that.

\speak{Stodolsky} This is not just a question to Jim; this is a question to anybody who can explain this to me, because first of all even if you want to use entropy there is some problem about defining a local entropy rigorously, right?

\speak{'t Hooft} Think of, when you do particle physics, how do you store the information? You say well, I'll surround the whole the thing with detectors and they say `click' one at a time. Those clicks are the real pieces of information. So imagine a black hole, it emits Hawking particles, for a long time you have detectors all around it and they go off every now and then. You say hey there's a particle there, there's a particle there, that's information. 

\speak{Stodolsky} It'll click whether I have an ordered system or a disordered system.

\speak{'t Hooft} Well that doesn't matter. A particle goes by and the detector says `click.' That was a particle. If you make a registration of all this, you have a way to distinguish different states the black hole can be in. So in another case you have detectors around a black hole and they say `click' at different moments so that's another state in Hilbert space. This is how you could, in principle, characterize the out-states, say, of Hilbert space. Same thing with the in-states. You have someone carefully monitoring the ingoing particles that make a black hole.

\speak{Stodolsky} He said the information is here or there, did you square something and subtract it from something else or what? Which quantities do you manipulate to find this local information?

\speak{Bardeen} The point is that if you just measure the particles escaping, the Hawking quanta\ldots

\speak{Stodolsky} You mentioned the particles, at infinity, or locally\ldots

\speak{Bardeen} at infinity, then that tells you nothing about the entanglement.

\speak{'t Hooft} You extrapolate back, as well as you can, where these particles were in the past. Whatever that is, you have a bookkeeping system saying those particles were there, there, and there; now in quantum mechanics, as you know, you have to make superpositions. Okay, fine, we do that. In principle there is a way of doing the bookkeeping of particles.

\speak{Stodolsky} We just agreed earlier that there is no such thing as a local particle.

\speak{'t Hooft} Well there is such a thing as a local detector going off, a detector saying 'click,' that would be the definition of a particle.

\speak{Stodolsky} Okay regardless of these profound discussions, there must be something you calculate when you say the information is here or it's there.

\speak{Bardeen} You calculate the Von Neumann entropy of the Hawking radiation.

\speak{Stodolsky} But that's a global quantity, Von Neumann entropy. Trace \(\rho\) log \(\rho\) of the density matrix.

\speak{Bardeen} If you assume that you started from an initial pure state\ldots

\speak{Stodolsky} Then the entropy is zero okay.

\speak{Bardeen} and you assume nothing really hangs around\ldots

\speak{Stodolsky} Just here you had some quantities, did you do some operation? So you could if it is on the horizon or if it's not on the horizon or is this just some intuitive way of interpreting the solution.

\speak{Bardeen} Certainly what I did was more intuitive. I think there are some certainly very deep issues from quantum gravity. Once you admit sort of significant quantum backreaction as affecting the overall geometry of the spacetime, either by evaporating the black hole or by something else, it seems to me that you're ending up with essentially a bunch of sort of classical histories which are part of the overall wavefunction but don't really interact with each other much, at least after a certain time. For instance imagine a Schrodinger cat experiment in which some quantum event which triggers something which causes some violent destruction, say blows up the Earth or something. You're creating a very macroscopic change from some quantum fluctuation and how you deal with that in the full theory of quantum gravity I have no idea. This is I think a very deep issue which has to be addressed.

\speak{'t Hooft} Shouldn't your question be: how do I characterize a state in Hilbert space? In particle physics we have states, states containing \(n\) particles each containing momentum \(p_i\) and so on, and we have a Fock space of particles by which we characterize all states that you can get. In the case of black holes there is an in-state, a Fock space of particles with momentum distributions or positions or whatever you like to use to characterize your state, and there is an out-state. The problem is, how do the in-states determine what the out-states are doing? Now you ask where that information is, well it's in my notebook because I have this list of momenta of particles that I've seen. That characterizes the state in Hilbert space. Now the question is, how do the outgoing particles form a Hilbert space and how are those states related to the Hilbert space states of the in-going particles? That is what we call the black hole scattering matrix. The ingoing things make a black hole, the black hole does all sorts of things, and then things come out.

\speak{Stodolsky} That's the same as the S-matrix in particle physics.

\speak{'t Hooft} Where that information is located is perhaps not so easy to tell but you can say the particles have\ldots

\speak{Stodolsky} Certainly in S-matrix theory there is no information any place.

\speak{'t Hooft} Well, there is lots of coefficients, but the in-state that you are looking at is characterized by a certain number of numbers. Well, you could also say the black hole has micro states and you have to put your finger on which micro state are we looking at, and that's information. Now conventionally you put that information on the horizon of a black hole. The black hole has micro states, the number of micro states is an exponential function of the surface area of the black hole; so you have to say which element of the micro states I am looking at. Micro states would be one way of characterizing the state that the black hole is in. By saying that there are ones and zeroes distributed over the horizon, you have to say exactly where those ones and zeroes are and that's the information that fixes the micro state of the black hole. Now there's a very complicated unitary transformation from those micro states to either the in-states of particles, all the momentum distributions of the in-particles, a huge Hilbert space of possible states which uses a basis, then you get out of that a particular state of ingoing particles. Then you ask the same thing about the outgoing particles. Now you ask about the unitary transformation that transforms the in-states in to the out-states. This is how I would like to see a theory of black holes being formed. My question is, how does this relation between the in and out-states for a black hole, how could I deduce that from knowing what happens at the horizon? That's why we say the information is on the horizon, because I need how these particles behave at the stretched horizon to figure out how the in-states could be related to the out-states. But that's only the beginning of an argument because, if you try to work out the details, you find how difficult it is to do this right. I say that the first step is to look at the gravitational backreaction that in and out going particles have on to each other. They interact gravitationally for sure. If the ingoing particles come in very early and the outgoing particles go out very late then there is a very very tiny region at the stretched horizon where they all meet. The minute displacement effect caused by one particle has a big effect on the other particle going out. That's why I said it all happens at the stretched horizon, where the ingoing particles meet the outgoing particles. But apparently that's a quite different language from what the speaker is now using.

\speak{Mersini-Houghton} If I understand correctly you are saying, if the information is at the horizon I am in trouble, I end up with a firewall basically. You are saying if information is located in the stretched horizon then you would end up with something very bad which would be very similar to a firewall. So you want to have information at the center rather than the stretched horizon and then, in the next slide, you are saying that doesn't help the problem, it brings us to square one. Either we lose information or we have a firewall and most likely we end up with both. But the way out is postulating that there is some chaotic foamy core inside and then you still end up needing an inward flow of negative energy. I just don't understand, the moment you, by hand, throw away the singularity and say well, I'll postulate there is no singularity there is a chaotic core, from that moment everything is solved. Why, on top of that, throw a second postulate like the Bousso entropy bound and then a third postulate of inward negative energy flow? Where would that inward negative energy flow come from?

\speak{Bardeen} There are vacuum fluctuations, that's where they come from.

\speak{Mersini-Houghton} But you would need just the right amount to make sure the Bousso entropy bound is correct.

\speak{Bardeen} There are a lot of very drastic assumptions made in that model, I'm not claiming that I have any at all solid argument that it works. Once this of course has existed for a long time it may be that the inner apparent horizon becomes quite singular if you have energies greater than the Planck scale and so on, it may be that there is no way to really continue the spacetime inside the inner apparent horizon, then you have to go to some string description or something. 

\speak{Mersini-Houghton} A more simple question, you have a statement there about the collapse not being the source of Hawking radiation. Do you mean the source as in physical location, or duration?

\speak{Bardeen} My understanding of Hawking radiation, and I think most people's understanding of Hawking radiation, it has nothing to do with a collapsing star. There are just vacuum fluctuations existing near the horizon which evolve and get distorted by the tidal effects. Certainly null geodesics are being pulled away from both sides of the outer horizon. 

\speak{Mersini-Houghton} But then how do you get an asymmetric flux, there is no collapse before but there is a collapse after?

\speak{Bardeen} Initially there is a given very small, potentially very short, wavelength fluctuation. Essentially you renormalize away their effect because on short scales the spacetime is Minkowskian, you have a local inertial frame. The vacuum, locally for very short wavelengths small compared to the scale of the curvature, there are fluctuations just like there are in Minkowski space. Now of course there is a problem of why does Minkowski space have zero energy density, or maybe not quite zero because of the cosmological constant.

\speak{Mersini-Houghton} No, my question was much simpler than that. There is no particle now, something happens, the star collapses, then suddenly we have a flux of particles. So clearly the collapse is very important to break the time symmetry. No before, yes after.

\speak{Bardeen} The vacuum fluctuations evolve, parts of them sort of leak through the potential barrier and go off to infinity, and at infinity they can be interpreted as particles by reference to Minkowski ground state at infinity. Near the black hole and inside the black hole they may not have any simple particle interpretation but they're going to evolve in some way which basically involves a given fluctuation, of course when we talk about fluctuations we're not talking about real fluctuations but just sort of different possibilities in the wave function for things happening, they evolve into the horizon. It seems to me the apparent horizon of a black hole is locally, on short wavelengths, indistinguishable from the Rindler horizon in flat spacetime. So for instance if you use a Hilbert space based on the Rindler modes it looks like there is something that becomes very singular, very high energy, going far backwards in time, but that doesn't stop us from thinking in the actual Minkowski spacetime that there's no energy there. It seems to me that the black hole horizon is locally no different, it's only when you consider wavelengths that are of the horizon that you start getting effects which deviate from just local Minkowski space.

\speak{Mersini-Houghton} [A question about accelerated observers in flat spacetime].

\speak{Bardeen} In flat spacetimes, if you consider a system of uniformly accelerating observers, you can define a quantum state based on splitting it up in to modes, the Killing time for those accelerating observers, those modes become singular on the Rindler horizon. But Rindler horizons are everywhere, any null surface in flat spacetime can be considered a Rindler horizon for some class of observers.

\speak{Mersini-Houghton} But then we would have very non-thermal radiation.

\speak{Bardeen} It doesn't stop us from considering Minkowski spacetime as basically something that's inert and doesn't create particles and so on. 

\speak{Spindel} I'm afraid what I will say is a rephrasing of a previous comment. If we think, for instance, about the Schwinger pair particle creation in a strong electric field, and if we compute the charge density in the vacuum state, you obtain zero due to charge conservation. If, instead, looking at the matrix elements, we look at the situation where we have a pair created with one particle here and the one with the opposite charge there, and we compute in-out matrix elements of the charge density, what we obtain is that the charges are localized on the past null cone and you see the creation of charge. In the same sense, I think, when you want to compute the backreaction in quantum gravity at first level, what you have to do is not an explicit computation of matrix elements but an in-out computation and use the in-out expectation value of the energy momentum tensor as the source of the linearized correction to the Einstein equation. What you obtain in such case is a complex metric but you need not worry about that because that complex metric will give a saddle point in the path integral  from which one we can compute for specific situations, for specific black holes, effects that are not the ones that you could describe when you just take expectation values in which case you obtain thermodynamical results. That's my comment.

\speak{Ford} While I've been waiting to ask my question, a comment has arisen on the intervening discussion. I think that they've been talking past each other. Jim is talking about a Hartle-Hawking state, Laura is talking about an Unruh state, one of them is time reversal symmetric and the other is not. By the way could we please stop calling these states vacuum, or at least the next time you say vacuum put in your paper a disclaimer `the term vacuum is purely of a historical origin.' Now my real question is, what is special about \(r=3M\) as opposed to any other value of \(r\) between \(r=3M\) and \(r=2M\)? Is that a maximum of some curvature invariant?

\speak{Bardeen} If you look at the mode equations for perturbations, say of an electromagnetic field in a Schwarzschild background, they have sort of a potential barrier which is partly a centrifugal barrier and partly associated with the spacetime curvature. For high \(l\)'s it's mainly a centrifugal barrier. So if you have a wave on one side it has some difficulty in getting through to the other side.

\speak{Ford} So this is something that because of the transverse direction\ldots

\speak{Bardeen} And the peak of that barrier corresponds essentially to the circular photon orbit which for very high \(l\) obviously\ldots

\speak{Ford} I know about that. 

\speak{Bardeen} So that's at \(r=3M\). I'm not saying that the particles are created right at \(r=3M\) by any means, you can't localize their creation to any particular radius. You just have vacuum fluctuations which are leaking through and you can't really interpret them as particles until you get out to some very large radius where you have a well defined vacuum.

\speak{Ford} From the point of view of someone far away, \(r=3M\) is close to the horizon. So to say that the pairs, the fluxes, are created close to the horizon would appear to be consistent with what you said.

\speak{Bardeen} What I mean by close is like a Planck length from the horizon. 

\end{dialogue}

\pagebreak
\section{Tuesday, August 25 2015 \\ \textit{\small Convener: L. Mersini-Houghton} }

		\pagebreak
		\subsection{The Information Paradox \\ \textit{\small Stephen Hawking} }
	    
%				\pagebreak
				%auto-ignore

Forty years ago I wrote a paper on the predictability in gravitational collapse in which I claimed there would be loss of predictability of the final state if the black hole evaporated completely. This was because one could not measure the quantum state of what fell in to the black hole. The loss of information would have meant the outgoing radiation was in a mixed state and the S-matrix was not unitary. The paper was very controversial. It was rejected by The Physical Review and accepted only after much argument, and a delay of a year. Since the publication of the paper the ADS/CFT correspondence has shown there is no information loss. This is the information paradox: how does the information of the quantum state of the in falling particles re-emerge in the outgoing radiation?

This has been an outstanding problem in theoretical physics for the last forty years. Despite a large number of papers, see the [Almheiri, Marolf, Polchinski, Sully] firewall paper for a list, no satisfactory resolution has been advanced. I propose that the information is stored not in the interior of the black hole as one might expect but on its boundary, the event horizon, in the form of supertranslations of the horizon. This is a form of holography: recording the state of the four dimensional region on its boundary, the horizon. 

The concept of supertranslations was introduced in 1962 by Bondi, Metzner, and Sachs, BMS, to describe the asymptotic group of asymptotically flat space in the presence of gravitational radiation. The BMS group is a semi-direct product of the Lorentz group with supertranslations. A supertranslation \(\alpha\) moves each point of future null infinity a distance \(\alpha\) to the future along the null geodesic generators of future null infinity \(\mathcal{I}^+\) while keeping \(\theta\) and \(\phi\) on the two sphere unchanged. In other words, the retarded time \(u\) is replaced by \(u' = u + \alpha\). The usual time and space translations form a four-parameter subgroup of the infinite dimensional supertranslations, but they are not an invariant subgroup of the BMS group. 

Listening to a lecture by Strominger on the BMS group at a workshop this summer I realized that stationary black hole horizons also have supertranslations. In this case the advanced time \(v\) is shifted by \(\alpha\), \(v' = v + \alpha\). I discussed my idea with Malcolm Perry and Andrew Strominger. 

The null geodesic generators of the horizon need not have a common past endpoint and there is no canonical cross section of the horizon. I take the tangent vector \(l\) to the horizon to be normalized to agree with the killing vector: time translation plus rotation, on the horizon. 

Classically a black hole is independent of its past history. I shall assume this is also true in the quantum domain. How then can a black hole emit information about the particles that fell in? The answer, I propose, is that the information is stored in a supertranslation of the horizon that the ingoing particles caused. 

In his recent paper, 'Chaos in the black hole S-matrix,' Polchinski has used a shock wave approximation to calculate a shift on the generator of the horizon caused by an ingoing wave packet. Even though the calculation may need corrections, this shows in principle that the ingoing particles determine a supertranslation of the horizon. This, in turn, will determine varying delays in the emission of wave packets along each null geodesic generator. 

The information about ingoing particles is returned, but in a chaotic and useless form. This resolves the information paradox. For all practical purposes, the information is lost. 
			
		\pagebreak
		\subsection*{Discussion}
				%auto-ignore

\begin{dialogue}

\speak{Rovelli} Is the supertranslation changing the classical metric or the quantum state, and if it is changing the quantum state how is it doing so?

\speak{Hawking} Sorry, what did you say?

\speak{Rovelli} If supertranslations act on the horizon the question is: does it change the classical metric or does it change the quantum state? In the second case, how does it change the quantum state?

\speak{Parker} I certainly don't fully understand yet what Stephen is saying, but somehow I suppose it would have to be related to supertranslations on \(\mathcal{I}^+\) that's left in the outgoing radiation that goes to \(\mathcal{I}^+\) from the horizon backwards when the black hole evaporates. 

\speak{Rovelli} Gerard, there seems to be a relation with your way of viewing what happens because the incoming particles affect the manner of the outgoing particle that's emitted and that's how information gets out.

\speak{'t Hooft} I would like to remark that this is indeed the way I've been looking at black holes quite some time ago. In my paper on the scattering matrix twenty years ago I do exactly this. From the ingoing particles they give a translation in the \(v\) variable, the outgoing particles give a translation in the \(u\) variable, these translations must be \(\theta, \phi\) dependent that's very important, I haven't seen that here in his talk.

\speak{Rovelli} I think that's what Stephen meant, that \(\alpha\) is depending on \(\theta, \phi\). These are the supertranslations right, this is local in the sphere. 

\speak{t' Hooft} The translation depends on \(\theta\) and \(\phi\) and as such affects the wavefunction of the ingoing particles or conversely the wavefunction of the outgoing particles. If you see how that connection goes you get a unitary S-matrix. However, it works in a too-large Hilbert space. We still have the problem of getting technically the things right. The particles also seem to obey the algebraical rules as a closed string theory in that the outgoing particles look very much like closed strings. That's because in such a description the particles behave like insertions in the bulk of a string, that of course is an insertion of an external closed string particle. That's how the in and outgoing particles seem to behave, as if string theory is a way to address this problem perhaps in it's own language. The strange consequence of this is that all particles are exclusively characterized by their contribution to the ingoing momentum because their momentum causes the supertranslations and so it is as if you characterize the particles only by their momentum and not by an other internal property like baryon number or god knows what. All that must be going down the drain at the Planck scale. 

\speak{Hawking} A supertranslation delays the emission of wavepackets thus it affects the final state. The idea is that the supertranslations are a hologram of the ingoing particles. Thus, they contain all the information that would otherwise be lost.

\speak{Rovelli} I understand this, if there were two scalar fields, different, couldn't two different kind of particles generate the same supertranslation and that's still having information loss.

\speak{Bardeen} So I'm going to give my take on this, people can discuss it and maybe Stephen can comment. As I understand it, a particle falling in will shift, if the particle falls in to a large black hole with mass very large compared to the mass of the particle or wavepacket or whatever, the supertranslation shift will be very tiny compared to the timescale of the mass, the light crossing time of the black hole, what that means then it seems to me is that the shift in the timing of propagation of the vacuum fluctuations along the horizon which eventually will give rise to the Hawking radiation, will be very tiny compared to \(M\), and as a shift in advanced time there is no redshift involved in that timing so when the Hawking radiation is eventually produced, the wavelength eventually becomes on the order of the radius of the black hole then it will still have a very tiny effect on the timing of the emission of the Hawking quantum. But the Hawking quantum is some sort of wavepacket which is spread over a time of quite a few \(M\) probably, or several \(M\), and that means then that any shift in that wavepacket by a tiny fraction of that will essentially be unmeasurable. Which I think is my take on it because the only way you could ever measure a tiny shift is by having many many repeated experiments where you prepare things in exactly the same way and measured the statistics of the timing of the detection of individual Hawking photons from repeated experiments. Since you can't repeat the experiment it seems to me in principle that there is no way you can actually retrieve that information even if in principle it's there in the tiny modification of the overall wavepacket. Anyway, people are welcome to challenge that if they wish or comment further.

\speak{Davies} Does the argument go through for a black hole deSitter? 

\speak{Hawking} Sorry, what did you say?

\speak{Davies} Does the argument go through in the case of a black hole deSitter spacetime? So not asymptotically flat, but when there's a cosmological constant.

\speak{Hawking} It applies to black holes in any background. 

\speak{Ford} I wanted to follow up on Jim Bardeen's question. I'll talk for a moment, but then I'm going to formulate a yes or no question for Stephen. I understand what Jim was talking about, the concern about the size of the wavepackets and I think there's a related issue that I was thinking of and that is spacetime geometry of horizon fluctuations. So far what we've heard deals with a fixed classical background but we know that of course there are going to be fluctuations of the background and I'm wondering how sensitive that will be to that. Maybe to speed things along I'll pose this as a yes or no question: will quantum fluctuations of the background geometry be relevant here? At infinity. I'm talking about will they be relevant for the outgoing radiation and the resolution of the information paradox. 

\speak{t' Hooft} If done well the calculation should take such fluctuations of the background in to account. The ingoing particles have a gravitational field, that's the backreaction, by which they drag along that trajectories of the outgoing particles such that the wavefunction of the outgoing particles now depends on the wavefunctions of the ingoing particles. So if you take that in to account you get a scattering matrix which is unitary but in a Hilbert space which is still too large. The reason for that is that it's very hard technically to take in to account the transverse gravitational forces. If that could be done correctly we would get exactly the scattering matrix which describes black hole formation and evaporation.  

\speak{Bardeen} My understanding is that Stephen is saying that it is the fluctuations in the background geometry are important and the fluctuations induced by the infalling particle are important. If you have a large black hole the induced fluctuations are going to be very small in terms of the metric or whatever. That means the supertranslations induced will be extremely tiny, the \(\Delta v\) will be infinitesimally small compared to \(M\) and it seems to me then that makes any hope of trying to measure the effect on the outgoing Hawking radiation essentially impossible.

\speak{'t Hooft} That effect blows up exponentially in time.

\speak{Bardeen} Well I don't see why they should grow exponentially in time. 

%1:13:56
\speak{Rovelli} I'm confused. One issue is how hard or difficult it is to measure something and a separate issue is where is information going.

\speak{Bardeen} The information could be there in terms of these very shifts. In terms of imagining somebody say at large \(r\) trying to extract that information, to actually measure that information.

\speak{Rovelli} Sure, so you're not challenging what Stephen is saying? You're just saying that if the information is scrambled in that manner and goes out in this time delay then it is not easily recovered. That's what your saying? But that's what Stephen is saying as well.

\speak{Bardeen} I don't see how any observer can actually measure those tiny time delays by trying to detect the Hawking quanta. I just don't see any possibility of that.

\speak{Rovelli} We don't see how any observer could recover the information about an encyclopedia after it has been burned. But this doesn't raise these concerns about the loss of information when you burn a book. So they're two different questions here, aren't they?

\speak{Bardeen} So in principle the information is there but in practice I don't see any way to retrieve it. 

\speak{'t Hooft} As Stephen says, the calculation is a classical one. All you have to do is calculate the classical gravitational fields of ingoing particles and you can compute the effect they have on the outgoing particles and that blows up exponentially in time. So you just wait a few seconds and then any ingoing particle will completely modify the spectrum of the outgoing particles. In other words, the outside observer will detect particles coming from a black hole by means of detectors, detectors saying click or saying click not click at all sorts of places. Whatever that observer detects will be completely effected by an early ingoing particle no matter how weak, it could be a single photon, a single neutrino, it's sufficient to completely alter the spectrum of all outgoing particles if you wait long enough. So the effect is always very big, it explodes exponentially in time. 

\speak{Bardeen} I just don't believe that. At least if what Stephen is saying about supertranslations is correct, the supertranslation is a translation in advanced time which is constant and doesn't change.

\speak{'t Hooft} No but you have to look at the coordinates used by distant observers. So the translation blows up exponentially in one direction and decreases exponentially in the other direction. So I think it's the \(v\) coordinate which blows up, well one of the two coordinates blows up and the other one shrinks exponentially in terms of external time. So no matter how small the supertranslation is it will become dominant in due time for one of the observers.

\speak{Rovelli} Gerard, if nothing falls in to the black hole later on the incoming particles are the ones that collapsed in the first place. Is that right?

\speak{'t Hooft} Yes, the incoming particles is the entire set if ingoing particles which make the black hole, which give rise to the emergence of a black hole in the first place, yes. You take them all together. The way I would phrase this is that you add one extra particle and you ask `what does that do to the spectrum of the outgoing particles?' You look at the quantum state of the whole thing, how is this quantum state modified by adding one single ingoing particle no matter how soft it is? The answer is yes, it has an effect on the outgoing particles, but maybe you have to wait for a while before the effect exponentiates sufficiently to become visible in the outgoing spectrum.

\speak{Rovelli} There's something that always disturbed me in the usual story which is the following: matter collapses, that's the usual classical story, and very rapidly all information is radiated away except for mass and spin and charge. What you're saying is that this is not true as far as the Hawking radiation is concerned because Hawking radiation is affected strongly by the details of the collapse to start from. So it's not true that all information is radiated away with the first oscillations of the black hole, but is actually still affecting the Hawking quanta which are emitted much later, which are  received much later.

\speak{'t Hooft} I'm sure there are many different ways of formulating what actually happens. My preferred way of seeing it is consider a black hole formed by some collapsing object, whatever that is, and that black hole sits in a special quantum state, it will produce a quantum state of outgoing particles as well. That's the S-matrix. Now what that state is I don't know and I find difficult to calculate. But what I can calculate much easier is if we take a small modification. So now I consider a small modification in the in-state by adding or removing one particle for instance. That small modification will produce a small modification in the out-state but this I can calculate because I can calculate the gravitational effect of that extra ingoing particle which was there or was removed there. So that gravitational effect modifies the spectrum of the outgoing particles ever so slightly it seems, as Jim said, but if you wait long enough that effect becomes very big so even the most minute change on the in-state makes a big affect on the out-state. All that should be described by a unitary matrix if done correctly, but to do this correctly is very difficult technically and that's why this procedure hasn't really shown up much earlier in the literature as the answer, because we are unable to do all the technical calculations correctly. They are very hard. But the basis is there, the basis is just look at the gravitational field of the ingoing particles, look how it affects the outgoing particles, here you are you get the scattering matrix. It's unitary, yes, but in the wrong Hilbert space. So, you still have to work at that. The Hilbert space is wrong because there are certainly effects which have been ignored, which are not the supertranslations but the other translations. Not the non-supertranslations. If you ignore that you get only part of the truth but not the complete truth.

\end{dialogue} 

		\pagebreak
		\subsection{Black Hole Memory \\ \textit{\small Malcolm Perry} }
	
%				\pagebreak
				%auto-ignore

First of all I'd like to thank the organizers, KTH, University of Stockholm, and Nordita for their marvelous hospitality providing us with a very stimulating meeting or at least so far stimulating. What I wanted to do was to carry on a little bit from what Stephen Hawking said this morning as he was describing the work that Andy Strominger, myself, and Stephen were doing. I thought the best thing to do right now was to describe a little bit about what the BMS group is and how it is related to extra black hole charges. 

So we're all used to the idea of the information paradox coming about because black holes only have a limited amount of hair given by the mass, the charge, and the angular momentum of the black hole. But we need to first of all see why this is and to see how you might go around trying to change this. What one usually does when talking about black hole space times is to think about stationary or static black holes and ask what the degrees of freedom are that are associated with them and it has been well known since the proof of the uniqueness theorems back in the 1970's that they are categorized by \(M\), \(J\), and \(Q\); the mass, the angular momentum, and the charge. Or, if you want to be a little more precise about this the \(l=0\) and \(1\) degrees of freedom of the gravitational field which give rise to the mass and the angular momentum. We could also add to that the linear momentum of the black hole if we wanted to, and for the electromagnetic field we have of course the electric charge (we could have had a magnetic monopole charge, if we wanted to) but in each of these cases you could generalize it to other gauge fields where you can have electric charges or magnetic charges. One reason why this is kind of limited is that because spacetimes that you're looking at are static or stationary these charges are things that you can measure at \(i^0\), at spacelike infinity. So these are things which are sort of fossilized in the spacetime. They're there forever because we're dealing with stationary or static black holes. These are the sort of an irreducible set of things the black hole is described by. However, in any realistic situation one is at a different position, one sits at \(\mathcal{I}^+\) and typically one will observe what is going on by looking at rays or matter coming out of the spacetime, of course \(\mathcal{I}^+\) is where astronomers live, and you would be observing a black hole or perhaps a collapsing star, who knows what, by looking at what happens at \(\mathcal{I}\), not what happens at \(i^0\). And for that reason you really want to be considering what the degrees of freedom are going to be at \(\mathcal{I}^+\) not at \(i^0\). At \(i^0\) the asymptotic symmetry group of the spacetime will be, at least for asymptotically flat spacetimes, will just be the Poincare group. But, as we've just discussed, you need to consider what happens at \(\mathcal{I}^+\) because that is where the dynamics of the situation is going on. One isn't going to be able to understand the evolution of a system without asking what happens up here [at \(\mathcal{I}^+\)]. So that's what the BMS group is going to describe and that's what I'm, going to concentrate on for the rest of this talk. 

So what exactly is the BMS group? The way to describe it is to think about the metric on the spacetime near \(\mathcal{I}^+\). There's going to be a retarded time coordinate \(u\) which measures how far up \(\mathcal{I}\) you actually are, there will be a radial coordinate \(r\) which tells you how far you might be from some central object, and there will be a set of coordinates which we will call \(z\) and \(\bar{z}\) which parametrize the two sphere, that represent sections of \(\mathcal{I}^+\). Near infinity you expect spacetime to be flat and so you expect if you try to ask yourself what is the metric you would start of by writing something like
\begin{equation}
ds^2 = -du^2 - 2dudr + r^2\gamma_{z\bar{z}}dzd\bar{z} + \text{correction terms}
\label{metric}
\end{equation}
which is just flat space written in \(u,r,z,\bar{z}\) coordinates where \(\gamma_{z\bar{z}}\) is the metric on the unit two sphere, plus corrections. The first correction that you come across will simply be due to the mass. There will be a term that looks like 
\begin{equation}
\frac{2m_B}{r}du^2
\label{eq:}
\end{equation}
where \(m_B\) is usually referred to as the Bondi mass and is related gravitational mass of the stuff inside. The second correction term tells you that this sphere here will in general be somewhat warped, it will be a term that looks like, again a sub-leading term
\begin{equation}
r\left(C_{zz} dz^2 + C_{\bar{z}\bar{z}}d\bar{z}^2\right).
\label{eq:}
\end{equation}
This is a sort of warp factor of the sphere and it's damped by one power in \(r\). And there will be some extra stuff
\begin{equation}
D^{z}C_{zz}dudz + D^{\bar{z}}C_{\bar{z}\bar{z}}dud\bar{z}
\label{eq:}
\end{equation}
and then higher order stuff that, at least for the time being, we're not going to pay any attention to. These objects \(C\) and \(m_B\) are related to physical things going on in \(\mathcal{I}\). \(m_B\), as I said is Bondi mass, but the Bondi mass can evolve because spacetime a dynamical object and the evolution of the Bondi mass is given by
\begin{equation}
\frac{\partial m_B}{\partial u} = \frac{1}{4}\left[D_z^2 N^{zz} + D_{\bar{z}}^2N^{\bar{z}\bar{z}}\right] - \frac{1}{4}N_{zz}N^{zz} - 4\pi G\left(T_{uu}^{\text{matter}} \text{ as } r \to \infty \right)
\label{eq:}
\end{equation}
what this is telling you is that the rate of change of the Bondi mass is given by this term here and these two terms \(T_{uu}^{\text{matter}}\) simply represent the flux of energy through \(\mathcal{I}\) this object here \(N_{zz}\) is the Bondi news function and represents the flux of gravitational radiation through \(\mathcal{I}\). 
\begin{equation}
N_{zz} = \frac{\partial}{\partial u} C_{zz}.
\label{eq:}
\end{equation}
So this the situation in which the spacetime can evolve, in this particular case this equation tells you about the mass loss. You could write other evolution equations, you could do it for \(C\) you could do it for \(N\), but they would be more complicated, they would represent the flux of other objects through \(\mathcal{I}\). But now we are in a position to be able to say what the BMS group is. The BMS group is, as Stephen told us earlier today, composed of the semi-direct product of the Lorentz group with the supertranslations, here we are only going to be interested in supertranslations and the idea is to figure out what objects, what coordinate transformation preserves the asymptotic form of the metric. The metric as \(r \to \infty\) is preserved by (I'll leave out Lorentz transformations because they're not really relevant to this) supertranslations and in terms of coordinates this is just simply the shift of \(u\), the retarded time, by any function on the sphere \( f\left(z, \bar{z} \right)\)
\begin{equation}
u \to u - f\left(z,\bar{z}\right).
\label{eq:}
\end{equation}
You also need to make changes in \(r\) and \(z\) and they are given by
\begin{equation}
r \to r - D^zD_zf \\ z \to z + \frac{1}{r}D^zf .
\label{eq:}
\end{equation}
This capital \(D\) is a covariant derivative with respect to the unit two-sphere with metric \(\gamma\). So this is the asymptotic symmetry, you can immediately see from this that this group is an infinite dimensional abelian group. 

Another way of looking at the same thing would be to ask yourself what would be the infinitesimal, the vector field, that generates this transformation. We'll call that vector field \(\xi\), and it's going to be something like
\begin{equation}
\xi = f \frac{\partial}{\partial u} + D^zD_zf \frac{\partial}{\partial r} - \frac{1}{r}\left( D^zf \frac{\partial}{\partial z} + \text{c.c} \right).
\label{eq:}
\end{equation}
So there's a nice vector field that is associated with this and roughly speaking pick any real function \(f\left(z,\bar{z}\right)\) will generate a transformation, and that's why this thing generates an infinite dimensional group. The fact that it is abelian follows from the fact that all you're doing as adding \(f\) here in an additive fashion, and this gives rise to the fact that it is abelian. You can see that this object changes both the Bondi mass and the warp factor because if you compute the Lie derivative with respect to \(\xi\) of the Bondi mass it's going to be
\begin{equation}
\mathcal{L}_{\xi}m_B = f m_B .
\label{eq:}
\end{equation}
That's a physical change in the spacetime that leaves the asymptotic form invariant. Similarly you can change the warp factor:
\begin{equation}
\mathcal{L}_{\xi}C_{zz} = fN_{zz} - 2D^2_{z}f
\label{eq:}
\end{equation}
This tells you that if there is gravitational radiation going through \(\mathcal{I}\) it will change \(C_{zz}\) but it corresponds to a supertranslation. So that is part of the physical meaning of the BMS transformation.

You can do a bit better than that. There are two things you can do, you can say `what is the effect of, say, some gravitational radiation going through \(\mathcal{I}\)?' Let's suppose that we have some gravitational radiation going through \(\mathcal{I}\) and [on \(\mathcal{I}^+\)] we put an initial point and a final point [further along \(\mathcal{I}^+\)] and the radiation is concentrated in that region [between the points]. What would be the effect of that radiation? Well the easiest way to work this out is to think about this function \(C_{zz}\), the warp factor for the metric. The simplest thing to to do is to observe that it obeys the equation
\begin{equation}
D^2_z C_{zz} - \text{c.c} = 0.
\label{eq:}
\end{equation}
That implies that there is a scalar potential for \(C_zz\), call it \(C\). So 
\begin{equation}
C_{zz} = -2D^2_z C.
\label{eq:}
\end{equation}
If you perform a BMS transformation you just shift this potential by \(f\), \(C \to C+f\). 

You can do a bit more than that, you can for example suppose that you have gravitational radiation [between the initial and final points on \(\mathcal{I}^+\)] and ask yourself what kind of BMS transformation do you actually find as a result of some gravitational radiation going through \(\mathcal{I}\). We can write down a nice little formula for that which I think most simply is written in the following way: the change in \(C\) from before any radiation happened to after there was any radiation is given by
\begin{equation}
\Delta C = \int d^2z' \gamma_{z'\bar{z}'} G\left(z,\bar{z}, z',\bar{z}'\right) \left( \int_{u_\text{initial}}^{u_\text{final}}du T_uu + m_B \right)
\label{eq:}
\end{equation}
with \(G\) some kind of Green's function which I'm not going to write down, and where \(T_{uu}\) contains two pieces: the effective little bit due to gravitational radiation, and the bit due to matter content
\begin{equation}
T_uu = \frac{1}{4}N_{zz}N^{zz} + 4\pi G T_{uu}^{\text{matter}}.
\label{eq:}
\end{equation}
What this shows you is that if you allow gravitational radiation to flow through \(\mathcal{I}\) then it's affect is just a BMS transformation, so spacetimes which are flat asymptotically are not necessarily all the same they will differ by BMS transformations. One way in which you could observe this is by what's usually called gravitational memory. This was discovered in the context of gravitational wave astronomy.

Suppose you have the Earth and you had some satellites that were sitting around the Earth [forming a circular ring] perhaps equally spaced. Suppose that these satellites are somewhere near \(\mathcal{I}\) and they were equally spaced and you allow gravitational radiation from somewhere to flow through the system for a short time and ask yourself what will happen to these satellites. Well, of course, after the gravitational radiation is gone the spacetime is still flat, but it's a different spacetime, it isn't the same one you started with. After gravitational radiation has passed through that region of spacetime these satellites will have moved in various directions reflecting the fact that this function \(C\) has changed. So the initial and final spacetimes differ by a BMS transformation. In this example I chose the satellites to be inertial, but you could do things in a different way if you wanted to. You could do the same for clocks on these satellites, it could be that they start out all synchronized but by the time they have moved they will be displaying different times.

So that's one way in which you can think about BMS transformations, but there's another way which makes contact with an idea about charges at infinity. So you can have BMS charges, how are these going to be defined? Well the simplest way of doing this is to think back to the work of Peierls, but subsequently many others, it started with him as he wanted to construct a covariant way of looking at Poisson brackets. What he did for an arbitrary relativistic field theory was to give an algorithm for constructing a covariant form of the Poisson bracket, one that would not require you to pick a particular time coordinate. This was followed in the case of general relativity by Ashtekar, Crnkovic, Witten, and subsequently many others, one of the most important for us was Wald, Lee, and Zoupas. They had a covariant way of describing Poisson brackets, that means that they invented a symplectic form on the space of all metrics and first derivatives. The form of this function is quite complicated, so I won't write it down in detail, but just tell you what it is, we'll call it \(\Omega\) as most symplectic forms are called \(\Omega\), it depends on some background metric and it depends on two perturbations on that background, \(h\) and \(h'\). 
\begin{equation}
\Omega(g,h,h') = \int_{\Sigma} t^{\alpha} d^3x \sqrt{\gamma_{z,\bar{z}}} \left[ h \nabla h' - h'\nabla h \right]_{\alpha}
\label{eq:}
\end{equation}
where \(\Sigma\) is any spacelike surface, \(t^{\alpha}\) is a unit normal to it, and \(\gamma\) is the induced metric on it. If \(h\) and \(h'\) obey the linearized equations of motion this object [the current inside the square brackets] is divergence free.  This enables you to define a charge. You can't easily fix the charge but you can easily fix the variation of the charge. Suppose \(g \to g + h\) and \(h'\) was the gauge transformation associated to the vector field that generates the symmetry,
\begin{equation}
h'_{ab} = \nabla_a\xi_b + \nabla_b\xi_a.
\label{eq:}
\end{equation}
Then you will have an object which is essentially a conserved charge. What \(\Omega\left(g, h, \nabla \xi \right)\) then represents is the variation in the charge due to a perturbation \(h\) corresponding to the symmetry \(\xi\). 

That tells you that for any supertranslation you are able to define a conserved charge which takes your favorite \(\xi\), that means you choose your favorite \(f\), decide you're going to vary the metric by adding, say, for example, \(C\) to it given by \eqref{metric} and this gives rise to a charge which you can compute. So what you discover is that this spacetime, rather than being static, even though the black hole itself might be fairly static, there will be a whole collection of BMS charges which characterize the spacetime beyond the \(M\), \(J\), and \(Q\) that we had earlier. This could be a black hole spacetime, it couldn't really be a static black hole spacetime it would be a spacetime that is evolving, but there will be a whole collection of charges which would change as functions of time (as the black hole evolves). So what we've done is to construct an infinite number of charges that relate to the black hole. Well that's just the BMS charge at infinity. We heard this morning about the BMS charge associated with event horizons. Perhaps I shouldn't say event horizons, I think I really mean isolated horizon because an event horizon requires a global construct whereas it seems to me we should be thinking about local constructs here. 

A clue that such a thing is possible, at least this is how I thought of it, Stephen may have his own way of thinking about it, but this is the way in which I thought about it when listening to Andy Strominger, we go back to our picture of the Earth surrounded by collections of satellites. Of course, I said the satellites to be out at \(\mathcal{I}\), out near infinity but in actual fact it doesn't make the slightest bit of difference where they are, you're just looking at them from Earth so you could really sort of let them go down any null ray you like and the physics of what's going on wouldn't change. The calculational details will change because what we did for the BMS group relied on being able to expand things in powers of \(1/r\) nicely and had some kind of asymptotic region and controllable path series expansion telling you what happens as you get towards \(\mathcal{I}\). But that doesn't change the physics of it, you could just simply bring these things in and at some point, well if it was the Earth you'd stop when they crashed but if it was a black hole you would have to stop when you got to the event horizon. So you need to ask what kind of structure would you have to have on the event horizon?

I'll talk about isolated horizons. These are null surfaces with topology \(S^2 \times \mathbb{R}\), what I mean by null is any normal to it is going to be tangent to it, so the corresponding vector would be \(l\) and I would like the convergence of the null geodesics associated with \(l\) to have zero convergence. What that means in practice is that 
\begin{equation}
\Theta = h^{ab}\nabla_al_b = 0
\label{eq:}
\end{equation} 
where \(h^{ab}\) is the metric on the \(S^2\). Under those circumstances you can write down what you think the metric is going to be in general terms it will look something similar to the way things look at \(\mathcal{I}\), one could say there is something times \(du^2\) then there will be something involving a \(dudr\) and then there will be some stuff that tells you about what happens on the sphere and distortions of it:
\begin{equation}
ds^2 = U du^2 - e^{2\beta}dudr + g_{AB}\left(dx^A - W^Adu\right)\left(dx^B - W^Bdu\right)
\label{eq:}
\end{equation}
\(U\), \(\beta\), \(W\), and \(g\) will be functions of coordinates, but they are such that on the horizon \(U = W^A = 0\). Then you can ask yourself what is the corresponding symmetry which preserves the form of the metric in the same way that the supertranslations preserves the asymptotic symmetry for an asymptotically flat spacetime at infinity. The answer is, the corresponding vector field is 
\begin{equation}
\xi = f(X^A) \frac{\partial}{\partial u} - \left(\int_{r_{\text{Horizon}}}^r g^{BC}\partial_B f dr' \right) \frac{\partial}{\partial x^C} - \left( \int_{r_{\text{Horizon}}}^r W^B \partial_B fdr' \right) \frac{\partial}{\partial r} .
\label{eq:}
\end{equation}
Those are generators of BMS transformations on the horizon. Similarly, you can construct a collection of BMS charges on the horizon. What this tells you is that, rather than black hole being dead objects specified by small number of quantum numbers, this tells you that there are in fact classically an infinite number of charges which are essentially conserved objects, and for that reason it gives you a potential way out of the information paradox. That essentially is the observation that led to what Stephen was talking about this morning. 

Various questions emerge from all of this some of which were hinted at earlier. The first is that this is an entirely classical calculation it's not a quantum mechanical calculation and to understand the entropy of the black holes coming from Bekenstein-Hawking entropy formula, one really should be doing a quantum mechanical version of this calculation. The second problem is that the way was expressed was just in terms of perturbation theory what one should really be doing is calculating BMS charges at past infinity and asking how the whole system evolves as you move to the future where there is both a black hole horizon and \(\mathcal{I}^+\). The third question, which should have been asked this morning, but didn't seem to have been, especially given that there was a lot of talk yesterday about it, is that black holes made of baryons and black hole made of anti-baryons look the same. Whereas in this kind of picture where things arise from gauge charges that could not possibly happen. The answer is perhaps controversial, but relates to what Gerard 't Hooft was saying this morning, it restricts the kind of models that would be compatible with this kind of picture. Baryon number is a global charge, and as such it is different to a gauge charge. In string theory there are no global charges. Baryon number will result from a collection of other things which are associated with gauge charges having to do with how one generates the standard model from fundamental string theory. So, in string theory you would not face that problem. But, if you just simply constructed a general arbitrary model then you would. 

Another question that is important is, is this enough charge to resolve the information paradox? Or perhaps too much? I can't really give the answer to those questions, however for gauge fields, not just the gravitation field, there is an analog of the BMS group. So, for example, for the Maxwell field there is an analog of the BMS group at \(\mathcal{I}\) which would enable you to say something about the electromagnetic hair of the black hole not just its gravitational hair.

Lastly, and related to a point also made this morning by Paul who asked about black holes in other space times, since this effect is associated with horizons, not necessarily just what happens at \(\mathcal{I}\), there will be this effect at whatever kind of horizon you have embedded in whatever kind of spacetime there is. At any form of infinity, there will be some analog of the BMS group, whether it's a timelike infinity or, perhaps unpleasantly, a spacelike infinity which you probably don't really want to contemplate. But, also there will be an analog of the BMS group at cosmological horizons which we have not scratched the surface of yet. 

So as you can see this is sort of a brief sketch of what is going on and plainly there is ample scope for rather more work and firming up this to make sure that it really does work properly. 

Thank you for your attention. 

		\pagebreak
		\subsection*{Discussion}
				%auto-ignore

\begin{dialogue}

\speak{Ford} If I understand, what you've described is a way of classifying classical gravity wave perturbations of a black hole, you say there are an infinite number of conserved charges, that's a way of classifying the infinite number of ways you can gravitationally perturb a black hole. That is all classical. It seems like there's still a big jump from that to seeing how these charges or perturbs leave an imprint on the outgoing Hawking radiation. Do you have any ideas about that?

\speak{Perry} None that I can discuss at the moment. Will visit Strominger late Sept to start a draft, Andy will visit Cambridge end of Oct to finish the draft.

\speak{Duff} You said that M, Q, and J are the only quantum numbers that classify a black hole, but in the case of supersymmetric theories the extremal black holes form a supermultiplet and the other members of the supermultiplet could carry spin or non abelian quantum numbers. It's the fermionic hair that supplies those. Does that enter into your considerations?

\speak{Perry} Well, it has not entered in to our consideration. I'd classify this question as being part of `is there enough hair?' So certainly the supersymetric theories, it's presumably the case for some analog of the gravitino. Although I have not made any attempts to investigate what that might be. For supersymmetric gauge theories then of course since we know that there is an analog of electromagnetism then you'd have thought that there should have been an analog of fermionic superparticles of the photon in a supersymmetric setting. So I would think that the answer would be, yes, there should be such things, but we have not seen them yet.

\speak{Bardeen} Could you comment on the quantum cloning issue. You're saying that you're getting the information essentially on the horizon through these BMS translations and, on the other hand, the particle that's falling in is not curved very much crossing through the horizon, so it should have, essentially, all the entanglement with the outside. In some sense, there is cloning of information and on the other hand I think you could argue that there's no one observer that could detect both copies of the information because the information is coming out at late time and the Hawking radiation is spread out over essentially the whole future history of the black hole while there's no observer that can both measure that and also detect something inside the black hole. Do you have any comments on that?

\speak{Perry} I think that you have put your finger on the heart of black hole complementarity which I will confess that I have always found a little puzzling, perhaps I'm not unique in that respect. I don't know if Stephen is going to agree with what I'm about to say, but I find myself in Gerard 't Hooft's camp on this subject. It is presumably the case that the two different observers are telling you different faces of the same thing. How that can come about is absolutely unclear to me. The alternative is simply that the nature of reality is not as objective as you think. You may take your pick. Maybe Stephen would like to say something on the subject.

\speak{Stodolsky} Would these considerations mean that the radiation seen at infinity is different from the usual thermal source?

\speak{Perry} Do you mean would they be precisely Planckian? 

\speak{Stodolsky} Not only Planckian but the lack of all kinds of correlations and so on.

\speak{Perry} Well I think what would actually happen is that at the beginning of the evaporation process they would actually look fairly Planckian, but by the time you got to the end there would be significant deviation from that because that's where you would expect the correlations to lie. 

\speak{Stodolsky} So it's no longer a perfect thermal source?

\speak{Perry} It would not be perfectly thermal, there would be corrections, and in particular large corrections when the temperature became large.

\speak{Rovelli} Just a thought, the imprint you're looking at on the horizon is not just on the horizon it's also outside the horizon. In fact you can find it, the metric, somehow in the region outside the horizon just from what you said, it seems to me. You're looking at its value at the horizon, but there is a similar imprint at various radii.

\speak{Perry} But you have to have a surface on which there is a collection of symmetries which can you regard as a sort of boundary of the spacetime you're interested in if you want to define charges, otherwise yes. The point is, of course, that you need a surface, which would be a timelike or a null surface, but if that doesn't have any isometries, or any approximate isometries then you can't really find conserved charges on it. What you're saying is that, I think, that if you're very careful and calculated all the back reaction properly you should be able to account for processes like this simply by being very careful. That's true at the classical level, I don't know about the quantum level. The quantum level is something we have to investigate more.

\speak{Rovelli} Your answer is the answer of a mathematician, if I want to define proper charges in that case to take in to account where the information is stored, then I can use these charges. But physically the metric outside, of course if I am a great mathematician I can compute it, still has this information. Is there any sense for which what you said implies that information is precisely on the horizon as opposed to just outside the horizon?

\speak{Perry} The answer to that is not clear, however you can make it clear by doing calculations which are in progress which involve doing expansions of what happens near the horizon when you drop things in and asking what the charges are before and after you've done that. That's a calculation that's in progress. It could be, that you discover that the charges are not zero, in which case you would say it is exactly on the horizon or it could be that you discover that all the charges were zero in which case you would have to say that it's outside the horizon because you're dealing with conserved charges. So that's at least it's on my mind.

\speak{Ford} I would like to understand a little bit better the nature of these charges. So let's talk purely at the classical level for a moment. We have a Schwarzschild black hole and we perturb it a little bit with some gravity waves, now normally of course as seen by distant observer the perturbation will decay away exponentially. After a while the black hole looks almost exactly as it was before the gravity wave came in and perturbed it. You're telling us that there is in some sense a conserved quantity, an imprint of that, so that this conserved charge says that in some sense once you kick a black hole it's never the same again. Is that a fair statement? 

\speak{Perry} That's right.

\speak{Ford} But then that says that this imprint must be very very hard to detect classically because we're dealing with something that is decaying away exponentially in time. 

\speak{Perry} What you're asking is whether or not this is enough to solve the information problem?

\speak{Ford} I'm still trying to understand the nature of these charges even in the classical theory. I can see there is a formal construction of them, but I think I would be a little hard pressed to ask an astrophysicist how to construct something to measure these charges. Especially if our perturbation happened a very long time ago. Can you say anything about that? 

\speak{Perry} On the subject of how you would measure the black hole, I am not sure of the answer because that's a bit hard. If you want to measure the charges at infinity then experiments of the type of the satellite experiment, observations that we were talking about, should be sufficient. At least in principle if even not in practice.

\speak{Ford} What I think you're talking about there is the gravity wave comes by, there's some geodesic deviation which occurs, and as a result of that geodesic deviation the orbits then are different from what they would have been. 

\speak{Perry} That's exactly correct.

\speak{Ford} That remains the same. But in the case of a horizon, it's much harder to see what remains to be observed at a later time.

\speak{Perry} Well I think Gerard was talking about how you get around that particular problem this morning. I don't feel I could put it much better than he did.

\speak{'t Hooft} The way I see this is that you have to look at the states asymptotically when Hawking radiation comes out of a black hole it can come out in very many ways. You make a detection of particles, you have the detectors all around the planet Earth, or the black hole, much like you sketched, all these detectors detect particles, yes or no. Some of them go off some of them do not go off. That gives you a tremendous amount of information about the final state. Now whenever you tangle, ever so little, with the ingoing states in the black hole the outgoing states will be different. From that moment on you're looking at a different states which means that different detectors will go off at sufficiently late time. 

\speak{Rovelli} But this is not the classical GR that he's considering. This is in the full theory.

\speak{'t Hooft} Hawking radiation of course is not classical GR, Hawking radiation is a quantum effect. It is in that radiation where you see all these differences take place.

\speak{Rovelli} The question was about classical GR charges, if I understood it correctly.

\speak{Ford} I was just trying to understand that, classically we expect perturbations to decay exponentially in time. Certainly as a matter of principle a decaying exponential is never strictly zero, but it gets awfully close. So this conserved charge seems to be a very subtle quantity. Very hard to measure, even in the classical theory. 

\speak{Rovelli} If you have a sphere of satellites that stay just outside the horizon for a very long time and they have clocks, at some point they're all synchronized. Then they wait long enough then the difference between the clocks will read their charges.

\speak{Perry} That's fine except you've got to keep them there somehow.

\speak{Hawking} Can you calculate entropy?

\speak{Perry} We have not done so yet. The problem is that you see infinite number of charges here, that comes from looking at the classical phase space of the system,  whereas what you want to do to calculate entropy is to look at the quantum mechanical version of the phase space. That requires you to make a modified structure of the phase space, typically by deforming Poisson brackets in to commutators or something of that type. In this particular case, I do not know how to do this. Various people have made various suggestions. One suggestion was that you just simply put in some type of cutoff. But how you would arrange that cutoff to give you precisely the right thing in a way that didn't depend on how you did it is beyond me. What one should be doing is looking at the quantum version of phase space and then seeing if you can reconstruct the entropy from that without doing anything strange. As of this moment, I don't know how to do that, though I thought you did.

%\speak{Hossenfelder} Do you consider the scenario where the singularity is still there? Or that is kind of resolved? I'm a little worried that if you look at the quantum version that the gravitons that you have escaping to infinity have an entangled partner that falls in to the singularity and you get the same problem again.

%\speak{Perry} I have no idea about the singularity. Of course the singularity is an entirely classical construct whether or not it persists in the quantum nature of spacetime is unclear to me. I think ultimately your question really is about black hole complementarity again because you're asking about what happens to observers which fall in to the black hole rather than observers who stay outside. I find it rather hard to give a clear answer to that question.

\speak{Rovelli} The two metric that you foliate, the two-metric is null?

\speak{Perry} The three-metric on the horizon is null, the two-metric is sections of that which exist on the sphere. The same is true on \(\mathcal{I}\). The three-metric is a null, degenerate, metric but sections of it at constant advanced or retarded time will be two spheres.

\speak{Rovelli} Is exactly a two-sphere, is a geometrical two-sphere? Not a topological two-sphere?

\speak{Perry} Yes.

\speak{Rovelli} So the charges are not written in this two-metric?

\speak{Perry} In this particular case, this could be any metric on a two-sphere, the case of what happens at \(\mathcal{I}\), then you can use a unit two-sphere and it wouldn't be a problem. Here you've got to be a bit more general because you don't really know what shape the horizon is, it could be squashed for example, then it wouldn't be the same thing as a two-sphere, it would be conformal to it.

\speak{Rovelli} I'm confused. Suppose we have a Schwarzschild in to which things have fallen, different charges mean that the geometry is not exactly a two-sphere? 

\speak{Perry} Different charges would mean the geometry is not exactly, it will be deformed away from whatever it was when you started.

\speak{Rovelli} So this is exactly the picture that the entropy Stephen Hawking was asking about. Count the possible deformations of the two-sphere. 

\speak{Perry} Well classically there is an infinite number of those.

\speak{Rovelli} Classically, right classically, which of course are infinite. In the old way, I'm not doing this anymore, but in the old way of loop quantum gravity the counting was the number of quantum states, the number of quantum geometries which they are discrete because of Planck length, with a certain condition. The condition being that the total area is fixed. I wonder if there is any\ldots The old loop quantum gravity calculations were based on the idea that you can view the horizon as a shaking sphere. Classically it just rapidly goes to an exact geometrical sphere, quantum mechanically it keeps shaking.

\speak{Perry} You're basically saying that you could excite the horizon by making it vibrate. The entropy is due to the thermodynamic nature of all those vibrations. 

\speak{Rovelli} Yes.

\speak{Perry} You would fix a temperature, then ask what is the entropy given that then energy is whatever it is?

\speak{Rovelli} Yes, exactly.

\speak{Perry} Maybe.

\speak{Rovelli} I'm not asking about that, I'm asking if there is a similar story here. In a sense you're viewing the information stored in to those geometrical deformations of the horizon.

\speak{Perry} Yes. We have had some discussions about that, or at least Andy and I have, I think Stephen had discussions along those lines in April. That has not yet been made precise. It's the kind of thing you might try to do but we haven't completed that yet.

\speak{Mersini-Houghton} Does it matter if you have an eternal black hole or a black hole that formed from a collapse?

\speak{Perry} Well an eternal black hole would be static, it's been there forever and it's unchanging, those seems to be the things that we're not really interested in. Here one is really interested in the question of black hole formation and evaporation, or at least black hole evolution. Eternal black holes are a little bit different. We could have a discussion about zero temperature black holes and entropy counting due to things like how you construct things with those particular quantum numbers out of branes, that works fine, but it only really works for zero temperature black holes. Those are static and the charges corresponding to the brane charges are always measured at \(i^0\). So in some sense those aren't really what we're interested in.

\speak{Mersini-Houghton} Can you also use this approach of supertranslations for deSitter space?

\speak{Perry} Yes. There is an analog of the BMS group on cosmological horizons.

\speak{Mersini-Houghton} So in that manner you would be able to mine information for what's beyond the horizon? 

\speak{Perry} Perhaps. Or what went through it would be a better way of putting it.

\speak{Mersini-Houghton} So we shouldn't expect thermal spectrum for deSitter space?

\speak{Perry} Well maybe for pure deSitter space you would. But deSitter space that contains some stuff\ldots

\speak{Mersini-Houghton} Pure deSitter should have zero entropy because it's a pure state.

\speak{Perry} deSitter space which contains some matter of some kind, which is more or less deSitter space, then the answer is yes.

\speak{Mersini-Houghton} But if you change the temperature of the spectrum would that imply a sort of decay of the cosmological constant?

\speak{Perry} I don't see how that would come about.

\speak{Ford} Can you define these charges on any null surface? Or does it have to be a horizon or \(\mathcal{I}^+\)?

\speak{Perry} I think it has to be something that has some type of symmetry group associated with it so that you can define a vector field which generates that symmetry and its charges would be associated with those vector fields. So the answer is no, not any null surface.

\speak{Freese} If at inflation you have something non-thermal, can you test this?

\speak{Perry} That relates to the question about deSitter space again. All I can say is that if you have deSitter space then on the cosmological horizon there is an analog of the BMS group. You can have a black hole in deSitter space, there is an analog of the BMS group on its horizon. I don't really know about anywhere else.

\speak{Mersini-Houghton} So that would also provide a distinction between radiation from a black hole and radiation obtained from an accelerated observer since the latter wouldn't have the symmetries.

\speak{Perry} Well an accelerated observer has a horizon, the Rindler horizon, and if you are going to ask whether that has a symmetry group associated with it the answer is going to be yes. There must be some similar thing that tells you about things that may have fallen through the Rindler horizon say, for example, inertial objects compared to the accelerated one.

\speak{Ford} I thought you just told me I couldn't take an arbitrary null surface? But surely I can take a null surface and find an observer who is asymptotic to that. I don't understand the relation between your answer to my question and your answer to the last question. 

\speak{Perry} This is something special about the Rindler horizon.

\speak{Ford} Well I can take any null surface and find some observer who is asymptotic to that, and then that surface becomes a Rindler horizon. A moment ago, you said you could not have conserved charges associated with any null surface. Now you seem to be saying that a Rindler horizon would have conserved charges. There seems to be a contradiction, or am I missing something?

\speak{Perry} I don't think for an arbitrary null surface there is a collection of observers that have that as there horizon.

\speak{Ford} Why not? Just take some timelike observer who is accelerating closer and closer to the speed of light so that observer is asymptotically approaching that null surface. Doesn't that surface become a Rindler horizon for that observer?

\speak{Perry} You're asking the opposite question, aren't you? You're saying, given a null surface can you always find a classical observer such that it's their horizon? I'm not sure if that's true. Is it?

\speak{Ford} It seems to me that the answer should be yes. Fairly obvious that you can take some timelike observer, an observer who's worldline asymptotes to a given null line. Then that makes that line a Rindler horizon for that observer.

\speak{Perry} You need a whole surface.

\speak{Ford} Take a family of observers. I see a contradiction between your two answers, unless there's something left out here. 

\speak{Perry} I reserve the right to think about that. 

\end{dialogue}
			
		\pagebreak
		\subsection{Black to White Hole Tunnelling: Before or After Hawking Radiation? \\ \textit{\small Carlo Rovelli} }
%				{\small See \verb|Rovelli.pdf| file found in the slides directory which can be found at \url{physics.unc.edu/~dnmorse/Hawking_Conference_Slides.tar.gz}.}	
	
		%	\includepdf[pages=-]{./Slides/Rovelli.pdf}
	
%		\pagebreak
		\subsection*{Discussion}
				%auto-ignore

\begin{dialogue}

\speak{Stodolsky} We now have some remnants around of these expanding moons and they're made of what? of ordinary matter?

\speak{Rovelli} Yes, in a sense this is completely conventional physics. So ordinary matter, ordinary general relativity, ordinary quantum mechanics. There's nothing non-conventional here. The hypotheses are that the were primordial black holes and a common idea in cosmology is that they were made by just fluctuations of whatever was there, typically photons because a typical moment where they could be created is when there is a lot of photons around. So it's photons coming in and what would come out is just photons again. Plus whatever other standard physics.

\spin Just for clarification, the formula \(m^2\) you want to recover from covariant loop quantum gravity and so you would like to calculate the path integral in this region 3 and somehow calculate the scattering time shift or something like that and to arrive at an exact expression with a prefactor.

\rov Exactly, exactly.

\spin The main difficulty is because you cannot evaluate this fully, or you have to go to a semi-classical approximation of the path integral? What is the main difficulty?

\rov There are two main difficulties. One is that loop quantum gravity transition amplitudes are nice and beautiful when you write them as a theoretician, but when you start trying to compute with them are a nightmare. Just to do this integral, this complicated integral, this is an integral of \(SL\left(2,\mathbb{C}\right)\), these are Wigner matrices, and these are the Wigner matrices of \(SL\left(2,\mathbb{C}\right)\) which are complicated things, which are hypergeometric functions and so when you actually try to do with that, we don't know how to do the integrals. The other difficulty, more conceptual, is that these amplitudes are given order by order in some approximation so you're actually taking these two surfaces and approximating them, sort of discretizing them in some sense, and computing the transition amplitude with increasing approximations. So the question is are we sure we can do that with a simple enough truncation, or do we have to go up with the truncations, we don't know yet.

\spin I see. 

\rov So that's conceptual. This is all finite integrals, these have been proven to be finite, the reason we can write these finite expressions is that we are in a truncation and the idea is that the degrees of freedom that matter are large scale degrees of freedom not small scale degrees of freedom, so you can do that.

\spin You described the transition from a black hole to a white hole with emission of energy in terms of the dynamics of the null shell that will constitute the black hole. What would be the time reversed process?

\rov Very good.

\stod What was the question?

\rov `What is the time reverse of the process?' is the question.

\stod For a white hole to become a black hole? It's not the same?

\rov Yes, exactly. The idea here, the Hawking radiation and the black hole they're both time oriented. In fact, by itself, the Hawking radiation is a dissipative phenomenon in the usual picture. You transform mechanical energy and heat. The idea here is that if this phenomenon is faster than the Hawking radiation, then the Hawking radiation is a small correction over this phenomenon, so we can neglect it. The picture I gave is neglecting Hawking radiation. The phenomenon is completely mechanical, it is quantum mechanical, but it is mechanical, it's not thermodynamical, and it's exactly time reversal invariant. The idea is that a shell collapses, bounces, comes out and this is a time reversible process. It's like if you take a ball, you let it go, it comes down, and then you ask how does it come up? Well it comes up exactly in the way it went down, but time reversed. 

\spin At the mechanical level, the motion of the shell, it's clear. The point that I didn't see is the extra emission of energy during the transition and all that.

\rov No no, the energy that comes out is the energy that goes in. There is no extra emission of energy. The energy of the explosion is the energy of the incoming shell. Energy is conserved and the phenomenon is time reversible. There is no other energy besides this one. The thing that comes out is the shell that went in, with it's own energy. Of course, in the real situation, on top of this, one should include the Hawking radiation which is small, \(M^3\) so it's \(M_{\text{Planck}} / M^*\) so it's many orders of magnitude smaller, but it's not time reversal invariant. It's a dissipative phenomenon. So it's like if I let a ball fall it doesn't really come out the same manner because a little bit of energy is left in the heat of the ground and \textit{that} would be the Hawking radiation effect. A small effect on top of a big one. 

\dow Do you have any model that goes beyond this spherically symmetric ansatz that you've used here?

\rov We wanted to do Kerr because Kerr is supposed to be more realistic, in the universe everything is turning, but we haven't done it yet because this is recent. But I think it has to be done. The other thing, I think to start believing this we should understand what happens not with a single shell, but with two shells, so with a continuum of matter or at least two shells. To put it very naively, if I have two shells do they come out like this or do I come out like that [with their order reversed]? I don't know.

\dow Even Kerr is very symmetric, the thing is when you don't have an event horizon you can't appeal to the uniqueness theorems in solving equations of motion. Then you open up a Pandora's box of possibilities. You would need not to just look at examples based on very symmetric situations, but you would have to see how perturbations might affect that.

\vid There is a possibility that this bounce ends up in a very un-isotropic situation. Let's say that the collapse is not exactly the same as the expanding part of the process, exactly because of what you say of the chaotic nature of the collapse it could be very likely that the resulting metric in the end is something very un-isotropic. What is interesting to me is that probably on the phenomenological point of view the characteristic signature should be more or less the same.

\dow Your construction relied on being able to to do the matching between the white hole and the black hole, but in a non-spherically symmetric, generic situation I don't know what confidence do you have that you'd be able to do such a matching.

\rov The first part of the matching is mathematically to be done sort of, but physically, it's trivial. The matching here is trivial. Imagine you have a deformed shell, you let it fall, whatever it does it does. Outside, there will be whatever it is and inside there will be whatever it is. Physically, there is always a solution of the matching here, so the question is if we can time reverse it and if we can think there is an elastic bounce.

\vid In loop quantum cosmology, there is this result that the universe can be coming from a contracting phase and then expanding nowadays. There the results are much more widely explored and in particular the result holds in the case of Bianchi IX, Bianchi I, and so on so this gives confidence that in the case of the black holes maybe we can do something similar.

\whit I wanted to ask about the two shell case. Suppose we have one shell that we want to represent a shell with a large mass so it would form a large black hole and then much later in time we want to send in a very small shell, basically a test shell, from \(\mathcal{I}\). In principle, if I tried to send it in so that it intersected your hole's central line after the \(\delta\), the way you did you matching would it have to bounce out from the middle of nowhere? If you go back to the plot you had up with the cut and the matching.

\rov No it just goes through. That would break the time reversal invariance of the process of course. If I have a big ball and a small ball, the big ball comes out first. If I send a shell here it just goes through, bounces here, and goes out. It is a test shell. Maybe I misunderstood the question.

\whit I see, so you'd basically create a new bounce process in flat space at the top?

\rov Yes. What is dangerous, if I send it just here it comes very close to here then, it's strongly blue shifted by coming down here so it might take away energy from this and move it out, in fact we're looking at the stability of the solution with respect to perturbations of this sort.

\ford The fact that this is not exponentially suppressed seems very surprising here. I'm trying to understand, I can see that you've put things together with matching but that looks maybe a little artificial. Suppose you try to do this in a path integral approach where you start with an initial configuration and you sum over amplitudes. Normally you'd naturally expect to get exponentials unless maybe there are such a very very large number of configurations that all these exponentials sum up to some other function, but usually tunneling amplitudes are exponentials. Can you comment a little but more on that?

\rov Yes, we don't have a strong answer to that. We don't have a strong argument saying no, this is not what is going to happen. We have a weak argument which is the following: if you do the saddle point approximation of the integral you get exponential of the action gives a strong depression of the probability, but then you have a phase factor which counts the number of states you can tunnel through in a sense. Now there's an argument by Mottola which says that in similar situation there is an almost exact cancellation between the two because you have a lot of states to tunnel through once you consider the full geometry. Remember in the path integral, there isn't just a spherically symmetric metric, you're summing over all. So if you want to know how many states there are available there, roughly it's the number of states that the path integral assigned to the black hole which is the exponential of the entropy. And the exponential of the entropy is again \(M^2\). So you have two exponentials of \(M^2\) which come with the opposite sign. Now if you say `alright can you Carlo prove this mathematically and convince me solidly?' No. If you have objections I'll shut up. That's an indication that we sort of used, what suggested to us \(M^2\) is anoter argument.

\ford I can certainly see if you have a very large number of intermediate states then they can sum up to something other than an exponential, but they have to be things that go between your initial configuration your final configuration, that seems to be the missing part here.

\rov Absolutely. It's not exactly tunneling in the usual sense. It's a phenomenon which is, strictly, related to tunneling, but once you write it down it's not the standard tunneling. One way of viewing this is that in computing tunneling probability you compute a dimensionless number, which is exponent, times a typical time of the process which usually you add by hand sort of. Right, if you want to know a Uranium atom when it decays, you sort of tunnel through the potential barrier and you get the exponent of some number, but then you have to multiply by something with the dimension of time and you just imagine that the thing is bouncing back and forth and you use the frequency of the bounce. That's what people do in nuclear physics, very naively. Here there is no natural time scale to start with, and the transition is not out of a stationary solution, the time comes in very indirectly from the outside. It's a funny game because remember that the time I'm talking about, which is the proper time of an observer here, it's just a proper time of an observer here, it's not something which is there. It's related to a geometrical quantity there which is the radius of a point here. And it's related logarithmically. All this to say that one cannot take the usual intuition of tunneling and transform it directly to this case. Which, of course, is not an answer therefore it's obviously not that. But it's not obvious that it's that either. 

\stod I had a sort of similar question. In ordinary tunneling you have a barrier and a particle confined in some region and you can say it wants to tunnel because of the uncertainty principle energy that tries to get through the barrier. Is there some similar understanding about why the gravitational field wants to tunnel? What is it that's fluctuating? Why does it want to get through the barrier?

\rov Because it doesn't know what to do here. If I use a classical theory there is a line here where the curvature keeps increasing. If I follow a line up here, classically the curvature keeps increasing. On this line the curvature becomes Planckian. It doesn't want to be Planckian. It's like squeezing something beyond the Heisenberg relations. It wants to make a transition to something else.

\stod So it is again a kind of uncertainty energy?

\rov Yes, except that this is played in a slightly different manner in which we don't have clear control. So yes, it's Heisenberg relations forbidding it to evolve to \(r=\infty\) here, or if you want to squeeze the shell too much here at some point it's going to increase the probability of making a transition to something else. The only something else by symmetry that it can be is the same thing going out.

\end{dialogue}
			
		\pagebreak
		\subsection{A new Quantum Black Hole Phenomenology \\ \textit{\small Francesca Vidotto} }
				{\small (Slides from the talk can be found at \url{physics.unc.edu/~dnmorse/Hawking_Conference_Slides}.)}			

	%		\includepdf[pages={1,2,3,5-}]{./Slides/Vidotto.pdf}

%		\pagebreak
		\subsection*{Discussion}
				%auto-ignore

\begin{dialogue}

\speak{Novak} You mentioned here at the end the possibility if maybe seeing these models through gravitational waves, how do you see these? Or how do you model these?

\vid Well it depends of course if the explosion is totally symmetric, I don't think there will be a signal. We need the explosion to be asymmetric, but in principle I expect to have gravitational waves associated with the explosion and imagining that in the future we will be able to capture very well gravitational waves one possibility is to see coincidence between one of the signals, whether in the low energy channel or in the high energy channel, and the gravitational waves. I don't have more insight about this because I am still waiting for gravitational waves to be detected.

%\speak{post-doc} I have a naive question about the spacetime diagram that you're drawing and what's going on at the point \(\Delta\). So it naively to me looks like that point is something like a timelike version of a conical excess singularity.

%\vid Yes, I don't think there is any conical singularity there.

%\speak{post-doc} How do you resolve that? It seems like you started with a spacetime before you peeled apart the old black hole spacetime. You started with a spacetime that had too much angle going around in the timelike direction.

%\vid My picture of it is that everything is moved there. There was the other picture in which the contracting one and the expanding one are joined together and then the idea is that you can smoothly connect these two region with a quantum region. 

%\speak{post-doc} Even when you consider the physics right around that point, the classical tip point?

%\rov Why do you think it should not be smooth?

%\speak{post-doc} I'm just worried that it may look like the timelike version of a conical excess because the spacetime that you started with before you pushed the two regions up and down, you've covered more than once.

%\vid Yes because of the double covering that we had in the Kruskal diagram.

%\rov [tears paper and wraps in to cone] This is a conical singularity, you can easily smooth it out.

%\speak{post-doc} And so you're saying that you have smoothed out that tip by adding the quantum space in between? Okay.

%\vid I don't feel offended if you have also questions to Carlo, we can both answer.

\kief For the formation of the primordial black holes you have the usual scenarios, so you don't differ from these? Theirs starts from an inflationary phase and derives the power spectrum, on which the details of the formation rates depend, is this what you use? But there it seems that there are only very few primordial black holes being produced because we have the spectral index smaller than one, point-nine-something, from the Planck data. 

\vid You mean the density contrast? 

\kief If you create primordial black holes from the fluctuations that emerge in inflation, the result depends on the power spectrum. It depends on the spectral index, \(p(k) \propto k^{n_s}\) and \(n_s\) is the scalar index. The smaller the \(n_s\) is, the more unlikely it is that you have primordial black holes. So my point is, I think this is a nice scenario, but you need the primordial black holes. 

\vid Sure, I am pretty new to the subject because I have started to get interested in primordial black holes very recently because of this work and so I had the chance to chat, for instance, with Ilia Musco and John Miller who have done very recently some studies. In their scenario, the formation of primordial black holes is very likely, in their case it is just because of density fluctuations. What they found is that the density contrast is enough to be 0.45 in order to have the formation of primordial black holes with a very wide spectrum of masses. The measured one is 0.5, so this is below and so this allows a rich phenomenology of primordial black holes. This is the picture in which I used to think about the formation of primordial black holes. In the last work that we have done with Barrau, Weimer, and Bolliet we mentioned the fact that their could be different possibilities for primordial black holes. For instance, in the case of exotic objects like cosmic strings and so on you can have also the formation of primordial black holes and this is not being taken in to account in the study that we have done so far. Maybe it's something we can do in the next work or maybe I can suggest it to people who like cosmic strings. Carlo, please, if you have more comments.

\rov One of the reasons people have been interested in primordial black holes is as a component of dark matter. At some point it was a hope, I believe somebody still has the hope, that at least a visible component of dark matter could be black holes.

\kief Yes it's true, we have long ago also discussed a model where primordial black holes contribute to dark matter. But there we had to use particular models, not the simple scale free models of inflation. For example, if you used the Starobinsky model of 1992, where you have a discontinuity in the derivative of the inflationary potential, you can create many primordial black holes. So there are models.

\vid You can have the production peak on the mass that you prefer also.

\kief Right, but if I go to the very simplest inflationary models, then somehow it seems at least that there are not enough primordial black holes.

\rov Right, the question raised is how many there are and there are these results which are that there are not enough, but they're not enough for dark matter. For being detected if they explode, they could be detected even if there are much less than the ones needed to account for dark matter or for a part of dark matter.

\vid I think that something that is very exciting to me in this model is the fact that the physics of primordial black holes changes completely because all the studies that have been done so far, for instance even the recent papers by Barnacka, they all constrain the primordial black hole using Hawking radiation. So in this case Hawking radiation is negligible and therefore the signal associated with them changes completely and new phenomenology maybe also new physics for cosmology and for the physics of structure formation can appear. I think it's open. If I could also add one more comment about this, I was mentioning this during my talk, that the best studies that we have so far excluding primordial black holes as candidates for dark matter they are coming from micro-lensing, for instance the Kepler mission, and its very interesting the fact that they don't observe black holes, objects provoking the micro-lensing, that are in the range of between \(10^{21}\) and \(10^{23}\) kilograms. As I said, if you believe this model all the black holes up to \(10^{23}\) kilograms they have already exploded so of course you don't see them.

\ford I think you said that the observed radio bursts tend to have circular polarization. Does your model say anything about polarization?

\vid Not yet.

\ford Obviously there is an obvious astrophysical explanation which would be synchrotron radiation, so in some sense you're competing. 

\stod Could they be rotating black holes?

\ford Yes, or could there be some kind of angular momentum involved that would give a sense of rotation.

\vid We know that most black holes are rotating at their maximum possible speed so it's very likely that we have to take this in to account in our model. Thank you for suggestion, this is something that we have to do.

%\speak{Lund} What can your model say about the lifetime of the white hole after the black hole has bounced? 

%\vid This was a confusion that was appearing also before, the white hole and the black hole are the same thing in this model.

%\speak{Lund} I see, so you're referring to the bounce process, the white hole in the classical sense that is the time reverse of the black hole.

%\vid The lifetime that we are computing is the time between when the horizon of the black hole forms and when the trapping surface of the white hole disappears. I wanted to add something because this was not coming out from the previous talk of Carlo, there is this picture of tunneling. For me the inspiration for this work came from loop quantum cosmology in which you model the time with some scalar field, and this is in the case of cosmology so here you have the scale factor, this is an expanding solution and you have a contracting solution for the universe and what you see using the effective equations of loop quantum gravity is that the points associated with the quantum equation goes this way. There is a point in which the quantum trajectory goes from one curve to the other and you can see it also in the numerical simulations. This is what we call tunneling in this case, there is a moment in which all the matter stops collapse and because of quantum effects instead you have repulsion, you have an explosion.

\spin I don't understand. In the wave function of the universe, you have two branches and you cannot say that you are on one branch and that you follow a trajectory to another one. In quantum mechanics when you have a stationary wave it's a superposition of incoming and outgoing waves and you cannot say that you are first on the incoming path and then you bounce and go on the outgoing path. I don't understand what you say.

\vid The idea is that you take a semi-classical wavepacket and you follow this semi-classical wavepacket back in time, it gets closer to the the big bang, but at a certain point you see\ldots

\spin The semi-classical approximation is not anymore valid if you insist on this point.

\rov But, that's the point. You use the full quantum equations, you evolve the wavepacket, and you see what happens. What happens is that instead of following the classical trajectory which would be to go smaller and smaller, it continues on the other.

\spin How could you put an arrow of time in this description? There is just two parameters, a scalar field and a geometrical factor. There is no time. 

\vid The scalar field is introduced as a clock in order to...

\rov It's an intrinsic time. Look there is half a century of discussion on that. The \(t\) parameter, it's gauge in GR so you just express the evolution in terms of relative evolution of one variable with respect to another one and you can arbitrarily choose...

\spin I agree that there is a long discussion about it, but I disagree with the interpretation of the proposal of the solution you have been talking about.

\vid I agree that this region is a region in which you are in the full quantum regime and instead of using the semi-classical equations you want to use the full quantum gravity theory, in fact you can do this and make this picture credible on the basis of the result in the full theory. I would be happy to discuss this more with you because I think it's related with some of your work.

\whit Do your calculations in loop quantum gravity give you any concrete information about the quantum region of spacetime that's involved? 

\rov Good question. No. Like the functional integral doesn't, in a sense. The transition amplitude is computed order by order in some approximation so it's a truncation, to a given order we just have a number associated to an initial state and a final state. Of course, given a classical metric you need the classical metric \textit{and} the extrinsic curvature, right? You need \(p\) and \(q\) initial and \(p\) and \(q\) final. Then you want to see what the probability amplitude of going from one to the other. That's like an overlap of wavepackets, it would be like you evolve a wavepacket very close to the wall then see what is the probability of being on the other side of the wall, in a sense. You don't have any information from this kind of calculation about the trajectory of the particle \textit{through} the wall. 

\whit But you could have some kind of expectation of the most likely position as it evolves. This is the kind of thing that Claus and Petr were able to do.

\rov Yes, so far I would say no. At least in the first order truncations we are using there is very little we can say about the intermediate states. Maybe to higher order, less truncations, a higher order approximation, you could say more but on the other hand the hope is that those are not needed because otherwise we can't do a calculation.

\vid If I can just say one thing. Just to be sure, we don't know exactly but for sure it's not singular, so there is no singularity there.

\whit I wasn't worried about that, I was worried about causal behavior where, say, \(r=\text{constant}\) lines are effectively spacelike or timelike. Because I have to change in that region, I wondered if you could explore that.

\rov Inside \(r=\text{constant}\) is obviously spacelike.

\whit Not everywhere.

\rov Inside, yes. Very much inside there so, very much outside not. Right, exactly. So, I don't know. No idea.

\whit In particular, it would affect what you would portray as the apparent horizon in this discussion.

\rov In a sense the interesting thing is not what happens at this point, at \(r=0\), it's what happened near the tip of the thing.

\whit Right, yes, I agree totally. So just to elaborate about \(r=0\), what you plot it looks as though the shell is going to \(r=0\) but in fact it's only getting to this minimum which you've calculated so it's spatially separated from zero. 

\rov Yes, \(r=0\) doesn't exist.

\whit Well, exist as something regular. So that raises another question, that means that around the edge of the shell you have a flat region that's everywhere flat. You must be able to actually join those flat regions together in some smooth way.

\rov You mean inside?

\whit Inside the shell, yes. Inside \(r_{\text{min}}\), which is very small, \(10^{-12}\)cm or something, but logarithmically it's far away from zero. 

\rov Yes, the gray region is flat so the past gray region and the future gray region in a sense can be joined.

\whit So they \textit{can} be, have you actually done that?

\rov The actual quantum region is only this funny shaped one there, so if I follow around \(r=0\)... let me put it this way: nothing prevents me to write the bounce just at finite \(r\) and inside have a continuity between the past flat region and the future quantum region. What I'm saying is that if I'm inside the shell I expect that basically I don't see anything, I see the shell approaching and then bouncing out. If I am very very small near \(r=0\), I just have my clock.

\whit The question is, when you try and match those two flat regions whether you'll basically have to be strongly red shifted or blue shifted from the shell in order to get the matching to be smooth. It's not totally trivial.

\rov In fact, in the paper we wrote a metric for the quantum region, and effective metric, which we don't want to give particular importance to but, nevertheless, we wrote it exactly to answer these questions like the question that she raised about smoothness. So we in fact have a metric filling up everything which is just written somehow smoothly joining, and in that metric which fails to satisfy the Einstein equations, in fact fails to have satisfied the energy conditions in the quantum region, basically goes smoothly to flatness when you go to \(r=0\) and the dramatic things in the region where you were mentioning, outside. Because the dramatic thing is the light cones are bent in before \(t=0\) and then quite rapidly they have to bend out.

\whit I think there is something dramatic happening at the bounce because there is one point in where I'd like to disagree with Francesca's diagram. I would argue that the apparent horizon, or that the trapped region, is bounded by the collapsing shell and the expanding shell and there's only one trapped region not two. On the outward going, they're bounded by the two parts that were the horizon.

\rov You need two, if you fill out the metric you need two by continuity, by topology. There is a topological thing that prevents it to be one. A trapped region is where \(r=\text{constant}\) becomes, inside it's spacelike and outside it's timelike. But \(r=\text{constant}\) can be spacelike in two different manners, \(r=\text{constant}\) is like that in Minkowski but it can be spacelike this way or spacelike the other way, and you can not go continuously from one to the other. To going spacelike this way, this is spherically symmetric, to spacelike the other way you have to go \textit{through} timelike. So, therefore, the trapped region upstairs and the trapped region downstairs must be separated by a region which is not trapped which is where the \(r=\text{constant}\) from spacelike goes back to timelike and spacelike again. We had a long confusion about that because that was what was preventing us to write this metric at first, because initially everybody had been thinking about a single trapped region. But a single trapped region is not compatible with a white hole and a black hole for the reason I just said.	 

\whit But it must be bound by the shell because once the shell goes inside the horizon it's bound by timelike...

\rov Yes, it's bound by the shell, I agree.

\whit So then it must go to the point where the shells intersect. The apparent horizon of each portion must go to the point where the shells intersect. 

\rov No no, the inner horizon can separate from the shell earlier if you allow the Einstein equation to be violated.

\whit If it's exactly Schwarzschild outside that wouldn't happen. What you originally were doing was mapping flat space and Schwarzschild and the only part that you were changing was this little quantum wedge. 

\rov Yes. Inside the quantum wedge, at some point the trapped region has to separate. We're still in the quantum region, in my effective metric, the inner horizon separates from the shell and comes in. The light cone on \(t=0\) must be symmetric. It can be neither this way nor that way. So \(r=0\) must be perfectly vertical in that diagram by symmetry. We are violating the Einstein equations inside of course.

\whit I think if you do the matching it doesn't have to be true that \(r=0\) is vertical, it only has to happen that it's not spacelike. It could still be symmetric even if it's not vertical. 

\rov This has been all the endless confusion we went through when we wrote this metric.

\vid I have a question for this outstanding audience, because probably the most interesting thing to do after finding this metric is not to go directly to rotating black holes but maybe to to trying to find a solution for not a shell but a dense sphere, so I was wondering if any of you have an idea for how this metric for a dense collapsing sphere should look like and if you have suggestions.

\stod Your shell is freely falling, there is no equation of state or anything?

\vid Well you can use the equation of state for photons because basically what is forming it is just radiation.

\thft You can also regard this sphere as a large set of collapsing shells. 

\stod Any way that these are formed in the early universe they're mostly neutrinos and photons and so forth, not much matter, so I don't know what comes out.

\rov Photons, that's what I expect to come out.

\vid Even though the explosion then gives some production of quarks. The high energy signal during the explosion can make the appearance of gluons. In fact, I forgot to mention this during my talk, in the last paper what we have done for the high energy signal was to use a code that is called Pythia, it's a standard tool for high energy processes. We have studied all the production of these particles.

\end{dialogue}

\pagebreak
\section{Wednesday, August 26 2015 \\ \textit{\small Convener: F. Vidotto} }

		\pagebreak
		\subsection{Black Holes as Open Quantum Systems \\ \textit{\small Claus Kiefer}  }
				{\small (Slides from the talk can be found at \url{physics.unc.edu/~dnmorse/Hawking_Conference_Slides}.)}			

%				\includepdf[pages=-]{./Slides/Kiefer.pdf}

%		\pagebreak
		\subsection*{Discussion}
				%auto-ignore

\begin{dialogue}

\ford Could you go back to the slide where you had the expression for your multi-mode squeezed state. You have that the squeezing angle is a linear function of time, I assume the Schwarzschild time.

\kief It is, yes.

\ford If you look at \(\psi(k)\), because it has exponential of the squared modulus of \(\phi\) in it, you can see it's decaying as the exponential of the square of time. That's, of course, exactly what you'd expect for an interference term in decoherence. I assume \(\phi(k)\) is the same as \(\phi_k\), isn't it?

\kief This \(\phi\) is the massless scalar field, I have small \(\phi\) and...

\ford That's not the same \(\phi\) then.

\kief Sorry if I was not clear. What we considered here was a massless scalar field capital \(\Phi\) on the background of the collapsing star, and for this \(\Phi\) we have studied the Schrodinger picture and this is the exact solution for the quantum state. Here if you look at that that is just the capital \(\Phi\), you start from a Gaussian and because of the massless scalar field it stays a Gaussian but it becomes a non-trivial Gaussian. This is a re-phrasing in terms of the quantum optics language. This small \(\phi\) is the squeezing angle, it has no relation.

\ford But if you have the squeezing angle, the phase angle, rapidly changing in time and say you look at an interference term and you average over that isn't hat going to cause that interference term to decay rapidly in time?

\kief Yes, I think this is just an observation that you have when you have decoherence. If you had maybe no rotation at all and you really had this ellipse sitting there then you would have to deal with the pure state. The point is that the pure state corresponds to the ellipse which has vanishing entropy. Then you see that the ellipse, the Wigner ellipse, rotates like that and then if you smear it out to this circle then of course you get an entropy. This is what decoherence gives you. This will be the disappearance of the interference fringes. 

\ford Right, and in a sense you already get decoherence at this level by just averaging over the phase angle. 

\kief This is true, but this is, if you like, a phenomenological description of the more fundamental thing, you also had Paul Davies' question, if you have, say in quantum optical system, the environment and the coupling and then you solve the full Schrodinger equation you have the full entangled state and you trace out over part of it, then you have the microscopic description. One can often show that at a heuristic level this corresponds to this smearing out of the Wigner ellipse.

\ford I guess the thing that confuses me about your description of the environment is, it's a little vague what the environment is and in particular what the coupling parameters are. Usually we expect decoherence times to depend sensitively on the coupling constants.

\kief Typically, it's one over the coupling constant.

\ford It looks like you really have, in fact, decoherence that does not depend on any coupling constant if you just average over this phase angle. 

\kief Yes, I admit that it's vague in that sense. But vague in a certain way because such situations are frequently discussed in quantum optics and you can, for example, base them... well there different levels. The most fundamental level, is you could do the calculations, is to have the full system having the full wavefunction and calculating the exactly the Schrodinger equation and tracing out part of it. This you can only do in very simple cases. The next level is to discuss the master equations because one can show that a huge class of such models gives you the same form of master equation which is a Lindblad form of the master equation so you have the usual unitary term, the \([ \rho, H] \) commutator, and then you have this \(L \rho L^{\dagger} \), and so on. This means that many details of this coupling, that of course I'm vague, I have not written down, go in to this Lindblad operators. From these master equations you can draw such general conclusions that it corresponds to the smearing out. Of course, if you go to some experiment, then you will have to go to the more microscopic description. For example, in cosmology we have done that for the quantum to classical transitions for the primordial transitions. There we have used really concrete models of fields and their interactions, for example coming from string effective actions, and calculated this. There you can also translate it in to this language which I have not done here, I have just used here this phenomenological language of smearing out.

\stod I'm not sure I understood correctly, but you showed that you were getting the number distribution according to the Planck distribution, you were getting the Planck result for the number distribution, and then you said something about that certain correlation functions are different. Does that mean that you get that the Hawking radiation is not purely thermal, it's different than a classic thermal source?

\kief What I mean is the following, here this particle number operator for mode \(k\) is \(a^{\dagger}_ka_k\), if you go to that pure state and you calculate that expectation value you get the Planck form. If you calculate, for example, \(\langle a^{\dagger}_ka_ka^{\dagger}_ka_k \rangle\) they don't factorize as they would for some ensemble. They have more information, if you could measure or you could calculate it and write it down then you would explicitly see that for such expectation values you have a difference between the canonical ensemble and this state. You could measure this radiation, you could see this is not thermal.

\stod Do you think that the Hawking radiation, if it's ever observed, is not truly thermal?

\kief I would think so, yes.

\stod Is this what the experts say, or not? I always read in the newspapers that it's supposed to be truly thermal.

\kief I think that this is the issue that we'll discuss here at the conference. Whether it's exactly thermal or whether there are some correlations that go beyond thermality. Certainly as you see from this, I mentioned at the beginning that also in the original calculation by Hawking, and I mentioned Parker's calculation from 1975, that he did not use a canonical ensemble exactly. He makes these calculations of expectation values.

\stod The reason that you think this is true is because of this squeezed state argument?

\kief Yes.

\whit First of all, can you explain why you described this as a two mode squeezed state rather than a three. Just from what you've said here I don't know how you can tell modes there are.

\kief Actually this is a result, it's not that this is an input. This here is the result of the calculation, so the quantum state. This state, how does it come? It's the solution of the functional Schrodinger equation with an initial state, so the initial state of course you need and this was at early times the Bunch-Davies state. Just the vacuum state for harmonic oscillators, in this collapsing background you get this state. This you can re-write in this form, this is just an algebraic re-writing in terms of these two parameters and looking at this and comparing it with quantum optics treatments you see that this is what is called a two mode squeezed state. A one mode squeezed state, if you start with the vacuum the one mode squeezed state is, if I remember correctly 
\[  e^{\tau a_k^2} |0\rangle = \left.\left| \text{1 mode} \right. \right\rangle \]
with some parameter \(\tau\). The two mode squeezed state is if you have a mode with wavenumber \(k\) and one with wavenumber \(-k\) then you have
\[e^{a_k a_{-k}} |0\rangle .\]
If you do that, if you start say with the harmonic oscillator in quantum mechanics with your ground state, applying this you get a Gaussian state that has this form.

\whit Sorry I didn't understand the terminology, you really mean two modes per \(k\) or two modes per frequency?

\kief Yes two modes. So \(k\) and \(-k\).

\whit My second question concerns what you were saying when you were doing this quantum gravity calculation where you had one mode, in a different sense, for the black hole and one for the Hawking radiation and you said you had a back-action between them. Did you use the semi-classical result for that back action?

\kief Do you mean the last part of my talk?

\whit Yes.

\kief No, there this was an exact calculation but the model itself was oversimplified. 

\whit What in particular did you use for the back action?

\kief Linear interaction terms between the oscillators to have it as an exactly calculable model. What we did was the following, and these are things that are often used in quantum mechanics, having the black hole degree of freedom described by \(x\) as an oscillator but having an indefinite kinetic term to have a property from the Wheeler-DeWitt equation impressed. This was important actually for the result, having a positive definite you would not have got it. Then we have an oscillator term for the Hawking radiation, to mimic it, and the back reaction you have then mimicked by a term \(\mu\) where \(\mu\) was a coupling strength. This is what was varied here in this. The \(\mu\) times \(x\) times \(y\), so a linear interaction term so that you start from a product state of oscillators then you can calculate this exactly because the interaction term is just bi-linear, otherwise it would not be possible. We have a full entangled state which is certainly not the correct state because they're oscillators, but to see some aspects, for example this indefinitnss of the kinetic term, and then it's a long expression and here we calculate then the black hole part by integrating out the Hawking radiation part, then the next slide just the opposite. 

\whit Basically \(\mu\) was treated as a free parameter?

\kief Yes it was, but as a free parameter of the model which gives the strength of the back reaction and it increases. Here is no back reaction so this means we have an initial Gaussian state so it will just evolve as a free Gaussian state, but then it becomes stronger and stronger then you see here that this very different picture emerges. 

\whit Would you expect that it would be more appropriate from a physical perspective to have \(\mu\) as \(k\)-dependent? Or not?

\kief Yes, of course. It would have to be \(k\)-dependent. There would be many steps needed to make this more realistic and one step would be to have it \(k\)-dependent, right. 

\dav You've explained very nicely how the decoherence with the environment might solve the information paradox, but we use information in two different contexts here. Most of the discussion takes place around retaining the phase information, or the coherence, of the quantum state, but we have other information-al labels as well, like the baryon number for example. Presumably, that still has to be violated in the decay process, and any of those other labels we might like to attach, the necessary gauge fields conserved. Is that correct?

\kief That is correct I would say, this is not considered here.

\dav Right, when we're couching this information problem it's as well to be clear because we often say `you can't tell what a black hole is made of, whether it's matter or anti-matter, or green cheese, or all that stuff.' But actually, the issue is not really so much about that, the issue is about retaining the phase information from the wavefunction. We still have to give up the other things. As John Wheeler long ago used to say, `A black hole transcends all of these things.' But Maybe it doesn't transcend the quantum coherence.

\kief My talk was really on this quantum coherence and the issue of the pure state to mixed state, or pure state to pure state issue.

\dav Right.

\kief You're right, I agree with what you said.

\thft As for the thermal nature of black hole radiation, it was pointed out that there is a way to get non-thermal Hawking radiation if you put a black hole in a strict vacuum, then it slowly emits particles so it's mass decreases, it's temperature increases, and since temperature isn't constant that isn't quite thermal. On the other hand, if I were to put a black hole in a heat bath then it would be in equilibrium with it's surroundings and then, of course, it would be perfectly thermal. It would be perfectly thermal if you put a black hole in a heat bath with the same temperature as the Hawking temperature, then it would be in complete thermal equilibrium then all we know about thermodynamics would then say the state has an exact thermal spectrum, the radiation that comes out. I don't know what kind of deviations you're thinking of. If there would be no interaction between ingoing things and outgoing things it wouldn't matter what you immerse your black hole in, it would continue to emit a perfectly thermal spectrum except there is a cutoff at high angular momentum and things like that. Also, if you look at the calculations it is not so easy to get non-thermal effects because also spacetime as you know, a complex spacetime has a perfect periodicity with the inverse temperature so that all field theories you do in there automatically becomes a perfectly thermal field theory. Only if you take back reactions in to account do you get deviations from that.

\kief Yes, I agree, it would certainly mimic thermality. It would be, in a sense, the case of what you have in quantum Brownian motion, in quantum mechanics where you also have a quantum particle, but put it in a heat bath and you can solve it at a quantum mechanical level, but effectively you see just that: the classical Brownian motion, but somehow justified from the fundamental quantum theory.

\kief So, if there are no questions, then maybe I can ask a question, if someone has more information about these quasi-normal modes in this context, or some opinion about the relevance of quasi-normal modes in the discussion of here.

\dav A thought occurred when you were discussing that. When we're thinking about the back reaction, is this going to excite those quasi-normal modes? Is that part of the back reaction calculation?

\kief Maybe, I don't know. My opinion is that one should take that in to account, those quasi-normal modes, they are really a fingerprint of the black hole. It's not if you are inventing some action from string theory or somewhere else, this is something that is there already and should be taken seriously. In classical relativity it's a standard issue to be discussed, there are many results. I tried, I found this result with the minimal noise temperature, but I did not get any possibility to really calculate an entropy.

\dav Another thought occurred to me, this may be irrelevant. Black holes have an intrinsic impedance, electrical impedance, it seems to me that's got to be related to this if you just think in terms of circuits and noise and impedance. 

\whit On the question of the quasi-normal modes, we could describe the perturbation, well some perturbed field in terms of an integral along the real axis in \(\omega\) if we have some state. But if we tried to deform that contour around the quasi-normal modes by pushing it down we end up with a contribution, so we can pick out the quasi-normal modes as we go down, but there are also contributions around the branch cut. It turns out that there are branch cuts not just at \(\omega = 0\), but at each one of the algebraically special frequencies. Nobody has a good sense of being able to say that the quasi-normal modes are complete in that since that you can describe every perturbation by a sum of quasi-normal modes. So I think that while it may be true we can understand things from the quasi-normal modes, if you excite the horizon then the quasi-normal modes are excited, they will evolve. It's not clear that they are enough, and it's definitely not clear that they give you all you want for the entropy.

\kief Thank you. Yes, Laura.

\mers A few questions, the first one is how much does the result depend on the scheme of coarse graining?

\kief You are referring to the earlier? Not the quasi-normal mode issue?

\mers Yes. If you were to choose a different environment?

\kief Good question, this also was related to the question about the details of the decoherence mechanism. I can say the following, this is only done at this heuristic level. We have done in a similar context, the primordial fluctuations in cosmology, we have done extensive calculations to see. There you can really do some concrete calculations that you have, for example in the fields some \(\phi^4\) coupling or others, of course depending on the interaction you get different results. But this whole class of models that you can discuss can be mimicked by Linblad type of master equations. If you discuss that and discuss the decoherence times there of course you'll get a dependance on the coupling strengths, you get a bit different result for the decoherence time, but all in all qualitatively the result is the same. So, it's rather robust. The details of the coarse graining do not enter unless you make some very peculiar coarse grainings. For example you could, of course, for some coarse graining that just squeezes the ellipses somehow in a weird way. I would say that here also it is rather robust though it's not calculated.

\mers Okay, because the mass/temperature relation of Hawking radiation is universal while normally coarse graining schemes have some degree of fine tuning built with them from the choice of what you call the environment and system. I was trying to understand if obtaining Hawking radiation from this method is as universal, does it have the universal feature built in to it or not?

\kief Yes, it has this universal feature that as long as you go here, well I start from this model where you have this pure state here, this pure state that has this form in the language of two mode squeezed states, and for that if we have this rotation of the Wigner ellipse which happens on that time scale, so this is basically 14 or 17 times the Schwarzschild radius, which if you insert here some astrophysical models or even primordial black holes are very small numbers. We have a rapid rotation of this Wigner ellipse here, so you have here this picture that you have the \(p\) and the \(x\), well in quantum mechanics you talk about field amplitudes of course, and you have here the Wigner ellipse which corresponds to the pure state but then you have here the rotation. For any situation where you can not resolve time smaller than this the situation is indistinguishable from having this circle and if you calculate the entropy of that then you get the standard result which, if you integrate over all modes, gives you the entropy of the Planck temperature. The limitation in this simple picture is that if you could observe it, well first have a primordial black hole, you observe the Hawking radiation, and somehow you have to do measurements on a very short time scale and having correlation functions maybe of this type that refer to these small time scales. If you could do that then you could really see the quantum effect. 

\mers A related question to what was discussed yesterday morning as well, if we think of Hawking radiation as being obtained by this method and we have the wavepacket that in real spacetime would correspond to a black hole plus radiation, and you throw a particle in it and you expect the outgoing to be correlated to what goes in, how does that translate in your picture? Would that mean that the wavepacket would shift to a new classical trajectory in the phase space?

\kief You have the picture of having a particle described by a wavepacket putting in to a black hole?

\mers You obtain the solutions you have here and then you decide to throw a particle by hand in there, introduce a new particle that you throw in to the system. What would happen to the wavepacket solution?

\kief If I understood correctly, if you threw something in to the black hole then it would no longer serve as an environment in this sense I would say. It has a relevance for the black hole quantum state but not for this issue with the Hawking radiation probably, at least not at a qualitative level. Maybe some numbers would be changed. 

\mers And that would correspond to a new classical path of this wavepacket \(\psi(k)\)? So if you were to throw a particle in to there what would happen to it?

\kief Of course, it would increase the mass of it, to do it in full you would have to have the backreaction on it, this is certainly not included, but if you have just a test particle described by a wavepacket then, of course, it would have to be added to the initial state and it would basically modify the mass, at least on short time scales. I'm not talking about time scales longer than this Page time because if you assume unitary evolution then we expect that something happens at that time, but at least at times smaller than that it just increases the mass.

\mers The reason I ask is because that expression depends only on the mass, so the only way you would recover any information about whatever you throw in to the black hole.

\kief This is not yet a state that gives an entanglement with the black hole as a quantum object. This is mimicked later in this way, well here we just address the Hawking radiation in this background of the collapsing star, this is an exact solution. Here in these considerations there we start from the full Wheeler-DeWitt equation actually of the black hole, this was done for the dilatonic black hole, for the CGHS black hole because there the calculations could be done. There you could solve the Wheeler-DeWitt equation in this semi-classical way we just have not included technical details here, but I could certainly say something more. \(S_0\) obeys the Hamilton-Jacobi equation which gives you then the background, actually we chose then this solution so that it corresponds to this incoming shockwave CGHS black hole. There is a concrete expression for the \(S_0\). This then gives a correlation between the black hole quantum state and the Hawking radiation, but approximately only. We have no theory where we have exactly here the full quantum state of the black hole and the Hawking radiation. Then, if you have such superpositions and you integrate out the Hawking radiation, then you can have decoherence factors of this form, so for example the black hole and the white hole the non-diagonal terms of the density matrix which are something like \(D_{\pm} \approx e^{- \text{some number}}\) and then of course you have some infrared divergence because you have infinitely many modes and, if the volume is infinite, what you get actually is something like \(D_{\pm} \approx e^{- \text{some number} \times L/4GM^3}\) where \(L\) would be the length, the linear size of the box where you put in the black hole. Of course, you would trivially get here this goes to zero for \(L \to \infty\), but this is often the case for infrared divergences. You would have to think about some appropriate cutoff. This is what we did here, this is from the Wheeler DeWitt equation.

\mers The last question, tracing out the Hawking radiation to obtain the decoherence, is there any connection with the back reaction that I was talking about on Monday? This is a different method where everything is done in a minisuperspace and with the Wheeler-DeWitt equation.

\kief You mean the connection with your talk on Monday? I would have to think about it, we should discuss this. If I remember, you did not have the full quantum gravity right? You had just a collapsing star and some classical ansatz for your \(\tau_{ab}\), perhaps one could have a quantum picture of it, but I think it's not yet quantum gravity. We should discuss this.

\whit I'm a little surprised by this expression you wrote down for the entropy of the Hawking radiation. This looks as though you are treating the Hawking radiation as something in flat space, not as something that's living in the spacetime of the black hole. Is that correct? This is what Laura's calculation takes in to account, the fact that quantum states around black holes, especially near them, are very different from quantum states of radiation in flat space. You don't seem to be including that, is that correct?

\kief You are right, it does not follow from this simple coarse graining picture. I would say this a little simple to have the corrections from the vicinity of the black hole. If you had this quantum state given, wherever it came from, it would give you the local equivalence to this thermal state. The information about the black hole is somehow encoded in to this state. You would also get such a state from just having the Unruh radiation where there is no black hole if you calculate Unruh radiation in the functional Schrodinger picture, depends on the acceleration.

\whit This state is position independent, we would expect some position dependence around a black hole. The energy density can be negative close to the black hole even if it's positive far away, for example.

\kief But the quantum state in the Schrodinger picture only depends on the configuration space of the field, right, and on the parameters of the black hole. How should here some dependence come in?

\whit Let's go to the calculation, what did you actually do to get the entropy of the radiation? You end up with a volume, there's no volume in this, but you end up with a volume. Does the volume come from just putting in an ansatz of a flat space integration? Or are you using a curved spacetime integration to get that?

\kief To get that quantum state?

\whit To get the entropy.

\kief Well actually this is just found at this general level of having in quantum mechanics a two mode squeezed state of that form and doing that coarse graining. Using the analogy of the state \(\psi(k)\) with a squeezed state in quantum mechanics and then just taking this quantum mechanical state and doing this, then you get this which is actually a standard result. The analogy with the black hole has only been encoded in this squeezing parameter \(r\) and then the squeezing angle \(\phi\), but if you did not know that then you would just have an ordinary laboratory squeezed state and you would do that coarse graining. That's how it's found.

\whit It's treating that quantum state like a laboratory squeezed state in a flat spacetime.

\kief Yes, that's true.

\ford I would like to ask about the last comment on this slide. Given that your decoherence time seems to be very short, it's not clear why anything special should happen at the Page time. Why shouldn't this hold even after the Page time?

\kief The idea here is based on that you have this semi-classical Hawking radiation and then the black hole. Page starts also from the assumption that the full evolution is unitary and that after a certain time the Hawking radiation is no longer in this thermal state. 

\ford I thought that was in the context of a model where you think that there has to be some correlations to solve the information problem. It looks like if you've already had decoherence perhaps you don't need that picture. I'm not sure the Page time has any special significance in this model.

\kief Your opinion is that this discussion could extend even beyond the Page time? In principle sure, but for the quantitative details...

\ford It would appear if things have already decohered on a few times \(M\) it's not clear that the Page time has any significance any more. 

\kief If you, say, started this after the Page time so the black hole is already in a state far evolved. Page time is of order \(M^3\) and you start then with this consideration, this is what I had in mind. Of course if I had earlier decoherence the Page time is of no relevance but if you started later for developed primordial black hole.

\ford If you're not relying on correlations in the Hawking radiation to solve any problems then in a sense if you see a black hole of mass \(M\) it's past history is not that important, whether it came from the evaporation of a black hole of \(10M\) or it's a fairly recently formed black hole they should be essentially identical. 

\kief You mean it should be robust on the result?

\ford Yes, at least it would seem that way in this picture.

\end{dialogue}  

		\pagebreak
		\subsection{Particle Creation From Vacuum in Gravitational Expansion and Collapse \\ \textit{\small Leonard Parker} }
%				{\small See \verb|tHooft.pdf| file found in the slides directory which can be found at \url{physics.unc.edu/~dnmorse/Hawking_Conference_Slides.tar.gz}.}	
				
			%	\includepdf[pages=-]{./Slides/Parker.pdf}
	
		%\pagebreak
		\subsection*{Discussion}
				%auto-ignore

\begin{dialogue}

\mers The origin of Hawking radiation. How important is the collapse process to producing the particles, what later becomes the Hawking flux at \(\mathcal{I}^+\)?

\park Let's take an example, for example from Misner, Thorne, and Wheeler. There is a discussion of a collapsing dust cloud with a finite extent joined to Schwarzschild, with at least the first and second derivatives of \(a(t)\) continuous. That's why I call this gravitational expansion and contraction. Yes, in that model it is clear there is particle production and you can calculate it in a very similar way. Because of the collapse of the dust cloud, you get a lot of red shifting as Stephen did a very elegant calculation, much more elegant than I would have been able to do, to show that. He didn't use the actual geometry that you can solve for in that simple case, more general. There is a lot of particle production that is occurring and the particles that are being created are created in pairs as you approach the horizon before you go in, of course, there is a particle that goes out along the event horizon which hasn't yet appeared, and another one that is pair created and in order to conserve energy you have to have that go in and bring negative energy in to the black hole relative to infinity. You do have pair creation outside. Did I understand your question?

\mers Yes. There has been some of discussing about this, if I were to talk about my understanding I think the collapse is crucial because you need the changing gravitational field to get the mixing of positive and negative frequencies and end up with a particle production process, and also to explain the time asymmetry of getting a flux, there was no particles before, but then something happens and you have some particles later. Another argument, and we have been having this discussion with Jim Bardeen and Bill Unruh and Don Page for quite a while now, but the other argument is once the black hole forms we speak of a surface gravity and the temperature and an entropy of the horizon assuming the hole is in equilibrium with the surrounding radiation. That would imply that the radiation should already have  been produced in order to speak of a temperature of radiation and associate that with a temperature of the horizon. That has been the debate. There are others, and I'll let Jim speak about this, that take the view that radiation is produced after the black hole has formed. It would be good to know which one is the case. I personally don't see how a semi-stationary object can give rise to a flux of particles. Let's think of an eternal black hole for example, and I also don't see how that object would break the time symmetry, and why, and what meaning the temperature of the horizon of that object have if particles do not exist yet, if the radiation is not produced yet. But I should let the others argue about this.

\bard I think nobody questions that there is particle production during the collapse, but the question is how much and what fraction of the mass of the black hole was radiated due to particle creation during the collapse phase, as opposed to later on. What's your opinion on the order of magnitude, in terms of the Planck mass and the mass of the black hole, of the fraction of the star's mass that would be radiated due to pair creation during the collapse?

\park Let me just try understand fully your question. You're saying how did the Planck mass come in to it? You're getting to the point where the black hole is down to the Planck mass you're saying?

\bard During the collapse, up until the time the collapse is over and the star is inside the black hole, which is the usual picture, what fraction of the mass of the star do you think would be radiated during that phase of the evolution? As expressed in terms of the Planck mass and the mass of the star, or the shell?

\park So let's give them names, the mass of your star is \(M\) or something before.

\bard Say you have a star of mass \(M\) and then \(\hbar\) in GR units is the square of the Planck Mass, so I would guess that it's proportional to \(M_{\text{Planck}}^2/M\) which is incredible small for a large black hole. Would you agree with that or not?

\park Off the top of my head what you're saying sounds reasonable, but I'd have to sit down and talk to you about it I don't want to just give a top of the head answer. Since you two have had perhaps discussions about this it's premature for me to try to. You're saying you can get this out of just dimensional considerations?

\bard Basically.

\park Right, that seems reasonable, I guess. If I understand what you're saying, which I probably don't fully. 

\stod Your exponential growth on the previous slide, there's nothing to stop it, it just keeps going? Or is there some back reaction we have to take in to account?

\park I was thinking about that, in fact, of course, that's the question `how do we get reheating?'

\stod Even within the context of the abstract calculation, it just keeps going? Or have we made some approximation where there is some back reaction, perhaps, which has been neglected?

\park My calculation was just based on the ten dimensional deSitter symmetries. I've been considering the possibility for another mechanism for reheating which I'll just throw out there, I think that probably it's been considered by Stephen I would bet. In any event, if you have your exponentially expanding universe, I would think that you would have a certain probability for black hole production, black holes which as you're going they would be producing radiation, or evaporating, and the model I was hoping to think about, which you guys can probably do much quicker than I can, is where you have energy produced by the set of evaporating black holes which is created during this exponential stage. There is a number of little points to it that maybe I could get some advice from people about but it strikes me that as these black holes evaporate they could serve as a source of reheating at some point. You could get some calculation of the possibility that they could end the inflationary stage at some point and bring you in to a radiation dominated expansion. Here, I would like to ask you all a question, during this situation suppose that since the dark matter, we don't know what dark matter is, but one thing we do know is that it interacts gravitationally, or it seems to, it's sitting in galaxies and so forth, so it could be weakly interacting matter or whatever, but the point is that whatever it is it's going to be created by the black holes that get produced as part of the inflationary expansion. If you take that in to account you should be able to get an idea about what temperature the radiation would reheat the universe to. It might tell you something about the content of the universe because of the fact that even things which only interact gravitationally uniquely get created by the black hole. That includes dark matter. I think that could be important. This is just some thoughts I was having and I'd be very interested to know what's wrong with it, there's probably stuff wrong with it. 

\stod Presumably, your effects also produce dark matter right, you've only done the zero mass field or something, but in principle it should also produce dark matter.

\park Yes, I guess so. Here I'm thinking from black hole production, as far as I know there's no reason why you wouldn't also get dark matter created. It's just through gravitation. I don't know. I'd be interested to get feedback. Is it obvious what's wrong, is anything wrong with this possibility? It can give you numerical results which you can try to test. The presence of the dark matter, you want to get the right percentage of that, and you have a reheating temperature you want to get to to exit inflation, and you can actually try to put that together and figure out what happens. If it works.

\whit Dark matter is usually considered cold, presumably if there is a dark matter particle it is heavy and if the dark matter particle is heavy it's going to be emitted only at the very end of the evaporation of the black hole. So it can't be a dominant part of the total energy generated.

\park Here I'm talking about microscopic, very tiny black holes created during inflation.

\whit You could calculate what the number?

\park They evaporate at a very high temperature according to Stephen's calculation.

\whit So you could use your model to put a constraint on the number density of small black holes in the early universe because there's only a limited amount of dark matter available.

\park Things like that, yes, if we run in to contradictions in temperature, yes. 

\mers You studied the neutrino flux production in the 1960's, 1965 I think the paper? 

\misn You're asking about what?

\mers I think you studied a similar problem to what is being discussed here, but for neutrinos produced a black hole.

\misn I don't recall discussing particles produced by black holes, no.

\park I think Charlie looked at neutrino viscosity.

\misn Oh, yes, I played that game of neutrino viscosity, but that just assumed that there were neutrinos there and hoped that they would do something useful. That was in connection with Mixmaster business that never fit in to the real universe.

\mers If there are no more questions can we bug Malcolm to give us the answer to the question he promised to Larry and I yesterday? About deSitter, flat, Rindler...

\per The Rindler horizon is not an isolated horizon, it has expansion associated with it, whereas the horizons we were considering have zero expansion, so that's the explanation of what's going on in the case of Rindler. That deals with Larry's other question.

\mers So you would the effect on deSitter, but not on Rindler?

\per That's correct.

\mers Which means that we would be able to tell the difference between accelerated observers and pure radiation?

\per Yes.

\mers Malcolm, one last question: why don't the correlations between the ingoing and outgoing particles create a firewall problem?

\per I have not thought about this.

\mers Tomorrow. 

\end{dialogue}  

		\pagebreak
		\subsection{Gravitational Condensate Stars or What's the (Quantum) Matter with Black Holes \\ \textit{\small Emil Mottola} }
				{\small (Slides from the talk can be found at \url{physics.unc.edu/~dnmorse/Hawking_Conference_Slides}.)}			
		
%				\includepdf[pages=-]{./Slides/Mottola.pdf}

%		\pagebreak
		\subsection*{Discussion}
			
				%auto-ignore

\begin{dialogue}

\thft I wonder if you have a material where the density is constant? Does that mean it is incompressible and therefore the speed of sound is infinite? Suppose you impose the constraint that the speed of sound should not be bigger than the speed of light, then what would you get?

\mot In fact that was Einstein’s objection to Schwarzschild's second paper, he didn't like it because constant density meant exactly that, which violates, of course, relativity. I don't think this should be taken seriously. These intermediate states, which I went through the sequence of solutions, I don't think are physical. It was simply a model to get to where I want to get to, and once you get to the deSitter interior you don't have this problem because the deSitter system is a vacuum. There is no speed of sound that you can calculate that way. If you just take pure deSitter space I don't know what the speed of sound is, it's ill defined in fact. You have to put some material in, the vacuum itself has no speed of sound. I think I perfectly agree that each one of these members of this sequence is not sensible but I would like to take seriously the limiting case as a dark energy interior. 

\ford On your last remarks, I was a little confused about whether you were suggesting that all the things that we think are black holes are really these objects?

\mot Yes. I'm going to go way out on a limb and suggest that.

\ford You're thinking that all the things in the centers of galaxies are really these and not black holes? You solve the black hole problem by not having any black holes anymore. 

\mot It's nothing short of that, yes. I mean that's a speculation, to be sure. There was a comment the other day, I think in Laura's talk, about whether or not black hole horizons have been observed, and of course I'm intensely interested in that question, and I would say that the issue is still open. Astrophysicists think they can argue that there is an event horizon, but the argument goes something like this: We know something about accretion on several of these objects so something is falling in, but then if it were to hit a surface it should thermalize. Then we should calculate the flux and see what's coming out, and there's no thermal radiation and therefore there's no surface. That's roughly their argument. Of course that depends on what happens to the material as it hits the surface, does it thermalize? Does it get bounced back? Does it re-radiate? Or does it get absorbed? If it gets absorbed, or mostly absorbed, you won't see it.

\ford At least in classical relativity you can form black holes of extremely low density, so it would seem hard to see why the physics that gives rise to this special phase would apply if we tried to make sufficiently large black holes.

\mot Well I was going to try to explain that but I ran out of time. The point is that the vacuum polarization effects, as you well know, for example in a Boulware like state, always get large near the horizon no matter how big the black hole is. It's just a matter of getting closer, this \(\epsilon\) parameter gets smaller, so you have to make this \(f(r)\) smaller before it blows up. But no matter how small the curvature is this is a quantum correlation effect which has nothing to do with the local curvature ultimately, it has to do with non-local correlations which I think are described by the conformal anomaly. But I agree, it's a challenge to understand how this is formed, that's of course the open problem I'm working on. I think the possibility is there.

\spin Just a remark, it's not mine by the way, about the constant density, this remark came from the book of Misner, Thorne, and Wheeler, Gravitation, when you solve the Einstein equation even if you have a solution with constant density, you can imagine some specic matter such that it's always constant and the pressure is a mechanical pressure, it's not the thermodynamical one. What you need is an equation of state.

\mot Yes, here there's no equation of state, I just solved the equilibrium, hydrostatic equilibrium.

\spin So we cannot say anything about the velocity of sound if we don't have any equation of state.

\mot Well, okay. It's one reason I think the Schwarzschild solution was cast aside for many years, because the master, Einstein, didn't like it for exactly the point you raised. I think that simply doesn't arise in a deSitter universe. We don't talk about the speed of sound of dark energy, it has no speed of sound.

\duf Normally you think if you have a theory of elementary particles where the particles are heavier than the Planck mass they would qualify as black holes. Does your theory rule out those too?

\mot No, I have nothing to say about Planck scales. I'm not trying to solve the problem of quantum gravity all the way to the shortest distances, I'm trying to work in an effective language for large black holes.

\duf I'm sorry, I thought I heard you say there are no black holes.

\mot Well, okay, if you talk about Planck scale black holes I have no idea. I tried to say that the thickness of this shell is the geometric mean of the Planck scale and the Schwarzschild horizon radius. It's very small, it's like Fermi, for astrophysics but it's very large compared to the Planck scale and there we should still be able to do something like semi-classical methods. It's your favorite topic, the conformal anomaly, which comes in there. I want to not cancel the anomaly, at least at these low energies, I want to use it in the same way I have an effective in low energy pion physics which comes from QCD and the chiral anomaly, I want to use the conformal anomaly to tell me something about these effective quantum correlations. 

\louk Could you draw the conformal diagram of the spacetime?

\mot Not easily enough because this region is completely blown up [in the conformal diagram]. What you should do is put in a regulator, if I put in this \(\epsilon\) then the surfaces are really spacelike tubes and then I can draw a conformal diagram. If you take it literally to be null, I'm not sure how to do that, so it's an interesting little exercise. We can do it, but I haven't done it.

\louk Can you draw it with a regulator in place?

\mot Yes I think so, I think then it's just a star. It's a star. You have a region of deSitter and you have a region of Schwarzschild separated by some finite boundary which is spacelike.

\louk By some timelike boundary.

\mot Yes. If you take that limit it's kind of tricky, so you have to be careful.

\bard Do you have any ideas about what would happen astrophysically if something like this existed, in terms of what sort of signal you would see?

\mot Well that's what I discussed.

\bard If there was accretion, for instance.

\mot In order to talk about matter? That's what astrophysicists want to know is what happens with rotation, what happens with magnetic fields, and what happens with matter? All great questions and I've only scratched the surface on that, I'd love to hear some ideas. Basically these are speculations. I don't see any reason why the system could not support currents, for example, so the way black holes just absorb energy and the magnetic field collapses may not actually be the case if you have something like a real current sheet there. That involves putting the standard model interactions in to this surface which is going to take a lot more work, but I think it can be done in principle. The first thing I'm trying to do is actually construct, this goes back to your talk also Gerard, you mentioned the monopole and how you get a singular solution until you put in some extra term that regularizes it, and that's exactly what I'm trying to do: to put in the extra terms coming from the anomaly, essentially, and demonstrate a non-singular solution. Once you have a non-singular solution then we can talk about small fluctuations in a meaningful way and we can talk about interactions with standard model fields. At the moment, with this very simplified model, I can't do that. I don't have enough physics in it yet. The only thing I can say qualitatively is that if it's a closed surface and I hit it it should act like a soap bubble, it should oscillate at normal mode frequencies without an imaginary part, as in the quasi-normal mode description, and that is a stunning signature if you would ever see it in gravitational waves. 

\stod The properties of your condensate are just given by gravitational quantities? The Planck mass, speed of light, \( \hbar\)? Or are you waiting until you include the standard model?

\mot I tried to motivate the idea that a condensate could form even from QCD because we know that are such condensates in QCD, so that's sort of starting the process, but no, the full answer has to involve standard model fields, the conformal anomaly, and the formation of a condensate. They're going to interact with each other, it's not just pure gravity, it's not just pure standard model.

\stod The parameters of this condensate are not just pure gravity?

\mot Well, if this object exists at any scale then the condensate has to be density dependent, has to be able to adjust. So, yes, in that sense it's the gravitational effect that's dominating. It may be seeded by standard model processes but in the end this has to be a gravitational object, right, where I can discuss it all in terms of the gravitational field equations with some corrections, that's the goal anyway.

\thft According to general relativity there are many other ways to make a black hole, you could simply collide two very energetic particles and make a black hole. My favorite is a cloud made of television sets as large as the galaxy. You let these television sets attract each other gravitationally and while they go through their own horizon the cloud is still so dilute that all television sets are still working. In other words, ordinary laws of physics apply while they go through the horizon. All you use is the principle of general relativity, that coordinates are equivalent, and nothing more. And then you get already that they have to go through the horizon, am I assuming something?

\mot You're assuming classical matter. You're assuming essentially an energy condition. You're assuming that I have ordinary particles, that can be television sets, but they're just ordinary particles.

\thft The energy condition is trivially satisfied by those television sets, they're classical objects that we totally understand, and they're very dilute so the density is much less than that of water. And still you can make a black hole.

\mot The point is here that dilution is a local concept, it is not what determines whether or not this condensate forms. The question is whether in the preparation of the quantum state, which is not in your description with television sets, whether or not the quantum vacuum can produce vacuum polarization effects near to the forming horizon which can trigger this process. That's not in the classical picture at all, right?

\thft What would be wrong with the television set solution?

\mot What would be wrong is that that would leave out the quantum effects. There's nothing that tells you that quantum effects have to be small on macroscopic scales, because we have superconductors, right? What would be wrong with assuming that a cold superconducting system behaves classically? Because there are quantum correlations that are not behaving that way. 

\thft Okay, here we have a system of a substance that we \textit{think} we have completely under control, we understand everything about it. But now you're saying we understand nothing about it, something happens on large distances. I still can't quite grasp what it is.

\mot I grant you that's the challenge. I grant you. I was going to try to describe how the conformal anomaly and vacuum fluctuation could produce exactly that effect, for any scale. For this to work it has to work at any scale, except the Planck scale, but anything larger than that it has to be the same mechanism essentially. Here, I agree, I chose a sequence of configurations and adiabatically compressed them, I think this a reasonable thing to do also for stars. If you want to do something else, like throw television sets in, then I have to do a different calculation and show it, we're trying to do that right now looking at 2-D collapse. It may not be sufficient, we have to go all the way to 4-D because of the transverse pressure playing an important role. That's the question, whether or not vacuum fluctuations and the polarization effects, which I think are encapsulated in the anomaly effective action, can actually do that. Those are the effects which are left out, of course, from the classical picture. I agree that's a challenge, it has to be shown, it has not been shown yet.

\misn I think it's very intriguing to have these new theoretical objects that we could go searching the universe for, it would be great to find them, as against whether all black holes, so called now, could be explained this way I have very strong doubts. Particularly because, by now, you concentrated mainly on the equilibrium state and a little bit of perturbations away from that, but nothing substantial about how you would form them in real practice. On the other hand we've been worrying about conventional black holes for so long that now there's a lot of studies on how to form them and I think there are computer models, classically, classical computer models of ordinary type material found in large mass stars, how the core will collapse and so forth. Those all seem very plausible.

\mot There's no doubt, Charlie, if you take classical matter and classical relativity you will get black holes. There's no doubt.

\misn Okay. Once you have taken that, if the quantum mechanics doesn't come in and cancel some of the assumptions you've made under reasonable conditions where you think you know what's going on, the black hole will have formed and the quantum mechanics can't come in and undo the horizon once it's actually been formed. I think it would be very strange if all black holes could be substituted with these things.

\mot I think this is a different form of the question Gerard was asking. I mean, you're challenging me to show you, take some other situation like throwing matter in, Oppenheimer-Snyder, called them television sets or air conditioners it doesn't matter, the question is: how do the quantum effects arise near the horizon that can trigger something like this? That, I agree, requires a different calculation than anything I've shown you. It's a much harder problem.

\misn It also requires an imagination that the horizon can be sensed by the matter while it's trying to obey the laws of physics.

\mot Can be sensed by the quantum vacuum polarization which, remember, is from where it came. Namely, an initial state with a very very diffuse form of matter, almost Minkowskian. That's all you need.

\misn Well, supposedly there are ten-billion solar mass black holes in the catalog. Now for something like that the scale is about fifteen hours for the light crossing time and the curvature, predicted horizon, is much weaker than the curvature in this room, it's very modest.

\mot I keep saying that, the curvature has nothing to do with it, it's a non-local effect.

\misn I can't imagine how something thrown in is going to feel this non-local effect before the horizon is formed. 

\mot I can't show you the last half of the talk, but I'll be happy to tell you the idea. Now whether the idea works I'll freely admit I don't know yet. That's the challenge, but that's a hard problem. You agree, right, that's a much harder problem. It's easier to look for static solutions under a lot of symmetry than it is to solve a dynamical problem. We are thinking about exactly this problem, okay, I have ideas. I can tell you about them. I haven't solved it.

\free I worked with Cosimo Bambi on something even more speculative, this was motivated by a talk that I heard by Petr Horava in Berkeley where in string theory you could possibly have super-spinning black holes with the spin bigger than the mass, and the issues of event horizon and singularity I won't touch, but what we looked at is the observational consequences. What it affects is the shadows of black holes, if you have something behind the black hole then you've got this black region that you can look at. In that case, instead of having a sort of relatively spherical shadows you would have slivers and you could tell the difference. I'm just thinking about observational things. I don't think that would happen here, but then I also remind you that LIGO is coming up, the gravity waves might be another test. I don't know, just throwing all this out there.

\mot It's up there, yes. 

\stod When you drew those light rays, you mean light really goes through it? 

\mot Yes, light really goes through it. Of course that's geometrical optics, the question is whether actually photons would interact with the matter, then they could be scattered and it would be a more complicated situation. All I did there was solve the geodesic equation, so it's just geometrical optics without any scattering. How it scatters, of course, is a question about how it interacts with matter on the surface.

\bard Would you agree that if LIGO does see a gravitational wave burst with a waveform that agrees with the standard general relativistic calculations corresponding to a black hole merger that that would destroy your model?

\mot Alright, put me on the spot, yes. Put me on the spot. I know you're anxious to do that. I think we have to put these ideas to test, we have to stop talking, if you'll excuse the phrase, in a vacuum, and start making predictions and see if it works. On one hand there are theoretical issues that have to be addressed and that's very challenging, but on the other hand it's time to start putting these things to the test. If one doesn't see anything like this that, of course, argues against it.

\dow Is the horizon that you're considering an event horizon?

\mot No.

\dow How would define it?

\mot You're asking about the idealized model? Or you're asking about more realistically, if I regulate it with something finite? The thing I showed has, in a sense, a marginal horizon. It has exactly one radius, the Schwarzschild radius, where things become null. It's, if you wish, a marginally trapped surface but there are no surfaces inside which are trapped. Now if I regulate that a little bit by pushing it away from zero then there are literally no trapped surfaces. No horizon.

\dow It's locally defined?

\mot It's locally defined, yes, I don't need to talk about what happens at infinity. Of course this was a static solution so it's kind of trivial.

\ford Presumably in your model there is no particular limit on the angular momentum versus mass relation like there are for rotating black holes?

\mot I can't solve the equation for rotating yet.

\ford Okay, but astrophysically it is observed that most astrophysical black holes are very close to the extreme Kerr limit so that seems to be a piece of observational evidence in favor of the black hole hypothesis and against your hypothesis.

\mot Why do you say that this wouldn't have a very similar limit? I don't know.

\ford But you haven't shown that yet?

\mot No, I haven't show that because to try to generalize this to rotation is a much more complicated problem and it just hasn't been done. It did take fifty years to go from Schwarzschild to Kerr, so give me break. It's a hard problem. By the way, an interesting thing which can be done, it's tedious but possible, is to put slow rotation and just see how things behave just to get some feeling. That's a well defined problem. If I had a student I would try to get them to do it.

\stod How do you want to solve the dark energy problem?

\mot Ahh, you want me to get even more out on a limb and speculate. Obviously, for this to work I need vacuum energy to be dynamical. It's not a constant, it's not a cosmological constant. It's something which depends on the external conditions, on the boundary conditions.

\stod You want to have the condensate everywhere?

\mot Yes, so now you know you're in the condensate. Except there's matter and radiation so you have to rethink a lot of things. I haven't done that.

\thft The fact that light is de-focused, does mean that light goes faster than coordinate-frame light? The index of refraction is less than one, in a sense. That suggests that the speed of light inside this object is faster than the coordinate speed of light. That's not directly an objection, that's just an observation.

\mot I hadn't thought about that. That's an interesting observation. Can that be used for something? I don't know. The other thing I could mention is if I, I went too fast but, if you had this literal delta function of course it would still take an infinite coordinate time to reach the horizon. However as soon as you put in a finite \(\epsilon\) that time becomes logarithmic, the logarithm of \(\epsilon\), so there would be a time delay. In addition to the de-focusing, if you had a transient source behind you could compare the light signals going passed and the light signals coming through, there would be a significant time delay in the light coming through. Another kinematic effect of that.

\end{dialogue}

\pagebreak
\section{Thursday, August 27 2015 \\ \textit{\small Convener: P. Moniz} }

		\pagebreak
		\subsection{Did the Chicken Survive the Firewall \\ \textit{\small Jorma Louko}  }
				{\small (Slides from the talk can be found at \url{physics.unc.edu/~dnmorse/Hawking_Conference_Slides}.)}			
				
%				\includepdf[pages={1,2,5,10,15,18,22,25,26,28,30-}]{./Slides/Louko.pdf}	

%		\pagebreak
		\subsection*{Discussion}
				%auto-ignore

\begin{dialogue}

\bard My question is, where does the energy for the firewall come from?

\louk Right! 

\bard There's no obvious source for it.

\louk Right. It's hard to get a calculation going.

\dav Is it not the case that the correlations between \(A\) and \(B\) are destroyed by decoherence? If you have an entangled pair and one goes down the black hole and the other doesn't the entanglement is lost because of decoherence due to the Hawking radiation. That's the sort of simple statement I've heard people make. The same is true in deSitter space. Did I just get that wrong, or is that part of the story, is that simply resolved? Are you talking about entanglement? Well, you are talking about entanglement not just correlations. You're talking about entanglement between \(A\) and \(B\). I'm a bit confused, is this intended to be like a pair of particles that are entangled and not just a correlation between two different outgoing particles? Because \(B\) and \(C\) are both like outgoing particles. What are \(B\) and \(C\) actually? Sorry to keep asking all these questions together. So \(B\) and \(C\) these are blobs?

\louk I've drawn \(B\) and \(C\) as sort of regions on this spacelike hypersurface and \(A\) as a region on this spacelike hypersurface.

\dav I'm not sure what that means.

\louk The thinking is in terms of localized wavepackets and we think of them as the correlated particle pairs that appear in the two-mode squeezed states in the standard picture of the Hawking radiation. I am aware that it is difficult to make that sense of localization there fully rigorous as well.

\dav Is everybody here agreed on the simple statement, which is that if you have an entangled pair and one goes down the black hole you lose the entanglement by decoherence? Maybe even that is not agreed here.

\louk Well this is supposed to happen on \(\Sigma_1\), at that particular hypersurface and you're still supposed to have a pure state over there. Now if you're saying there is decoherence with some other degrees of freedom, perhaps.

\dav Yes. I think that's what the claim is. There is a particular calculation by Gerard Milburn that looks at this, and also for deSitter space. This is really single particle quantum mechanics rather than quantum field theory adding entangled pairs of particles in that background and looking at the fate of the entanglement and it decoheres at late times. The simple thing is that the Hawking radiation scrambles the phases. I hope I'm not garbling Milburn's result but I think that's the gist of it. It may not be directly relevant to what you're saying but whenever I see something inside and outside and you want it still to be entangled I'm not sure that that's correct.

\louk Thank you, I should look in to this.

\dav I'm just thinking out loud here, I'm sorry to interrupt you but it seems like a really good opportunity for us to have a discussion about this.

\stod I would like to make an extended comment about the remarks that seem to be the basis of many of the things we've been saying, which is that if you have a pure state you need some kind of mysterious correlations between different regions of spacetime. Let me give you a totally trivial example which looks equally mysterious but has nothing to do with general relativity, singularties, or anything. This has to do with the fact that, in studying multi-particle production, one of the most common and oldest models is, in fact, a thermodynamic model. You take the energy in some process and you distribute it equally over all states. That's, of course, the microcanonical ensemble and you get something which looks thermodynamic at the end. To reduce my example to the most absurd limit consider that I have a big complex system, like a uranium nucleus, and I've brought in an anti-uranium nucleus and I let them annihilate. I have uranium anti-uranium, let's say. These capture processes are usually in the S-wave. This is a single quantum pure state in the initial state, let's say they're spin-zero uranium isotopes, and it's just one quantum state. Now when they annihilate what'll I get? In the simplest model, two pions, three pions, going up to maybe 600 pions because I have 600 nucleons there, and maybe some remnant protons, and so on. This system will look pretty thermodynamic. Tell me, where is the mysterious missing information that I had a pure state in the initial state? You don't see it. There's no need to correlate different regions of spacetime or anything, this has nothing to do with general relativity. So where is the missing information? Well, there are some phase relations which we can't see between the two-pion state and the three-pion state and so on, obviously, because of the kind of detectors we use in this case we can't see those phases. But of course they must be there. Here something totally elementary happened in terms of quantum mechanics. There's no general relativity or anything involved. For example, if you did this with photons you might see something like Glauber states in the initial state and the final state where there are correlations between different number states. The fact that you start with a pure state and get a thermal looking state is nothing that mysterious, in my opinion.

\ford I had a question for Jorma. It seems like maybe part of what's going on is that you're using a very simple model for a detector, which is a two-level system. In particular, such a system can only absorb a finite amount of energy. In some sense you can't do anything too drastic to it. One question would be, could you generalize these to a slightly more complicated model such as a harmonic oscillator in it's ground state? A harmonic oscillator in it's ground state can, in principle, absorb an unbounded amount of energy. So this is really more of a comment that it might be interesting to see what would happen in your models for such a detector. 

\louk Those formulas that I showed they were done in first order perturbation theory for the Unruh-DeWitt detector so all the backreaction of the detector on the field is neglected there. It would be important to understand what the effect of that backreaction is. As you mentioned the harmonic oscillator, if you use the harmonic oscillator instead of a two-level system, and you still work to first order in perturbation theory, the answer is unchanged. The reason is that this linear coupling between the detector and the field in first order perturbation theory it only does transitions between neighboring energy levels. If you just plug in a harmonic oscillator rather than a two-state system and otherwise do exactly the same things you will get exactly the same formulas.

\park There was an old paper by Peleg and Bose and myself in which we actually did a two-dimensional model, it was relatively simple, but those were the days they had introduced these two dimensional models and I don't remember exactly, I would have to look up what went on, but we did have a kind of a related spacetime and the thing could be solved exactly and this involved the energy and momentum tensor. What you found was that you had to have, this is why I asked about the negative energy flux, at just the point at which this singularity disappears somehow in this model you get a flux from infininty, I think, give or take my memory, that goes right in to that point where you have the sharp right angle. That's an exact model. We'd have to look it up, it's in physical review or maybe PRL.

\end{dialogue}  

		\pagebreak
		\subsection{Black Hole Evaporation and Classical Gravitational Waves: Comparison of Calculation Techniques \\ \textit{\small Bernard Whiting} }
				{\small (Slides from the talk can be found at \url{physics.unc.edu/~dnmorse/Hawking_Conference_Slides}.)}			
				
%				\includepdf[pages=-]{./Slides/Whiting.pdf}
	
%		\pagebreak
		\subsection*{Discussion}
				%auto-ignore

\begin{dialogue}

\full It seems to me that most of the things you've been discussing are technical problems of numerical analysis rather than conceptual problems.

\whit No, I wouldn't argue that.

\full That's what I don't understand. You seem to be mixing them together in a way I don't follow. 

\whit A model problem like the one I suggested, of taking a fixed horizon, throwing in some negative energy, and getting a smaller horizon, that's not a numerical problem. It's not a numerical problem that we're technically stuck on, it should be a problem that we can basically calculate just on the basis of equilibrium states that we already know. That is not a problem with numerical analysis and the problem that Laura has I don't think we could argue that brighter minds will solve those numerical problems, I think brighter minds will find the correct physics so that the numerical problems which they have, of going further in to that part of the geometry, will be different and they'll be solvable. 

\full I don't understand, what physics is missing? Is the theory incomplete in some way? The macroscopic backreaction problem?

\whit Let's take Laura's problem as an example. She's putting some interior in by hand, okay? So she's talking about Schwarzschild outside and some interior she puts in by hand. That leads her up against some point where she can't evolve numerically. If she changed the assumption about what she put inside, would that still happen? 

\full But not everybody agrees that she has to make that kind of an ansatz in the first place. For example, in the two dimensional model, for the given background we have an exact unambiguous formula for the renormalized stress tensor.

\whit Right.

\full The problem, of course, in two dimensions is you don't have an Einstein equation so it's not clear what the backreaction problem means. But the vacuum polarization side of it would not be a problem. Could you go back to the previous slide, I thought I agreed and understood the statement that you're making. You say `we have one more unknown than equations and this situation has not changed in almost forty years,' then the next line is `different authors make different choice about how to deal with this missing information unless they use exact numerical data for the stress tensor.' So you're saying that if you actually write down the normal modes, do you not agree that there is a completely well defined stress tensor in these four dimensional problems, and progress toward finding out what it is was reported in Jim Bardeen's talk, for example?

\whit Of course, the places where exact calculations can be done are not evaporation spacetimes, they're fixed background geometries. While we can do those calculations I would argue that they're not enough. This model problem where we throw in some negative energy and look at what happens\ldots

\full Why do you have to throw in extra energy? Why can't we just take the problem that we have and try to solve it self-consistently? You have a stress tensor, you require that the spacetime now satisfies the Einstein equation with that stress tensor.

\whit What would you say we write down for the stress tensor, in a completely unambiguous way, if the geometry is evaporating? What should we write down? Is it completely un-ambiguous?

\full In principle yes.

\whit Yes, in principle.

\full It's a harder problem because you have a time dependent background, but in principle you could solve the quantum field theory problem in that background.

\whit Nobody has done a calculation where they can, time slice by time slice, with the evolving geometry, completely evaluate the quantum state and calculate it's stress tensor in order to go to the next step. Each one of those is basically an infinite exercise and nobody has done that. 

\bard It seems to me we can calculate the stress tensor for a stationary black hole geometry, right? Okay, now if you use that stress tensor as a source of the Einstein equations I think it's a very sensible to first order, the change in the geometry can be calculated. Now if that change is slow and well behaved then the fact that when you did the original calculation of the stress tensor you assumed the geometry wasn't changing shouldn't make much difference. If the change in geometry is hugely long compared to the dynamical time, relaxation time, of fields around the black hole. Now where there's a real problem, which I certainly don't know how to address, is when there's a large change in geometry. Then you get in to a situation where if you are thinking of evolving from some initial quantum state. Because the evolution is quantum you don't really know which classical geometry you should be perturbing around and therefore you have to go to some multiple history sort of picture, which I think is very hard to really understand how to do that in a proper way and I certainly don't understand it. But as long as you're dealing with slow semi-classical evolution then I think there's no problem. Back in 1991 I wrote a paper on the evolution of the black hole geometry and it's very straightforward. The geometry just stays Schwarzschild.

\whit Well, I know how to do that calculation, many people have done it. I would argue against your premise. The core of my argument would be, during the evaporation you say the changes are small, but with every change what happens is that \(r=\text{constant}\) lines which are timelike on the inside of the black hole become spacelike on the outside of the black hole.

\bard Yes.

\whit I argue that's not a small geometric change. 

\bard Why? You can use coordinates which are perfectly regular across the horizon, advanced Eddington-Finklestein.

\whit That doesn't really solve the problem. What's happening is part of the geometry that's on the inside is essentially now coming out, it's not a coordinate problem. I use coordinates to describe it simply, but it's not a coordinate problem. You're bringing spacetime from inside the horizon to outside the horizon and I argue that that's not small.

\bard Okay well yes, conceivably there could be some firewall inside the horizon which is going to do strange things, but on the other hand I think it's hard to argue very seriously that that's really going to happen in the real world.

\ford I just wanted to comment that I think one of the limitations of the approach that you're talking about is that this is the semi-classical theory assuming a fixed classical background. Usually it's stated that that should be okay as long as you're well above the Planck scale but I think we should be a little careful about that. Partly reiterating the point that I made after Jim's talk, that in fact the stress tensor is really fluctuating  and those fluctuations in dimensionless terms are not small, they're of order one. It's true that you can time average them, and I think that you get a very plausible picture for a very long time if you think of the expectation value of the stress tensor as describing the time averaged effect. If you ask about what's happening on time scales on the order of light travel times I think it's a little misleading. In fact, the effects of the geometry are undergoing quantum fluctuations around this mean which is of the same order as the change that the semi-classical theory predicts. Of course, as the black hole gets smaller that's going to become more important and it's something we should keep in mind, that you're using at best an incomplete theory in this framework.

\whit I would agree. I think semi-classical theories are incomplete, but we could at least try and understand what the semi-classical theory says. We may find out it was a silly thing to do, but we still don't know that answer.

\thft I just want to remark that I would definitely object if someone says that no new physics is needed. I definitely think that new physics is essential here because gravity is not a renormalizable theory. If this were Maxwell's theory we would know exactly what to calculate even if your problem has run away solutions, we can handle that in a quantized version of Maxwell's theory in principle. But we cannot handle it in gravity because gravity is not renormalizable which means at small distances everything goes haywire, including, in particular, if you try to calculate the energy momentum tensor and see how it back reacts on the metric. Things start to diverge in an uncontrollable way and that's the problem. That's the reason no one does this calculation, because it comes out nonsensical. Something has to be done about that, we have to see how to make gravity finite at the Planck scale basically, or renormalizable at least. But without that nobody can ever do these calculations, because they lead to misleading and nonsensical results. You see it all the time when you compare, say, the Hartle-Hawking vacuum and the Boulware vacuum and such things, these things diverge at the horizon and that's where all the difficulties come from.

\mot I just wanted to remark, it seems to me that we're talking a little bit cross purposes. Semi-classical methods are a reasonable first step, I think is what Stephen Fulling is trying to say. That is a difficult calculation to try to do the backreaction in the time-dependent fields, but one could in principle set that up. It's coupled pde's with a very complicated backreaction, you have to do the renormalization. I think the way to handle the divergences is reasonably well understood, that there are higher derivative terms because the Einstein action, as you know, is not renormalizable by itself, but there are techniques to remove the higher derivative terms which you would get even in flat space. Various things are being said, I'm just making a remark, people know all of this, but the real question I think is not technical, the real question is whether or not the semi-classical approximation truly remains viable all the way down to the horizon and whether one can trust just replacing the stress tensor by its expectation value and whether or not there might be corrections to that. But it's certainly a reasonable zeroth order thing and we did that in electromagnetism where you can just solve the semi-classical equations, for example in a background electric field, and let the current of the created pairs back-react on the electric field and one gets a reasonable evolution. The only new feature in gravity is that you have these \(R^2\) terms that you have to handle.

\thft The \(R^2\) terms are not just a solution, it's not that easy. Because then you generate unphysical configurations as well which you have to consider, you have to handle, so that's my claim that new physics is needed.

\mot Certainly new physics is needed to understand quantum gravity, no one is saying that's not true. I'm saying that over the years that people in this room, Leonard and so on, have developed methods to basically remove the unphysical degrees of freedom which are at the Planck scale. As long as you're interested in macroscopic objects, black holes the size of kilometers, it's not unreasonable to say that I'm going push the Planck scale physics far away from my problem and just solve an effective field theory, if you wish.

\thft There is also the problem of the micro-states which, of course, do occur at this very sensitive length and time scale, the Planck length. That's where these micro-states basically stop being micro-states. There you can't just ignore the small distance structure on gravity. New, and I believe very elegant, physics is needed to get everything in place there.

\mot I'm not going to argue against new physics or elegance, I'm just saying that one could imagine the semi-classical approximation as a first attempt and there are techniques to handle this problem.

\thft You'll get very incomplete answers then.

\mot You certainly don't get the full answer and I, in fact, am arguing, I did in my talk yesterday, that one has to look at corrections to it. The real question is: what are the nature of those corrections? Do we really need a full blow quantum theory of gravity? Or can we imagine handling these corrections in some controllable  way? It's like doing pion physics, do we really have to have the full microscopic theory or can we develop something like a chiral perturbation theory where we can handle the higher order terms in a controllable way? It's not easy, but it's imaginable.

\thft The theories we had in the past were all renormalizable in the very end, or could be made renormalizable. That's not the case with gravity unless you severely mutilate the physics.

\mot It's like pions, pions are not renormalizable.

\park I thought that's what Gerard is saying and that's not a problem that I or anybody has really solved.

\mot The full quantum theory?

\park Yes, quantum gravity could well be impacting these kind of things. I think that's what he's saying.

\mot We are not contradicting each other, I'm just trying to clarify. Various things have been said.

\thft We're not contradicting, I would put the emphasis somewhere else than other people do.

\mot Okay, fair enough.

\whit So Leonard, let me ask you. You said yesterday that you think you're technique you could show that you get Hawking radiation out of gravitational collapse. Even at the semi-classical level. Using your techniques can you show what goes in to the horizon. Can you show what the quantum state inside is?

\park I can't solve the problems of the world. That's a problem I think where we don't know the full answer because we don't know the full quantum gravity. 

\whit But your analysis could treat it semi-classically?

\park Semi-classically? Maybe you could have problems semi-classically is the point. Due to the fact that we have divergences in the full quantum gravity theory which we don't really know how to deal with at this point, what the full theory is that would be renormalizable, for example. That's what I think Professor 't Hooft was talking about in his talk. 

\whit But in your calculation of particle calculation you didn't need to use quantum geometry to get that result.

\park That was because of the fact that that was a problem you could do because we knew what goes on, that's a very important point in fact, the fact that we looked at limits in which we already knew what goes on in Minkowski space, either experimentally or not, and we were also not dealing with real full problems of quantum gravity. We saw that already because what happens is that we had well defined states in Minkowski space, ignoring full quantum gravity, when I was a student I didn't really know what people referred to as quantum gravity. I always thought it had to be the full quantum gravity that is being talked about, but that's not a fully solved problem. So it turned out by having these limits you could calculate the particle production and what went on in between, we did find these infinities that we didn't know how to deal with in between. That was the whole point of having these intermediate situations that we did know how to deal with. Now in your situation that you're talking about, maybe there's some analogous thing where the problems of quantum gravity, full quantum gravity, would not come in at certain stages where you have well defined situations. If you could do that and avoid the really big problems of full quantum gravity that might be a possibility. I don't know, but it's a possible approach maybe. 

\ford  I want to comment also that I think there's an intermediate range between trying to have a full quantum theory of gravity where we solve the non-renormalizability problem and strictly using semi-classical theory with a fixed background. I think there is a reasonably well defined theory of small fluctuations where you imagine a weakly fluctuating geometry driven either by linearized quantized gravitational perturbations or by a fluctuating stress tensor. That has been discussed to some extent but I don't think it has been worked out as much as it should be. One of the reasons, I think, to think there could be new physics in this is  the classical concept of the horizon. It's not clear how robust that is because, after all, the horizon in classical theory is a discrete boundary between histories of rays that necessarily get out to \(\mathcal{I}^+\) and those that necessarily fall in to the singularity, and that's a very thin line. If you start allowing horizon fluctuations that line has to get blurred out a little bit. A little bit of work's been done on that, for example about ten years ago a student Robert Thompson and I wrote a paper where we asked whether or not small fluctuations would upset the original derivation of the Hawking effect. Would it be large enough to interfere with the outgoing modes that become the Hawking radiation? At least at the level of analysis that we concluded we decided that probably the fluctuations do not do that, so in some sense the simple picture \textit{is} relatively robust under small fluctuations. I think there is still room to look at that more carefully.

\whit Certainly the horizon is a rather odd construct and the problems we're really interested in, if there is evaporation, it's that there is no really event horizon or we're not interested in an event horizon, it's an apparent horizon which fortunately is something that can be determined locally. I would argue that with the right tools, and if we had ways of writing down geometries with apparent horizon behavior, we could make progress. Even if it's ad hoc, what we're doing now is we're behaving in ways that don't focus on that particular content. 

\bard Of course it's very true that nobody has ever done a calculation of the semi-classical energy momentum tensor for gravitational perturbations of a black hole. There's certainly conceptual problems even at the semi-classical level because there is no real gravitational energy momentum tensor, it's very gauge dependent. But again my own feeling is that, if one is willing to accept long term time averages, that there should be some sense in which averaging over some long period of time we can get a definite answer. Nobody's really done that so that is something that needs to be worked on.

\end{dialogue}  

		\pagebreak
		\subsection{Gravity = (Yang-Mills)\^{}2  \\ \textit{\small Michael Duff} }
				{\small (Slides from the talk can be found at \url{physics.unc.edu/~dnmorse/Hawking_Conference_Slides}.)}			
				
%				\includepdf[pages=-]{./Slides/Duff.pdf}

%		\pagebreak
		\subsection*{Discussion}
			
				%auto-ignore

\begin{dialogue}

\dav Is there anything in Yang-Mills theory that tells us why spacetime should have 4 dimensions?

\duf Is 4 picked out as a special number when you square and get General Relativity? No. You can have super Yang-Mills in 3, 4, 6, and 10 you can have super-gravity in 3, 4, 6, and 10, at least. In fact you can have it in up to 11 but it doesn’t pin it down to 4, no. If you could explain why Yang-Mills had to be in 4, yes. You could argue Yang-Mills is only renormalizable in 4 but I don’t think that’s a good argument because the gravity you get is non-renormalizable.

\stell You presented the way the supersymmetry transformations actually emerge on the field, for the duality symmetries you're giving us a rather abstract presentation of the group. Is there a way to actually get the duality transformations out of the field as well?

\duf You're referring to the dualities that live inside this non-compact \(G\), is that right?

\stell Yes, in particular.

\duf Well let's look at four dimensions that's more familiar. In maximal four-dimensional super-gravity there's an \(E7(7)\) and there are 28 photons and the 28 field strengths and their 28 duals transform as an irreducible 56 under the \(E7\), it's an electric-magnetic duality because it takes \(F\) into its dual. Kelly is asking can we see the origin of that from the Yang-Mills perspective? Well we looked at that and it's a bit of a puzzle because the only place we could see it coming from was the Yang-Mills dualities, on the other hand the Yang-Mills dualities are non-perturbative. The \(F\) into \(\star F\) operation in Yang-Mills is not one you can see in the classical field equations. On the other hand I can't see any other place where it could come from. That's the best I can answer at the moment, it looks like it's coming from the Yang-Mills duality but it's strange that a non-perturbative Yang-Mills symmetry is giving you a perturbative gravitational symmetry.

\stell You have the divisor group, \(SU(8)\) for the \(N=8\) theory, and then you have the coset generators. Does this tell you that the super-gravity theory should have the 70 scalars that sit in the coset?

\duf Yes, it tells you that. Let me be a bit more careful with what we actually derive. These gravitational theories have scalars that belong in a coset G/H. G is a global non-compact symmetry, H is a local maximal subgroup. In the case of N = 8 the global symmetry is E7(7), the maximal local subgroup is SU(8). Now when we do the squaring what we’re actually deriving is H rather than G to be honest. That’s what we can justify. If we make the assumption that it’s a homogeneous space that uniquely fixes G, but that’s an assumption. It doesn’t come out of the squaring, there’s an extra step. If you make that extra step then it immediately tells you that G/H is 70 and in the three-dimensional case it was E8/SO(16) so it was 256.  

\stod We have Yang-Mills theory so we can perform local isospin rotations, you're two isospin rotations do they somehow give general coordinate invariance or something?

\duf Yes. That's what we're saying. Let's go back to the beginning, for the fields we're saying the field is given by the square of two Yang-Mills with this spectator field sandwiched in between. For the gauge parameters something similar. The general coordinate parameter is what you would call isospin times a vector with a scalar sandwiched in between to soak up the indices. 

\stod I'm trying to visualize what these isospins are.

\duf Let's take the simple case where both left and right is \(SU(2)\). This is an \(SU(2)\) gauge field, this is an \(SU(2)\) parameter, and then I multiply them together and soak up the indices with this spectator that's in the adjoint of \(SU(2) \times SU(2)\) and I get a vector field which is local because this is local. That's the diffeomorphism of the general coordinate transformation.

\stod It's not the vector fields of the vierbein it's some other kind of vector field.

\duf It's the parameter, I'm talking about the symmetry here, this is the amount by which the metric changes under a diffeomorphism. Is there anything I can do to convince you more?

\stod I'm trying to visualize what it would be, what these isospins or whatever would be.

\duf I mean I have conceptual problems asking what does it mean to start with a flat space Yang-Mills theory and end up with a curved space gravity theory. But that's what you get. I'm sure there's much deeper questions about that, just doing the calculation this is what we get.

(Someone speaking softly.)

\duf Can someone... my hearing's not very good.

\whit He says you're getting flat space gravity not curved space gravity.

\duf Well, gravity to the linearized order. It's curved because there's a linearized Riemann tensor, as I showed you.

%\speak{post-doc} I see. So this was only at the linearized level?

%\duf The Riemann tensor is the square of the Yang-Mills field strengths. 

\stell Just to come back to this issue of the \(SU(2)\) indices, not in direct relation to what Mike's been telling you, but in somewhat analogous calculations of Zvi Bern, Lance Dixon, and company when they obtain super-gravity amplitudes, full super-gravity non-linear amplitudes from Yang-Mills they do it in a color-stripped fashion. They are multiplying left and right but then they actually strip out the color indices by basically assigning them in a four-point interaction, colors one, two, three, four and a fixed order. The details of the Yang-Mills group don't really come in to it, it just has to be more than Maxwell.

\duf Which \(G\) left and \(G\) right you choose seems to play no role in what you end up, they're arbitrary as far as we can tell. 

%\speak{post-doc} If you express the curvature in terms of the metric you will see that you get it only at the linear level. The curvature can be written as a function of the metric, you got the metric only at the linear level in terms of the gauge fields, in that sense this is a map only at the linear level. When you write the curvature in terms of the metric, the curvature is a function of \(g_{\mu\nu}\) but you did construct \(g_{\mu\nu}\) only \(h_{\mu\nu}\) which is a perturbation of the metric if I understand correctly. In that sense the expression for the curvature that is constructed, if you really write it down in terms of the metric...

%\duf Yes, the linearized curvature. It happens to coincide, as a matter of interest, with the curvature in the freely-falling frame. I've wondered if that is a useful observation in going to the next stage of non-linearity.

%\speak{post-doc} Second question, are these relations only at the kinematic level? Or can you also get them at the dynamic level?

%\duf Well that would be the goal, but at the moment we're being rather modest, we're just trying to derive symmetries. When we've understood the symmetries we'll then ask about Lagrangians. But obviously that's an important question.

\end{dialogue}  

		\pagebreak
		\subsection{Black Holes and Other Solutions in Higher Derivative Gravity \\ \textit{\small Kellogg Stelle} }
				{\small (Slides from the talk can be found at \url{physics.unc.edu/~dnmorse/Hawking_Conference_Slides}.)}			
		
%				\includepdf[pages={-29, 31-}]{./Slides/Stelle.pdf}

%		\pagebreak
		\subsection*{Discussion}
			
				%auto-ignore

\begin{dialogue}

\ford Have you looked in to any issues with black hole thermodynamics?

\stell Yes, I happen to have a couple of slides on that. Because we have the curvature squared terms in the action, the notion of entropy that one should use is Bob Wald's proposal for entropy, the integration over the horizon of the variation of the Lagrangian with respect to the curvature. The temperature is given by the usual relation to surface gravity and in that case, let's look at the \(R\) and Weyl squared theory, again with this form of the metric \(B\) and \(1/f\), we get the following expressions for the temperature and the entropy. The temperature is still given in terms of a nice geometrical quantity, the entropy you have to go and calculate in terms of the details of the \(A\) and the \(B\) functions but for the non-Schwarzschildality parameter, near Schwarzschild, you see we end up with this expression for the entropy, this \(\delta\) is the same \(\delta\) I was talking about before. Then you get the following graph of entropy, Wald entropy, versus mass. The solid line here is the new solutions and the dashed line is Schwarzschild. What you get from the numerical analysis here, the mass and temperature in terms of the entropy, you find that this does still verify, approximately, the first law of black hole mechanics. That aspect of thermodynamics seems to be in good order. About second law and so forth, well we were talking about his over lunch with Fay, I don't know what to say. I don't know if there \textit{is} a good second law of thermodynamics for these systems. Fay may have some comments on that tomorrow, she was of the opinion that there should be a second law of thermodynamics for this case. Is that a fair quote?

\dow Exactly about your case I don't know. What I do know is that for Lovelock black holes the second law is violated even at the classical level. Two merging Lovelock black holes don't satisfy the entropy increase law but, I was just looking online, Aron Wall has proved that at the linearized level higher derivative theories do satisfy a generalized second law. It's just when highly nonlinear events like black hole mergers occur that can't be described at the linearized level that there is an apparent violation. But if you restrict yourself to the linear theory then apparently there is a second law. 

\stell That brings to mind, I should stay, about stability there is a point that were discussing a bit in Gerard's talk. The ghosts, prima facie, look very bad but there are indications that at least for perturbative excursions away from flat space, or other positive energy solutions of the theory, they may not immediately lead to runaways. An example of this is Christodoulou's proof of the stability of flat space in GR which can be extended to GR plus a scalar and I have it from Toby Wiseman this even works if you flip the sign of the scalar kinetic term so that you make it a ghost. Still, flat space is stable. It's basically the dispersive nature of gravity that is causing this to happen. Of course general relativity with a wrong-sign scalar eventually can have instabilities but they would have to be non-perturbative. So that seems similar to what you were just saying.

\dav Just to follow up on that, in these higher derivative theories there are certainly cosmological solutions where the deSitter-like horizon will shrink with time, so violating the generalized second law, when you extend it to cosmological as well as black hole horizons. It's been known for some years. 

\stell Are you referring to Lovelock type gravities? I know there's some comments about that for Lovelock.

\dav If you just take these higher derivative theories and then just take the usual, I don't know with the Wald definition, but just take the usual horizon area definition for the cosmological horizon then you certainly have solutions where that will decrease with time. Then that opens the question about whether one wants to use that as a criterion to rule out those theories, in a general sense do we say we are so fixated on the second law of thermodynamics that, as Eddington said, any theory violates it should collapse in deepest humiliation? Should we use it as a criterion for ruling out such theories? Or should we only rule out the solutions which have decreasing generalized entropy and still retain the other solutions in the theory? It's a more philosophical question than a physical question.

\stell I'm quite aware of philosophical debates about this kind of thing, for example with Cliff Burgess who is very much involved with effective field theories. His point of view on effective field theories is you should \textit{only} consider the solutions of the original second-order differential equation with small perturbations and discard everything else. I don't know how to engage in that. He says he is rather Taliban on that subject, he wants to insist that this is the way you have to do it. To me it seems, for example Gerard's proposal, if you're going to have this Weyl squared term because it's renormalizable then, if that's your theory, you should take it seriously and look at what the theory has. You can't at that late point, decide to pick and choose the solutions you like and you don't. You can impose boundary conditions, boundary conditions that have no rising Yukawa's at infinity and so forth, but once you've satisfied those the theory gives you what it gives you.

\thft I want to play the Devil's advocate, if you have this gravitello you can produce negative energy particles and positive energy particles. I was worried about positive energy and negative energy states that together have zero energy and therefore should be very easy to produce out of nothing which would be an indication of the vacuum itself being unstable against productions of such a thing. You start with an infinitesimal field configuration that infinitesimally deviates from the vacuum and then that should blow apart because no energy conservation law stops it from growing.

\stell Absolutely.

\thft Did you find any of such phenomena?

\stell No, all I can say is, again, I refer to these studies of the stability of flat space in theories with ghosts where, even though there are negative kinetic energy terms in the action, nonetheless flat space can still be stable. That's the only thing I can respond to that with.

\thft Is it all stable or are there are cases for \(\alpha\) and \(\beta \) and so on where you can construct such configurations?

\stell I'm not claiming that I know that in this theory, the examples there are general relativity with a wrong-sign scalar. The massive spin two ghost is more complicated and nobody has tried to analyze that, but it's a possibility that the dispersive nature of general relativity is still powerful enough to not allow a runaway to develop in the way that you're suggesting. I should say, another person who takes this attitude is Andrei Smilga who has a series of papers on theories ostensibly with ghosts but which you can live with. It's basically the same idea, that you don't necessarily excite the runaway solutions just because you have a negative kinetic energy term. 

\ford It is interesting that there seems to be a black hole that has zero entropy in this theory. Does that have a nonzero temperature at that point?

\stell I think it does, yes. Well you just have the equation here so you have to solve that.

\ford Okay, yes. In any case there presumably couldn't be any kind of a microscopic definition of entropy in terms of state counting for this type of black hole because if its entropy is zero it could only have one state yet it's still a thermodynamic system that radiates. 

\stell Maybe one point of view in all this is that there are enough strange features of these solutions that you decide that, you really come to the conclusion that, you shouldn't allow this at all. Or maybe they're just artifacts of something like a string theory expansion which goes on forever and yet higher terms remove the un-palatable aspects of these solutions.

\thft Then the other possibility is to add a symmetry, such as local conformal symmetry, which could say yes, all these instabilities do occur but only in the gauge sectors which are equivalent to the vacuum so that we won't see anything of this, this would just be a gauge ghost. So if you promote some of those ghosts that you see as being gauge artifacts, then maybe you can get away with it.

\stell Except that the spin two ghost is not a gauge artifact. That would affect a scalar. By the way, the \(R^2\) scalar here is perfectly innocent, it doesn't produce any problems at all. In fact it's exactly what people want for inflation.

\thft But that is I think an example of this because you rescale the metric tensor to remove the \(R^2\) term. 

\stell Yes. Well so there is, at least at the linearized level, a reformulation with an auxiliary field introduced to get rid of the Weyl squared term in favor of a Pauli-Fierz type action with, however, the wrong sign. You can, similarly, introduce an auxiliary field to have second order differential equations.

\thft That's how I derived this gravitello by making this transformation.

\stell Yes, right, exactly.

\end{dialogue}

\pagebreak
\section{Friday, August 28 2015 \\ \textit{\small Convener: L. Ford} }

		\pagebreak
		\subsection{Quantum Damping or Decoherence. Lessons from Molecules, Neutrinos, and Quantum Logic Devices \\ \textit{\small Leo Stodolsky}  }
				{\small (Slides from the talk can be found at \url{physics.unc.edu/~dnmorse/Hawking_Conference_Slides}.)}			
				
%				\includepdf[pages=-]{./Slides/Stodolsky.pdf}

%		\pagebreak
		\subsection*{Discussion}
				%auto-ignore

\begin{dialogue}

\misn Just a comment that the first half of your talk on chiral molecules has been re-done by some computer chemists a few years ago. I'm sure you would be interested to read this. They go beyond the two levels, they actually deal with one hundred or more levels. They've got a chiral molecule which they chose to be deuterium disulfide for strange reasons, but it's chiral, it's a very small chiral molecule.

\stod And do they actually calculate the scattering amplitude?

\misn Using a master equation they calculate the evolution of the density matrix where that molecule is in a bath of helium.

\stod Okay, some of the things I've seen of this they've got mixed up between the real part and the imaginary part.

\misn I don't know. I only read reviews of the paper but I can give you the references. I'm sure you will find it amusing because it did say they had to invent new computer techniques and then run a computer for a month doing hundreds-by-hundreds matrices, or maybe it was thousands-by-thousands.

\stod This original experiment that we proposed with the left tunneling to the right is a beautiful experiment waiting to be done, a Nobel prize level experiment if anybody can do it.

\misn They chose this particular molecule with the hope that it could be done experimentally. Their estimate of the oscillation rate was I think 170 Hz or something like that. It looks like hydrogen peroxide but they didn't want to use hydrogen peroxide because in reality that's explosive when it's pure and so this wanted this disulfide and then they found, well, if we use deuterium instead of hydrogen some of these numbers get closer to what experimentalists might like. Anyway, they've done this work which seems to be an example of exactly what you're talking about but keyed to a more real world where they might do the experiments. 

\ford I have a question about the watched-pot effect. I'm a little bit puzzled. If I have a system with let's say a time-dependent wave function but where the amplitude for something, say, to have made a transition after a relatively short time is very small. It's certainly true that if I measure it and then restore it back in it's original state it's not going to decay. Isn't maybe a better way to think of it is imagine you have a very large ensemble of such systems. You have a large ensemble of systems that you've prepared identically so then when you start measuring them it's true that most of the time you'll put them back in the original state but there'll be some fraction you'll find, even after a short time, will have decayed. 

\stod Well, this is why I insist that the right thing for doing such a problem is the density matrix. There you've averaged over ensembles and there's nothing to worry about. There is one time-dependent density matrix and that answers all questions.

\ford But isn't the density matrix throwing away some information? I think you're missing something by looking only at the density matrix.

\stod No. By the way this reminds me of another comment I want to make that's not about decoherence. We had some discussions earlier where wave packets were coming up and here's a paper by me that you might be interested in where I explain that most of the time when you're talking about wave packets you actually don't need them. Talking about ensembles and wave packets is all unnecessary, you just take the density matrix and this answers all questions. In particular when you have a stationary problem there can be no coherence between different energy states. The thing I don't understand about the black hole problem is do we have a stationary problem really or not? That I don't know. This is not on this subject but you might be interested in this paper.

\dav I think if you couple the system to, say, a von Neumann detector then the interaction Hamiltonian with the detector is such that it's not like having an ensemble of freely decaying particles and you just sample them. The interaction actually knocks the thing back to the initial state with a very high probability. That's all been worked out in quite some detail. You have to take in to account that interaction.

\ford High probability, not probability one.

\dav No no, of course not, no. But in the limit of continuous measurements then it is probability one. By the way, didn't Wigner tell us that there is a superselection rule which means you can't have superpositions of wavefunctions of different masses?

\stell Is that a question?

\dav Do you know? I would have thought you would know the answer to that Kelly.

\stell No I do not. No.

\full I think the superselection rule doesn't say that you can't have a superposition of different masses, what it says is that a superposition is indistinguishable from a mixture.

\dav Right, okay.

\stell I just wanted to know the bottom line about biology. Are you telling us that the preference of one for another, for example in sugars, is really related...

\stod Is what?

\stell In biology. What's the bottom line of your comments? Is it really related to weak interactions?

\stod In biology, this is a question that often comes up, we can explain if you start with one chirality why you stay there but we can't, at the level of decoherence, we can't explain why you choose one rather than the other initially.

\stell Abdus Salam towards the end of his life had some comments on this but I don't think they were taken very seriously.

\dav He had a paper on it. The very last paper he sent to me.

\stell Yes I know, that's what I meant.

\dav The point is anything that breaks the chirality and lots of things that can break it and then if you iterate and amplify it doesn't matter how you started off. Doesn't have to be weak interactions, could be chiral surfaces, catalytic surfaces, for example. Or even polarized star light, these things would all give you an enantiomeric excess which could then, in principle, be amplified. Nobody quite knows the details though my collaborator Sara Walker did her entire PhD thesis on it.

\ford You didn't really say anything, you mentioned gravity briefly at the end, but you didn't say anything about black holes. Do you have any thoughts about decoherence in black holes?

\stod Well, there's lots of things I haven't understood in the discussion this week. In this example I gave yesterday it's easy to have a system that looks thermal but is actually a pure state, like in my pion gas. In the case of black holes one of the things I'm confused about, more in connection with this issue here, is that is the black hole really stationary? Or do I have to consider the whole process including when the black hole is gone and I just have the radiation at infinity? It seems to me if you do that you'll end up again with a pure state but that's just my gut feeling without any real calculation. Because what you're doing is throwing in something in a short time and then it's coming out spread out over an enormous time. That must imply some kind of very tricky phase coherence if you view it as the global process. Most of the things I've talked about are stationary or quasi-stationary problems. 

\end{dialogue}

		\pagebreak
		\subsection{Puzzle Pieces: Do Any Fit? \\ \textit{\small Charles Misner}  }
				{\small (Slides from the talk can be found at \url{physics.unc.edu/~dnmorse/Hawking_Conference_Slides}.)}			
				
%				\includepdf[pages=-]{./Slides/Misner.pdf}

%		\pagebreak
		\subsection*{Discussion}
				%auto-ignore

\begin{dialogue}

\free We have advanced LIGO taking data very soon and they're expecting to be able to see collisions of black holes. They expect to see gravity waves from colliding black holes. I'm wondering what can we learn from that other than the fact that these are giant objects colliding. Is that the only thing that matters or do you learn something about black holes? I don't know the answer to that question.

\misn My guess would be that things will go as expected and you don't need any quantum mechanics and that GR will work perfectly well for solar mass black holes or intermediate mass or huge black holes. All those things probably can live their lives essentially as handled by the numerical implementations which one hopes is giving a picture of what the actual solutions of the Einstein equations would look like. Whatever is going on deep inside the black holes where their could be quantum mechanics or new laws of physics at high energies or who knows what. Whatever is going on in there will be irrelevant to what we can see.

\whit Just a follow up on the question, well a comment to the question. If we had two astrophysical objects that inspiral and collide and we compare the case where those two objects are both black holes, or one black hole one neutron star, or two neutron stars of equal mass, then the initial inspiral wave would be the same but the final merger and ringdown waveform would be different. So from observation we can distinguish between  astrophysical states that contain black holes and ones that don't. A neutron star would be different so we could learn something about that. In the neutron star case we actually would like to be able to learn something about the equation of state in the interior which, again, is telling us something quite quantum mechanical. So we may learn something about nuclear matter from seeing a gravitational wave.

\misn Not only that but in the case of if they were two black holes, you hope to be able to check some of the details from the ringdown times, these special frequencies that are associated with black holes when they've been disturbed. If you have enough signal to noise ratio to begin to see that kind of stuff, it will have severe tests on whether classical non-quantum relativity is in charge of the situation.

\mers Getting back to the very difficult heavy-duty topics, the entropy and the EPR tests, when it comes to the thermodynamic entropy I completely agree. Of course we all know that gravity is a negative capacity system so it would do the opposite of what normal would do, which is a positive heat capacity system. But we do have an out-of-equilibrium  thermodynamics which is well defined so I was a bit confused why can't we use that? Although the system, as you said, will be out of equilibrium, the self-gravitating system will be out of equilibrium, but we can still define a thermodynamic entropy for this out-of-equilibrium system correct?

\misn I'm not familiar enough with that, I know the field exists but I've never studied it. If it can handle the things, great, I'm just not familiar with it. Obviously you're right in saying that that is an appropriate tool to attack these problems with.

\mers And the second question the EPR: using that on a black hole setting, which sounds extremely interesting and that's related to the comment Paul made, if Hawking radiation can break this entanglement for two spacelike separated events that are correlated with one another wouldn't we run in to the same trouble of firewalls? Because breaking the correlation that energy has to go somewhere, it has to build up somewhere. If we wanted to run experiments to test EPR on the horizon of a black hole and we wanted to break this correlation to do the auditing part, wouldn't we end up with surplus energy which is incredibly blue shifted near the horizon?

\misn In the usual EPR there is no energy considerations, if you do measure things in the compatible directions so you can check them, or the incompatible directions so that they're independent of what the expected measurements are, all those things take place with exactly no penalty of energy which way you do it.

\mers But in this case you are very near the horizon where half of the events are beyond the horizon and you are just trying to obtain information being outside.

\misn Yes but it's only the information that's changing, it's not the mass or the velocity or any other property of the object that fell in, it's only the phase of the wavefunction is somehow getting lost it's correlation with something outside. 

\dav These calculations have been done, Milburn I mentioned, but also Robert Mann at the Perimeter institute, looking at entanglement with Dirac fields in a non-inertial frames. I think the conclusion is that entanglement does not survive one of the pairs going down the black hole. As I said earlier, it had better not survive otherwise we've got additional information about the interior of the black hole over and above what is given to us by the no hair theorem. I don't know how that fits in with what we were hearing from Stephen and Malcolm earlier in the week about BMS hair but that's classical.

\mot I just want to remark that correlations on the apparent horizon giving rise to effects is basically exactly what I was talking about in my talk, there's no reason why there can't be macroscopic quantum effects even on the scale of an apparent horizon of a large black hole. It doesn't violate causality because of course it came from some initial condition in the original collapse in the same way EPR doesn't violate causality.

\misn It doesn't violate causality but can it provoke something happening locally? That was my question. Does it know enough to provoke the kindling of this firewall?

\mot Yes, well I wouldn't call it a firewall because that's a different term, but there can be such effects. That's what the vacuum polarization in certain states actually shows. Now you can argue about whether those states are going to be produced and that's of course a dynamical calculation that has to be done, but that doesn't violate any fundamental principles in quantum mechanics.

\ford I wanted to comment on your remark about energy conservation. I think the lesson from, say, dark energy is that energy conservation is fine but it's just that the fundamental energy conservation is local, the covariant divergence of the stress tensor being zero, that's the thing that is still really valid. Of course it doesn't make much sense to talk about a global integrated energy being constant. We don't really know whether the dark energy is \(w=-1\) or some other value and of course the local conservation law is very relevant in trying to probe that, what the nature of the dark energy really is observationally.

\misn I always like to think that dark energy is like the early universe where there was also a false vacuum and an exponential expansion that went on for some time and then somehow it hit a limit and all this wonderful expanded space turned in to all kinds of matter. Well nature likes to play a theme in variations, it gave us gauge theories with E\&M and then after we studied things a little farther we found gauge theories all over the place. If nature did a big inflation a long time in the past maybe we're just entering another inflation on a different time scale, instead of having the time scale be the crossing time of a meson or something like that the time scale will be scale of the e-folding of the present universe under the so-called dark energy and therefore instead of a heartbeat being a second a heartbeat will be \(10^10\) billion years or something like that. There's plenty of time so interesting things could happen they just happen more slowly. 

\ford I think it's worth recalling that the observational evidence for inflation doesn't necessarily prove that it was deSitter inflation, after all power law expansion models with a sufficiently high enough exponent also fit the data just as well as deSitter expansion.

\free Anything superluminal.

\ford deSitter is simpler so that's what people assume but we don't really know that. Actually I thought I'd maybe throw in another comment, you recall of course many years ago Roger Penrose had an idea about gravitational entropy being associated with Weyl curvature. I think that idea has been pretty much discredited. Paul and I and Don Page wrote a paper against that, but of course I think part of his idea, what motivated him, was thinking of stochastic gravity waves. Even if you can't put gravitons in a box, you can certainly imagine the universe filled with some bath of stochastic gravitational radiation and it's perfectly reasonable to attach an entropy to that.

\free I wanted to comment back on this question about the future of the universe with a cosmological constant, if it is really a constant then we do die a heat death because the universe keeps cooling off until eventually you reach the temperature of the microwave background and at that point, anything living has to metabolize, there's no way of eliminating the heat anymore. So if it's a constant we do die a heat death. Now if it's not a constant, then you can sort of eek along a probably unpleasant existence to our taste, but anyway you can keep going. 

\stod I want to make a comment about your first point, which you didn't really mention, which is: can we ever detect a graviton? The graviton and the photon are degenerate systems, they both have zero mass, so one should be able to mix with the other. There's a paper by myself and Georg Raffelt, we were mainly concerned with axions but we also treated gravitons, where in a magnetic field (which you need to make up the missing spin) the photon can turn in to a graviton and vice versa. Conceivable a high energy photon, or maybe better a high energy graviton, coming from someplace could produce a very high energy photon. Of course the rate is rather small until you get near the Planck energy and you need large regions of uniform magnetic fields, but it's not conceptually impossible.

\bard I think the whole question of the entropy of black holes and what it really means is something I think is far from settled. We have the Bekenstein-Hawking entropy, which I would think represents basically some ignorance about what's inside the black hole in a very coarse-grained sense. We also have some sort of Von Neumann entropy of the black hole measuring the number of quantum degrees of freedom that are inside the black hole and I don't think those two entropies in general have much to do with each other. Perhaps the Bekenstein-Hawking entropy  represents a maximum possible Von Neumann entropy but even that's debatable. At least for a young black hole the number of quantum degrees of freedom excited above vacuum is presumably much less than the Bekenstein-Hawking entropy. Anyway, I'm just throwing that out and if people want to disagree that's fine.

\end{dialogue}  

		\pagebreak
		\subsection*{Discussion with Malcolm Perry}
				%auto-ignore

\begin{dialogue}

\ford I think it might be useful to have a preliminary discussion. I know I'm confused but I'm not sure on what level. Let me see if I can, the way I understand it, recapitulate, Malcolm, what I think you said on Tuesday. Go back first to the BMS group for asymptotically flat spacetime. As I understand that's a way of encoding all the information that could possibly escape in gravity waves. Is that a fair statement? You have an asymptotically flat spacetime in which gravity waves could be radiated off to \(\mathcal{I}^+\).

\per What happens in gravitational radiation in asymptotically flat spacetimes is that as gravitational radiation passes through \(\mathcal{I}\) it induces a BMS translation. The particular BMS transformation in question encodes the information from the Bondi news function, but I would imagine that there are other things that it could encode, well, could be encoded, that I did not include. I think that there is more information in gravitational radiation than just found in the BMS transformations. I'm not absolutely certain of that. 

\ford The discussion of supertranslations in that sense is the discussion of at least some of the information carried by gravity waves to \(\mathcal{I}^+\).

\per Yes that's correct, that's absolutely correct.

\ford As I understood, what you were telling us though, the corresponding supertranslations on the horizon are encoding information that could go down the black hole.

\per Yes, I think what we should do is take one step back first. At \(\mathcal{I}^+\) there are going to be a collection of charges which contain some of this information. There is a similar collection of charges on the horizon. That's a classical statement, but if you ask me about quantum mechanical information I do not know how to make the link between that and those charges.

\ford Let's first understand it at the classical level. As I understand it the charges on the horizon are encoding information about waves that enter a black hole.

\per That's absolutely correct.

\ford The supertranslations, or the collection of conserved charges, are telling us about things that went across the horizon. As you pointed out, there is a certain sense in which black holes have an infinite amount of hair. Now of course people like John Wheeler were aware of this sort of thing, their statement about black holes having no hair was referring to what observers far away would be able to actually observe after a finite time.

\per At \(i^0\), yes.

\ford That is, that the effects on the horizon apparently exponentially decay as seen by the distant observers. I think the question I'm wondering about is what is the next step? If you want to use this in trying to solve the information problem then somehow this information... I can understand formally that the information, gravity waves for example, going across the horizon is encoded there. How does that get out?

\per I think that the next step to try to take would be to attempt to answer what Stephen asked during my talk which is, can you use this to calculate the entropy of a black hole? I have spent the last day or so making attempts at doing that. I have not really had a great deal of success at the moment. The basic problem is that there is an infinite amount of hair. That means that you're dealing with systems with infinite numbers of degrees of freedom, and one thing that is clear is that black holes do \textit{not} have an infinite number of degrees of freedom, the entropy is a nice function. It's bounded by the mass squared which is faster than most things,  but nevertheless it is bounded by that. In order to get an answer that looks like the area it seems, at least what I had to do to get any answer at all, would be to apply some kind of length scale cutoff on the horizon.

\dav Could you put it in a box?   %6:30

\per No because the reason for the divergence is the fact that the number of modes on the horizon has no bound. 

\whit Technically we have the same problem if we consider a box of thermal radiation. The number of states we can have at arbitrarily low energy is arbitrarily large. And yet there is a finite amount of entropy in radiation at a finite temperature.

\per Yes that comes about because you've got a Boltzmann factor which cuts things off for you. Here it is not clear that you do. The problem could be resolved by imposing a length scale cutoff on the horizon. Putting it in a box isn't going to help because putting a black hole in a box isn't going to do anything about high angular momentum modes on the horizon. Essentially the angular momentum on the horizon is classified basically by spherical harmonics. Those are unbounded in \(l\), \(l\) can go up as much as you like, and it's that divergence which causes a problem. But nevertheless if you think about how these modes behave on the horizon they'll have nodes, they'll oscillate, and if you thought there had to be a cutoff because you couldn't have too many oscillations, that is to say there was some kind of length scale on the horizon, then you could cut it off and get a reasonable answer and it would be proportional to the area. However I don't know that that's a sensible thing to do and the problem with length scale cutoffs is to make sure that you have a sensible way of doing this that doesn't violate Lorentz invariance. Fay is more of an expert on not violating Lorentz invariance with finite length schemes than I am, so maybe she could tell us a bit more about that. The only way in which I could justify, or hope, that this is not as bad as it sounds, to have a length scale cutoff, would be to say, well, this only is working on null surfaces and they're going to be a bit special. That is very ad hoc.

\ford One thought that occurred to me, it's unfortunate that Gerard 't Hooft isn't here now because he made some comments, I think, after both your and Stephen's talks which were related I think to the idea that he had perhaps twenty years ago studied of interactions between ingoing modes and outgoing modes leading to a well defined S-matrix. As I understand, his idea is that if you have ingoing modes that enter the horizon, which would be essentially what you're dealing with in your formalism, those will scatter in some way off of outgoing modes, and that could leave some imprint and hence some correlations in the outgoing Hawking radiation. The part at least to me it seems is missing, that I would like to, if anyone understands it, clarify, it seems like there is still something carried across the horizon that is not encoded in those modes. I think Gerard is trying to define an S-matrix, he was hoping to find a theory in which black hole evaporation could be defined by an S-matrix which would then clearly be a unitary process, but until you can say what happens to the modes that go across the horizon and that information, then you're still dealing with the same problem, as Jorma talked about in his talk, of the correlations between the things that cross the horizon and outside the horizon. There still seems to be a paradox.

\per I'm not going to disagree with you. I have no idea how information crossing the horizon can be completely preserved. For it to be completely preserved would require you to violate the no-cloning theorem, certainly some of the information that is passing through is going to get in. I don't see how you can get all of it. But nevertheless I think the matter should be investigated further.

\ford I thought the whole idea of yours and Stephen's approach is to solve the information problem.

\per What we have done is to identify an infinite collection of charges which should assist us in understanding what is going on. I don't know that it's a solution to the whole problem, however I must say that Stephen is rather more optimistic than I sound.

\ford I guess what I'm doing is I'm expressing a little bit of skepticism that this program can work unless you can deal with what happens to the information that goes across the horizon.

\per That's fine, I mean I'm not going to object to that.

\mers Malcolm, if you have a supertranslation invariance of some scattering matrix why do you need to introduce a cutoff? Unitarity is already built in there, the moment you have a cutoff you break unitarity, and all the other bad stuff that goes with it. Cutoff dependence and all that.

\per Well the question that I was trying to answer was, can you use these thoughts to generate the correct entropy for black holes. What I was saying was that unless I introduced a cutoff I did not succeed in doing that, but I didn't really want to introduce a cutoff. I'm not quite sure where that leaves us, it may just mean that the right calculation hasn't been done yet.

\whit Can you associate these charges corresponding to the supertranslations with vibrations at the horizon?

\per The answer is quite possibly, because they are basically classified by the spherical harmonics and that is precisely the same kind of thing. Maybe.

\ford I thought that was what you were telling us on Tuesday, that essentially these charges are ways of describing the classical perturbations.

\per Yes, they are the classical perturbations due to ingoing stuff. But as he points out, those excitations will die out.

\whit On the horizon?

\per On the horizon.

\stod So they die out asymptotically but I'm not sure what happens to them on the horizon.

\dav They die out because they're damped.

\per Well they're just damped because of the usual sort of arguments about quasi-normal modes. So they all die out.

\ford That's the reason that it's plausible to say that black holes really have no hair, because there all these infinite number of degrees of excitation that might be excited when the black hole forms but they decay exponentially.

\dav Through gravitational radiation.

\bard It's a combination of emission to infinity and in to the black hole. There's radiation both ways.

\dav Can you define a friction for the surface of the black hole? Some sort of intrinsic coefficient of friction?

\per You're asking about some kind of dissipation?

\dav Yes.

\per Well there is for electrical resistance so there probably is for other things.

\dav That's what I'm thinking. I've never seen it discussed.

\per Do the membrane paradigm people discuss it?

\dav Yes I was thinking the membrane people should like this.

\per I don't know very much about that.

\ford Bernard isn't it right, if you excite a normal mode it's going to decay?

\whit I'd like to make a distinction between two separate cases. One is where you excite the quasi-normal mode outside the black hole and it definitely decays in time, I'm not so sure what happens if the quasi-normal mode is excited on the horizon by the infalling photon or whatever it is. For example if it's excited on the horizon and we ignore black hole evaporation for a moment, we just work with these classical charges and this classical geometry, excitations on the horizon, which we acknowledge can exist, are not going to radiate to infinite because the horizon is null. They can radiate to the interior, possible, but they can't radiate to the exterior.

\dav Can we connect this with what Charlie was just saying? If the particles that are falling in are too feeble to ever emit one graviton then they're not going to excite the modes on the black hole, is that right? So if Charlie falls in doing this [waves hands] it's not going to make any difference.

\ford That only makes the problem even worse because then in some sense there's less way to encode information in the horizon isn't it?

\dav Right. 

\per I think what you're asking for is to ask what effect does this have on the Hawking radiation that is going out. That's really the question that you want ask, well it's one of the questions you want to ask, at this stage I do not know the answer to that question.

\ford I think Gerard's given a partial answer, or at least tried to give a partial answer to it, though at least to skeptics like me it's not clear how that could be a complete answer. How that can preserve unitarity by itself. 

\bard It seems to me this mechanism of these supertranslations is a way of preserving some of the information at least about the classical perturbations, stuff falling in to the black hole, without it decaying. The perturbation in the affine parametrization of the geodesics lasts forever, it doesn't dissipate. It's hard for me to see how it really encodes all the quantum information that you would need to reconstruct a pure state in the exterior for one thing. Then what one would hope is that somehow, I certainly don't have any detailed picture, but somehow the vacuum fluctuations which will eventually become Hawking radiation in part, as they redshift and as their wavelengths become the order of the radius of the black hole, some sort of phasing of those is affected by this supertranslation distortions of the geodesics, the generators of the horizon, and that then can induce correlations in the outgoing Hawking radiation. That's a very handwaving argument.

\per The answer is to put some meat on to that, isn't it?  You've got to be able to do that calculation, that's what needs to be done.

\bard That's, I think, highly non-trivial but yeah it certainly would be important to understand that.

\park The string theory formulation, though not entirely applicable to all black holes, that at least gives you a very concrete picture in some situations of D-branes and vibrating strings.

\per Yes, so what Leonard is talking about are essentially zero temperature black holes. These are ones for which you could imagine an internal state made of branes and essentially you can get the entropy by counting up the number of different configurations consistent with those charges. That's a very concrete realization. Of course that really is in the limit in which \(G\) is gone to zero because you're using a flat-space argument to list the configurations nevertheless it does reproduce the correct entropy formulae. It sort of works when you perturb away from \(T=0\) and I suppose that that's, at some level, some evidence that there is some microscopic structure associated with these things but in that particular case it doesn't seem to have anything to do with any horizon. After all when you think about things like that there are no horizons to think about. The picture that you're talking about also works particularly well for black holes in dimensions 4 and 5 where the picture is quite clear, outside those dimensions it doesn't work too well. But I don't see how that fits in with these configurations at all. It's a bit of a mystery really.

\park I was wondering if it was just possible to combine these kinds of concepts.

\per Well maybe because in one case there is a clear statistical mechanical interpretation, even if it's a bit obscured by the fact that the model doesn't seem to be entirely coherent. In this particular case it looks like, again, you have got a series of configurations where you can identify some statistical mechanical system but in this case, as I said, I could not satisfactorily reproduce the entropy formulae. Maybe that's just a defect of trying to do it here when too many other things are going on.

\mers Still, back to the basics and trying to get at the most basic picture. Can you play this supertranslation symmetry game in any background? Or do you need the asymptotic flatness to even define the BMS?

\per What I needed to make anything work nicely was a concept of an isolated horizon. That's a null surface with topology \(S^2 \times \mathbb{R}\), it's a null surface and the null geodesics that are outgoing, I assume that there is a concept of outgoing, have zero convergence.

\mers So you are not defining the BMS group at \(\mathcal{I}^+\)?

\per \(\mathcal{I}^+\) is a little bit different because that's sort of at infinity, but cosmological horizons are okay, Rindler horizons are not because they are not \(S^2 \times \mathbb{R}\).

\mers Can you explain why the dS space is okay and the Rindler space would be different? That's important, that's a statement about Unruh's accelerated observers detecting something completely different from black hole radiation.

\per I can't at the moment.

\mers Okay.

\whit Two questions: one is in the case of isolated horizons, which just exist for a finite amount of affine time, can you still get an infinite number of BMS?

\per Yes because the BMS transformations are classified by spherical harmonics.

\whit The supertranslations?

\per Yes, the supertranslations. You see it's just a real function of the coordinates on the sphere, so if you want that to be real valued those are just the ordinary spherical harmonics.

\whit The second question is about not when you have isolated horizons but when you have an apparent horizon instead of a null surface. In the case of an apparent horizon, which isn't null, how does any of your argument go through at all?

\per I have no idea, I haven't tried it.

\whit Thank you.

\bard Would you agree that the same sort of supertranslation memory occurs in basically any null surface in a certain sense. Any sort of object crossing, say, a Rindler horizon in Minkowski spacetime, will generate supertranslations in the Rindler horizon.

\per Yes well that's a followup to Larry's question about the Rindler horizon, I'm not sure about Rindler horizons. There has to be something different about those.

\bard I don't see how. What I would say is different is that the vacuum fluctuations on the Rindler horizon stay vacuum fluctuations, they don't become Hawking radiation. It's only because, for in the black hole case, they become Hawking radiation that they really have any physical consequence.

\ford It's worth remembering though that the things that become Hawking radiation are really modes not on the horizon but just outside the horizon.

\bard Well it's definitely close.

\ford They have to be close to the horizon but they're not strictly on the horizon or they never would get out. They have to be, strictly speaking, modes outside the horizon. Those are the ones that you need to imprint if you're going to produce correlations in the Hawking radiation that will carry away information.

\bard Hawking radiation at late times is associated with modes which have highly sub-Planckian wavelengths within highly sub-Planckian distances from the horizon at early time. To all practical purposes they're on the horizon.

\per I should say that it is very early days.

\ford I didn't mean to put you on the spot, I just thought that you raised so many interesting questions in your talk that haven't been resolved and we have a few minutes, it was useful to air them and see if we could make any more progress.

\mers And the firewall question is the other one that I know you don't have the answer to that yet.

\per I have nothing to say about firewalls.

\end{dialogue}

		\pagebreak
		\subsection{The Generalized Second Law and the Unity of Physics \\ \textit{\small Fay Dowker}  }
				{\small (Slides from the talk can be found at \url{physics.unc.edu/~dnmorse/Hawking_Conference_Slides}.)}			
				
%				\includepdf[pages=-]{./Slides/Dowker.pdf}

%		\pagebreak
		\subsection*{Discussion}
				%auto-ignore

\begin{dialogue}

\dav A couple of points: one is that what you've said could be the jumping off point for the next great phase of investigation, which is to look at black holes in systems which are far from equilibrium. There's been a lot of recent advances, not in anything to do with black holes, but in non-equilibrium statistical mechanics and the second law. There is something called Jarzynski's inequality, the work of Jeremy England is particularly noteworthy I think. I think there's been a really major advance in this field conceptually in the last five years or so; the other point I was going to make is you mentioned in passing about rotating black holes and there's no Hartle-Hawking. I can remember Roger Penrose saying a long time ago if you put a rotating black hole in a box then it eventually will reach equilibrium and there will be some rotating pattern of radiation. There will be a state, anyway, in which it can be in equilibrium. I suppose that would be a sort of analog of the Hartle-Hawking state, whatever it is. I don't think anyone has worked it out. 

\whit I guess I'm asking for clarification, it looks to me as though you're saying [Aron] Wall shows that if the generalized second law holds, and we have smooth Lorentzian metrics, then we must end up with black holes.

\dow No. You're saying that if the generalized second law holds then you can't have smooth Lorentzian metrics, there has to be a singularity. There has to be null-geodesic incompleteness, the proof goes forward just as Penrose's original singularity theorem.

\whit Right, he's arguing that we must have singularities.

\dow Yes he's arguing we must have them.

\whit Which I would say is a black hole.

\dow Oh I see, right, if we assume that cosmic censorship holds then there must be what I would call black holes. Because people have been using `black hole' to mean non-singular things I didn't want to assume to say `black hole' means that there is a singularity. I would agree with you.

\whit A corollary would be that if, say, Stephen and Malcolm are correct and we now understand that there is no information loss in black holes, and we understand that they can evaporate completely, then the generalized second law can't hold.

\dow Well, the generalized second law would have to mean something else I suppose. In some sense this is almost a fine grained generalized second law, in the sense that the coarse graining that you have to do to get this result is the most objective coarse graining you could possibly imagine, which is just to trace out what is unobservable behind the horizon. It's sort of objective in a certain sense that the coarse grainings that you usually have to do when you're proving a second law are not. But if there's no event horizon and you therefore don't trace out over anything, then the second law kind of holds trivially in some sense because the state is just pure and it remains pure and the entropy is zero and it always is zero. I think the onus is then on people that think that the evolution is always unitary to explain it is that they mean by a generalized second law.

\whit The proof that you demonstrated of the generalized second law required a horizon.

\dow Yes. In this work, so for example showing that the generalized second law doesn't hold for the dynamical horizon, I think there he's just assuming that you do the same thing. That you trace of the states of the matter inside the, in this case, dynamical horizon. You choose the area of the dynamical horizon to be the entropy of the black hole, the non-singular thing, and you so that that quantity, the sum of those two things can decrease. You do what you would do if it were an event horizon and you show that that quantity will decrease.

\whit In that calculation was the dynamical horizon taken to be spacelike or timelike?

\dow A dynamical horizon is, Malcolm do you know? I think it's spacelike actually, I think it's spacelike, the dynamical horizon.

\ford In your discussion you didn't say anything about assuming an energy condition, I think it must have been implicit in the proof of the generalized second law that you sketched. You pretty much have to have some kind of assumption about energy conditions. You could have a pretty clear counter example, that is if you could create a pure quantum state that had unlimited amounts of negative energy in it then you just shine that negative energy on to the black hole and you evaporate it. You basically decrease the area without necessarily a corresponding increase in entropy someplace else. Could you say where the energy condition, where does it slip in?

\dow Are you assuming the first law?

\ford Well the proof of the second law, I think there must be an energy condition somewhere. Not sure where. It doesn't have to be a local energy condition, it could be an averaged energy condition.

\dav Is it somehow implicit in the Hartle-Hawking state?

\ford Well it could be in this case. But I think more general, the generalized second law has to say something about energy conditions. I'm pretty sure that [Aron] Wall's proofs have some type of an average energy condition as one of the postulates.

\dow I think that's not true, I think that the generalized second law in that case is an assumption rather than a conclusion.

\ford Yes, in that case, but in the cases where you try to prove the generalized second law starting from other assumptions, I think one of the assumptions has to be something that involves an energy condition or an equivalent to avoid the counter-example that I just gave. 

\dow So what would be violated in that case? That's a good question.

\ford I think with the Hartle-Hawking state there probably isn't any problem because the matter here is just thermal radiation which necessarily automatically satisfies the energy conditions. This was kind of getting to the next point I wanted to make that when you're using the generalized second law, for example to restrict wormholes, I think another way of saying what's going on is that both the generalized second law and restrictions on your ability to build wormholes depend on quantum field theories having restrictions in them. Call them quantum inequalities, that is that they're, in flat space you can prove fairly rigorously, they're restrictions on the sample time-integrated energy density along geodesic observers' paths. Those are fulfilled by quantum field theory and of course they carry some of the same content that the generalized second law has. They're what prevent you from giving the blatant counter-example that I gave. But they also show why you can't, for example, build traversable wormholes, at least macroscopic traversable wormholes, because you would violate these inequalities.

\dow I agree with that, I think one could ask which of the two conditions is stronger. Perhaps an assumption that the generalized second law holds perhaps implies these average energy conditions.

\ford I think you can work it both ways, there's some equivalence, but if you want to try to prove the generalized second law starting from, say microphysics, then I think you have to go through some restrictions on energy. That kind of brings me to the next point that is the quantum inequalities have been proven fairly rigorously, at least in flat space and certain other spacetimes, for free fields. They haven't really been proven for interacting fields so I think there is still a question then, I mean we expect them to hold, but I think that still is an open question, to what extent the generalized second law on the one hand or the quantum inequalities on the other hand will hold for general interacting field theories. I don't think that's really been explored well enough yet.

\vid I would like to make a comment in the same direction of what Larry just said, usually when you work in trying to do singularity solution in quantum gravity, at least in loop quantum gravity, you start from the full theory in which you have a fundamental spacetime discreteness. Then you work out your effective equations and you find a new term, this new term is a modification of the gravitational part of the Einstein equation. Once you went to semi-classical well you can decide to move this new term on the matter part and say, oh okay everything goes as if there was a violation of the energy condition. In your fundamental theory, if you do things properly, you have never violated the energy condition. It's just that when you want to describe things out of the regime in which you originally worked, then you can think as things were violating the energy condition. I am of the impression that here there is some confusion about the domain of application of the theorems that we are talking about because, so there is in this state an equality between the energy condition and the generalized second law. But it seems to me that this is something valid in the domain in which the quantum field theory in curved spacetime is valid but if you want to describe bouncing cosmologies, bouncing black holes, then in that specific region you should use a full quantum gravity theory.

\dow Probably none of these questions can be fully resolved until we have a full theory of quantum gravity. Even questions of exactly what quantum field theory in curved spacetime means can't really be resolved until we have a full theory of quantum gravity, but if one pins ones hope on, or is guided by or inspired by the unity of physics, then hoping to get out a generalized second law, having it be true in the full theory, I don't mean that it means anything in the deep quantum regime because, of course as far as we know, to even talk about entropy increase we have to have some notion of entropy at one time which, for now, we just have to associate to some hypersurface, and entropy at some other time which, for now, we have to associate to some hypersurface, those are semi-classical concepts. What I mean by saying that the generalized second law holds in the full theory is that when we have the full theory, in a state in which the semi-classical approximation holds, then using the full theory in that situation will give us the generalized second law. I don't mean that it has any meaning or you can make sense of increase necessarily, maybe it does in fact. But it may be that we can't even make sense of it in the deep quantum regime where there is no spacetime at all to talk about. But I suppose what I'm advocating is that the generalized second law can be used as a guide, so for example if you have a non-singular black hole model then can you try and produce what I would call a gross violation of the second law, in other words a perpetual motion machine, that would be, I would consider that to be a bad thing. You could just say the generalize second law just doesn't hold, okay, you might be able to live with that but I think being able to produce a system which is in perpetual motion that's a bad thing. For example Ted Jacobson and Aaron and another collaborator argued that in Einstein ether theories where you have more than one metric, in bi-metric theories for example where black holes have two horizons defined by the different causal structures of these two different metrics, then in such theories you can produce perpetual motion machines from such black holes. That seems to me to be quite a black mark against such theories, I personally would think that was something which dis-favored such models.

\whit I'd like to come back to the dynamical horizons again, I think it may bear on some of the comments we've just had. I think you're right if we're dealing with classical general relativity and matter does obey the energy conditions so that positive energy can fall in to the horizon, but not negative energy, then the dynamical horizon will be spacelike. But as you pointed out earlier, quantum fields violate the energy conditions and so we can have near a thing that is forming a black hole, we can have negative energy density. If some of that falls in to the black hole the apparent horizon is no longer spacelike and becomes timelike. My question would be, it would be much more interesting to see if, in the case of timelike dynamical horizons, they can show that, in other words, if we allow the black to hole collapse instead of just expand, to evaporate instead of just grow indefinitely, is the generalized second law recovered? You're saying if we allow it to expand indefinitely then it doesn't hold, but if we allow it to evaporate is it possible that it does hold?

\dow My understanding is that that's ruled out by Aaron's theorem.

\whit But how? It depends whether he is using dynamical horizons which are spacelike or timelike. 

\ford Could I comment on that? I think this is a point where it is very important, as Fay pointed out, that you're talking about the full event horizon and not the apparent horizon. That's very crucial in this. You give a simple example, that is, quantum field theory does allow you to have local negative energy, but at least if we're dealing with quantum field theory in flat spacetime the magnitude and duration of the negative energy fluxes are highly limited. What's more they have to be followed within a fairly short time by compensating positive energy. If you create a negative energy flux you can send it in to the black hole but then you must send in a larger amount of positive energy within a fairly short time. In fact the longer you wait the more positive energy that it would have to be. That's what Tom Roman and I called the quantum interest conjecture. It was actually proven so it's more of a principle. Now here you can see how that saves the generalized second law because the true event horizon does not respond locally to the instantaneous energy, it's a global construct. In this teleological sense, the true horizon knows the future, it depends upon the future positive energy that's going to come in because it's the future history of the rays that in the end are just barely trapped. That future history is dependent upon the positive that is going to come in after the negative energy. So in fact, as long as you use the full horizon for your generalized second law you're okay, but you would run in to a trouble if you tried to use the apparent horizon. 

\whit If black hole evaporation occurs so that there is no event horizon then the only theorem we can...

\dow Why do you say that Bernard? Here's a picture of black hole evaporation.

\ford You have the thermal radiation too though.

\whit Well that's a black hole evaporation with a remnant.

\dow No, no there's no remnant. It just disappears. It is completely regular, there is no remnant. Nobody knows whether that's right, do we know it's not right? Then there's an event horizon because the boundary of the causal past of \(\mathcal{I}^+\) is this.

\whit Okay so if we have singularities then we can use this theorem, if we have no singularities we can't have an event horizon. If the evaporation leads to a completely regular spacetime so we don't have singularities...

\dow I suppose the general idea is that if there is no singularity and then I suppose the overall idea is that there is no way to define a generalized second law. You would have to say that nature is malicious right, I mean this you would have to say `yeah yeah, you thought that was so at some stage of science but actually it's not.' That would be malicious. Nature is subtle but not malicious I would say. Without a singularity, without an event horizon, I think the idea is that there is no generalized second law.

\whit In the kinds of geometries that Francesca was looking at those apparent horizons exist for the duration of the evaporation which can be many times the life of the universe, but technically there is no event horizon. There is no singularity there is no event horizon.

\dow Then I think that the claim is that any attempt to define what you mean by a generalized second law to associate entropy to the object, the object which has this horizon, any attempt to do that and have a generalized entropy that always increases, fails. You could just say, so what?

\whit That's what I'm trying to understand. In those geometries that she described, the apparent horizon there are parts of it which are timelike. In particular they're timelike during evaporation. I'd like understand if the theorem that you quoted from Wall, Aaron Wall, holds for timelike dynamical horizons or only for spacelike dynamical horizons?

\dow I don't know, we can look and see.

\whit I consider it important because if it holds for timelike dynamical horizons, then this entropy that we can associate with hidden behind this dynamical horizon, could allow the generalized second law to be restored. First of all defined, then to hold.

\dow We can see.

\ford As long as you have the large positive entropy in the thermal radiation going outward then it may be okay for the black hole entropy to decrease. That's the usual picture of the Hawking evaporation.

\whit Right.

\dow If the total dynamics is Unitary then eventually the radiation will all be in a pure state.

\ford Then everything is zero.

\dow Then everything goes back to zero so that's sort of trivial, it's violated.

\whit Well we could imagine the quantum field states that are causing the collapse not to be pure states, to be mixed states, so we could allow that there is entropy to begin with. So it wouldn't necessarily go back to zero.

\dow So it would go back to whatever it was in the beginning, it would decrease to whatever it was in the beginning.

\end{dialogue}

\pagebreak
\section{Saturday, August 29 2015  \\ \textit{\small Convener: L Mersini-Houghton} }

		\pagebreak
		\subsection{Wrap-up Discussion \\ \textit{\small S. Fulling} }
				%auto-ignore

\begin{dialogue}

\full I don't propose to lead the entire discussion I just want to kick things off early in the morning when I'm awake, I seem to become non-functional from lunch on. I'm very grateful to the organizers, not only for allowing me to speak right now but for supporting me on the entire visit. I come as a information paradox skeptic. That is, I've never been convinced that there's really a problem, but I've also been acutely aware that that may be due to my ignorance because I have no strong opinions about quantum gravity at all. My own contributions to this field have been concentrated at level (ii) [below] and even that, as far as any relevance to the subject matter of this conference, stopped about twenty years ago.

\vspace{0.1in} 

I would like to review a little bit of history which is largely the history, of my own understanding of the subject. But I think it's chronologically accurate and of course I will welcome anyone else who wants to interrupt at some point. To start with, in the 1950's and early 1960's there was quite a lot of talk about the existence of black hole singularities being inconsistent with the conservation of baryon number. I never really understood what that was all about, it seemed quite clear to me that if you had a singularity, a particle could reach the singularity in a finite proper time and disappear and who cared? That just meant that the topology of the space was such that the particle disappeared. That's not inconsistent with conservation of charge or baryon number or anything else, in our theories conservation laws basically come from, let's write it here, local conserved currents, which under certain conditions can be integrated to give global conservation laws. There's no reason why, in a complicated geometry and topology of space, that this should give some quantity that is absolutely conserved if the structure of the space says otherwise. So I never understood why there was a problem.

\vspace{0.1in} 

\begin{figure}[h]
\centering
\begin{tikzpicture}
	
	\draw (0,0) -- (0,3);
	\draw (0,0) -- (2,2);
	\draw (0,2) -- (1,3);
	\draw (1,3) -- (2,2);
	
	\path [draw,snake it] (0,3) -- (1,3);
	
	\draw [->] (0.5, 2.2) -- (1.0,2.7);
	\filldraw (0.6, 2.35) circle (0pt) node[anchor=west] {\tiny(A)};
	
	\draw [->] 	(0.4, 2.2) -- (0.1, 2.5);
	\draw [->] 	(0.6, 2.4) -- (0.3, 2.7);
	\draw [->] 	(0.8, 2.6) -- (0.5, 2.9);
	\filldraw (-0.1, 2.8) circle (0pt) node[anchor=west] {\tiny(B)};	
	
		\draw [thin, dashed] (2,2) -- (1.1,2.82);
	\draw [thin, dashed] (0,1.9) -- (0.9,2.8);
	\draw [thin, dashed] (0.9,2.8) arc (140:35:0.14);
	
\end{tikzpicture}
\caption{}%
\label{Pdiagram}%
\end{figure}

Then we began doing quantum field theory on a curved background, without considering backreaction, and we get similar situations where to draw the famous diagram [Figure \ref{Pdiagram}] . We have some kind of flux created that goes out here [(A) in figure \ref{Pdiagram}], some kind of negative flux being created that goes in here [(B) in figure \ref{Pdiagram}], and if you want to consider a Cauchy surface for this spacetime when you push it as far to the future as you can, it's going to be like \textit{that} [dashed line in figure \ref{Pdiagram}] if you're going to stay outside the horizon. Everything is perfectly unitary as people say,  or if you are going to allow observations inside the black hole you can go in there and get something that is perfectly unitary. No problem. You get to the singularity in a finite time, but again who cares? Nothing, as far as I can see, is wrong or strange there. I never really thought much about this problem until recently when I suddenly began to realize why people are concerned. If you assume that by energy conservation the horizon must actually be shrinking and eventually it gets to zero size so that the singularity in some sense disappears, carrying lots of the ingoing stuff with it, then that does seem a bit strange. 

%\vspace{0.1in} 
\pagebreak

Let me tentatively draw a line across here and take a poll, is there a consensus that there is no paradox above this line?

\begin{enumerate}[(i)]
	\item Classical physics on a curved background.
	\item QFT on a curved spacetime.
	
	\noindent\rule{2cm}{0.4pt}
	
	\item Semi-classical gravity (with backreaction).
	\item Quantum gravity.
\end{enumerate} 

\park You mean down to that line?

\full Yes, no paradox down to [the line]. If so we have at least one consensus. 

\mers So you want us to take a vote if there is an information paradox or not?

\full Is there an information paradox within the theories at this level [above the line] or this level?

\mers Okay, let's take a vote.

\mot Can I rephrase the question, is there a way to get the Hawking effect entropy only above that line, in a way that makes sense and is completely consistent with statistical mechanics as we know it?

\full Again I plead ignorance. I'm not a statistical mechanician.

\mot That's a different question but it's part of what of the information paradox contains, where is the number of states that go in to the entropy formula? So you want those.

\mers Let's take two votes then. Let's go to the more general question of information loss and then go to the details.

\full Can I just respond to Emil's point. This Penrose diagram, which I drew here, refers to these physical theories above the line, so I should have drawn the line down like that. In this picture nothing has shrunk, so is there any problem with having lots of entropy there?

\mot I still don't know how to calculate it.

\spin Maybe this is not a paradox but a difficulty because the point is how should you define the Wightman function or, in other words, how will you define the vacuum state on a curved background. We know, for instance, if we use the Boulware state, it's perfectly defined outside a star, but if the star collapses we have to use the Unruh vacuum and all that. In general I think we do not know the precise answer how to do precisely quantum field theory on a general background.

\full Okay there are two questions there. One is the question what the quantum state is, what initial conditions did you choose, and I think that if you start with a stable star back here and then it collapses, a strong argument can be made that the exact state is well approximated in this region of spacetime by the Unruh state. The second question is what to do with it when you have it, can you actually do any calculations? That's a technical question. As long as we're simply not discussing backreaction, but simply taking quantum field theory on a fixed background, I think there is no problem. The conceptual problems were settled in the 1970's, the technical problems remain because these calculations are hard.

\park I gather what you're saying is that if you don't worry about the evaporation of the black hole the formalism is fine.

\full Just calculate the expectation values of stress tensors.

\park Once we get to the second half of your blackboard, then we do face the question of what happens as the black hole evaporates and what's left, do you get back all the information and so on? Things like that.

\full As I see it the big controversy surrounds level number three and the issue of whether passing to level number four will or will not resolve it. Are we agreed with that?

\whit I don't think I quite understand your position even on (i), it may not be that it's controversial but I'd still like some clarification. Just purely classically, forgetting all quantum physics, we can make black holes, classical black holes, different ways. Once we have a black hole we can't tell how it's made. At a classical level I would suggest that in some sense that implies there is a loss of information in having a black hole, because knowledge about how the black hole formed is not determinable from the classical matter outside it. I wouldn't say it's controversial but I find it difficult to vote in support of saying there's no information loss at level (i) in this sense.

\full Certainly if we're going to time-reverse the evolution we have to give some initial data. You turn the picture upside down, stuff is coming out of the black hole. You have to say what it is. It's not determined by the black hole.

\whit If you're asking is it deterministic then I would say yes.

\mers If I understand Stephen, if you include quantum fields on curved spacetime you will end up with Hawking radiation and that produces the time asymmetry in to the setup of the problem. You cannot reverse the situation. If you have classical physics only then you can always reverse.

\whit Right, but above level (iii) he's not allowing evaporation.

\mers He is allowing it, quantum fields on curved spacetime produces particles.

\whit No, level (i) and (ii) have no backreaction.

\mers Yes, backreaction is a different story. Just a quantum field on curved space would give you Hawking radiation.

\full You're arguing about the definition of evaporation. I have Hawking radiation but I'm not allowing it to change the background spacetime.

\ford Even for level (i) I think you need to be a little clear about what we're just talking about. If you have a classical black hole then in a sense you do have the paradox that Jacob Bekenstein identified. Namely, before there was any notion that the black hole had any entropy you could throw hot objects in to the black hole and decrease the entropy of the universe. Now of course Jacob saw a way to correct that by assigning an entropy to the black hole. So I think even at level (i) there is a potential paradox if you don't include Jacob's idea of black hole entropy.

\mot To follow up on that, it seems to me the logic of your position is: who cares? Entropy decreases because it goes through the singularity and I don't care about the second law at all, it goes in to some other space or god knows where, it doesn't matter to me on the outside.

\full I could shoot hot matter off to infinity too and say that it's no longer relevant.

\whit So on point (ii) you would argue that you have outgoing radiation but you have something totally unphysical. The total energy in that outgoing radiation is infinite, so this state that you put on this spacetime is nothing that we could build physically or would be interested in physically. You may be able to define it, but it's relationship to physics is thin. To say that it implies you have outgoing radiation in a spacetime is, I think, not possible. In a physical spacetime.

\full You're saying the Unruh state evolves in to something that is singular on the horizon?

\whit No, I'm worried at \(\mathcal{I}^+\). You say you have outgoing radiation there and the total energy\ldots

\full It's not infinite density. It goes on forever, eternally.

\whit The total energy going to infinity is infinite and we can't create states like that. 

\full If you don't have backreaction then you're not going to have energy conservation, in that sense. You still have energy conservation for the scalar field because of negative flux over here, but you're right there's an infinite amount going out.

\park Can I make a suggestion? The big question is what happens after the black hole starts evaporating and so forth, and you might say that the no-hair theorem is false because, at least in the way that Stephen Hawking is looking at things now, you're actually saying that there is some kind of hair on the event horizon of a black horizon. These sort of holographic remnants, you might say, of what went in. Well that's a violation, in a way, of the no-hair theorem. The no-hair theorem doesn't work in that case, you can't prove it. I think this is one way of getting a solution to the problem and it would be very interesting to see if it works. It's a very natural kind of solution, there are other possible solutions as well.

\full I'm about to turn matters over to other people because, as I said, I don't know much below here but I see Paul has something to say.

\dav I wanted to hear what you had to say about (iii) and (iv).

\full Oh, well what I have to say about (iii) and (iv) will not be at all conclusive but I come as an emissary from another conference. There were two conferences on this same topic this month and I believe I am the only person who attended both. You may think of me as the last photon that arrived before these two groups went over each other's respective horizons. The conference was two weeks ago at the Perimeter institute and was in honor of the 70th anniversary of Bill Unruh. Heavily represented there were people like Bill Unruh and Bob Wald who believe that information \textit{is} lost in singularities, even in quantum gravity, and it's not a problem. They were feeling so strongly on this that I got the impression that when I got here I would face the opposite party and the room would be full of people who were violent proponents of firewalls and fuzzballs. That is not what happened. There have been people talking about firewalls but I get the impression mostly grudgingly, trying to defend the more traditional general relativity and quantum field theory from the more modern speculative ideas. What I want to do is just briefly report my understanding of the other position as represented extremely well by Bob Wald and his previous talk at the Santa Barbara workshop and list the alternatives for dealing with steps (iii) and (iv). 

\park You're saying they argued that there is no information paradox?

\full That there is information loss and you shouldn't be upset by that.

\mers Stephen, before we move to (iii) and (iv) can we take a vote on: is there an information loss paradox with this open minded crowd here?

\full We're still up here?

\mers Yes. Let's take a vote.

\full Okay, is there an information loss paradox at level number (i).

\whit There is information loss but it's not paradoxical.

\full I tried to place the emphasis on the paradox rather than on the loss.

\bard There's certainly a paradox if you have Hawking radiation going out to infinity and you’re not having backreaction. That's a paradox.

\full That's a breakdown of the model, surely, but whether it's actually a paradox\ldots

\mers Let's take a vote.

\full Okay, vote on number (i), yes there is a paradox or no. On number (ii), yes there is a paradox, no?

\bard Repeat the statement please, I don't know what I'm voting on.

\dav I want to hear about (iii).

\full Is there a paradox at this level?

\duf How would you describe Stephen's first paper?

\full His first paper, I don't remember everything that was in that paper but I think the original paper was at this level [(ii)] and then there was later paper at this level [(iii)] which was the one about breakdown of predictability.

\dav Can we hear about what Bob Wald says?

\full Yes, this is Wald's hierarchy, for lack of a better word, slightly edited by me. I will write things in the opposite order from the order he gives them. Starting with zero which is his own position, that information is lost in singularities even at level (iv). His mantra is: the creation of mixed states from pure states by tracing out certain degrees of freedom is not a breakdown of quantum physics. It is a prediction of quantum physics. There is no reason why one should start modifying agreed-upon quantum physics and agreed-upon general relativity at the level of a macroscopic black hole where we think we understand things in order to try to avoid that.

\mers Can I ask a question there? Unitarity is also a prediction of quantum physics and that's violated by singularities.

\full No, when you go from a Cauchy surface to something smaller you should go from a pure state to a mixed state. That's what quantum physics says. It's not a violation. It just means you're looking at a subsystem. Let me go on to complete this list. 

\begin{itemize}
	\item[(0)] Information lost in singularities.
	\item[(0.5)] Baby universes.
	\item[(1)] Remnants.
	\item[(2)] Final bursts.
	
	\noindent\rule{2cm}{0.4pt}
		
	\item[(2.5)] Fireworks.	
	\item[(3)] Firewalls.
	\item[(4)] Fuzzballs, EPR = ER. 
\end{itemize}

The next one is remnants, where the idea is quantum theory modifies the collapse to such extent that the singularity is averted and you have something left over, Planck sized something or other left over. I would put up here something he didn't mention, but Paul has mentioned, which is baby universes, where the missing degrees of freedom are somewhere they're just not connected to our universe. Then there is what we can call `final bursts,' which says that something happens in the very late stages at Planck scale, or somewhere getting down towards Planck scale, that causes the black hole to explode, dissipate, or whatever. The partisans of this are, of course, Hawking since 2005, Bardeen, Hayward, Frolov, and others. I'm going to put another line. Then the remaining things are those that propose drastic modifications of physics at fairly early times, when we are looking just at traditional macroscopic black holes. We're going backwards in time here, the farther you are on the list the earlier and more drastic the modification is. So the firewalls go here, the fuzzballs and EPR = ER, wormholes, go there, and I think we should insert here Laura's fireworks. This [(2.5)] is a kind of burst, but it occurs much earlier than this[(2)]. Here is where the controversial line is. The position of Wald and Unruh, with which I'm sympathetic although I don't know enough to make a strong statement in that direction, is that all these things below the line are rather drastic modifications of what we think we know. These above the line are all, for some reason, regarded as physically unsatisfactory by the proponents of these below the line. I think at that point I should surrender the floor to other people who have more to say. Wald's position is this [(1)], but he also will allow one of these above the line but say 'I don't see that it solves any interesting problem.' He doesn't really say that these are impossible. My personal position is that it might well happen, for all I know that when the full problem is solved quantum theoretically, the outcome might depend on the initial conditions, say the mass of the system just as for supernovae. If you have a fairly small star you have a nova and it collapses to a white dwarf, if I remember correctly, if it's bigger you have a supernova and it goes to a neutron star. I don't see why you can't also have some systems that go to true singularities and some that go to remnants and bursts. Now I'm just speculating completely beyond my expertise and so I will sit down and shut up.

\ford I don't think I have any great pronouncements to make. I can spell out, very quickly, my position on many of these questions, the bottom line is I don't know. Early on, if you asked are there paradoxes at the classical level or at the level of quantum field theory in curved spacetime, I think not if you have defined things carefully. At the classical level, as long as you understand there must be a resolution to the Bekenstein paradox and black holes must have some kind of entropy, then there isn't a classical paradox. At level of quantum field theory in curved spacetime without backreaction, as long as you say there must be some backreaction someplace but we're not yet taking it in to account then there's no paradox. Where things get tricky is then when you include backreaction and you ask what's going to happen to the black hole, then things are a lot less clear. I think I would add one thing, though, to the list there and that is the issue: can there be correlations in the outgoing Hawking radiation? Certainly the typical thing that we might think of is, we take a piece of paper and write a message on it and then burn the paper. Atomize the ashes. Then nobody feels that that's paradoxical, where did the information go? It went off in phase relations between the various photons which we have no hope of catching them, but they're there. That could well happen with the Hawking radiation for some reason. If I understand things like Gerard 't Hooft's resolution, presumably the only way to use Malcolm and Stephen Hawking's ideas on supertranslations, is that these have to leave an imprint on the outgoing Hawking radiation. How that happens and whether there is a plausible mechanism, is not clear. That's, I think, kind of the level, of course for a full quantum gravity it's too early to speculate. I think that's about all I can say.

\mers I can ask a very simple question. Back in the 70's when you were all working on deriving black hole radiation from quantum field in curved spacetime, what is the origin of Hawking radiation? Is it produced during the collapse of the star? Is it produced long after the black hole has formed and the horizon and singularities are there? Is the collapse crucial, do you need that in order to produce particles?

\whit I think we tried to answer that in this very simplified model we did with Chris Stephens and 't Hooft where we collapsed a shell to some point and then, by acting as supreme beings we reversed the shell and made it go out. Almost everywhere this spacetime obeys Einstein's equations except at the bounce. What we found is as we allowed the bounce to go closer and closer to the horizon that the outgoing quantum state became more and more like a thermal state. In this geometry there was no horizon anywhere and there was no singularity, well there was a singularity at the point of the bounce, but there was no horizon. Even though the outgoing state became more and more thermal, a thermal distribution, globally we had a unitary evolution so there had to be correlations that we couldn't find, spread all over \(\mathcal{I}^+\) that preserved the pure state even though locally next to the bounce, the radiation became more and more thermal-like. I think we know the answer to that.

\mers Thank you. So the collapse of the shell was important and I know Philippe has a beautiful review of the subject, I think you calculated Hawking radiation in your review paper.

\spin I think it's a kind of re organization of the definition of vacuum according to the Cauchy surface you consider. If you remain outside and make only measurements outside the horizon obviously you have to trace on the inside Hilbert space and this produces an effective thermal distribution. It's mainly our definition of the vacuum that you need when you evolve in the collapse framework.

\dav What Bernard was saying, you can do the same thing with an accelerating mirror. You can pick one that gives you something very close to thermal output, and then just smoothly join it to a uniform velocity at the end. All is good up to that point. It doesn't seem to be connected with the horizon. Another telling case, I had a student many years ago look at uniform acceleration in a compact space. So there is no horizon because you go around and around and the light cones go around and around, again you get the thermal result. In that case you accelerate forever. You get the thermal result, but in that case it's thermal radiation appropriate to a compact space, you have discretized modes. 

\mers But the acceleration is crucial, one obviously could not get particle creation if the mirror was not in uniform acceleration or if the star is not collapsing.

\dav Right, but I must take you to task about your terminology which a lot of people conflate. When they say the origin of the Hawking radiation, what do you mean by radiation? Are you talking about particles, or are you talking about energy fluxes?

\mers Well the stress energy tensor is the only thing that makes sense in this case because particles don't have a meaning.

\dav Right right, a lot of people just flip from one language to the other which confuses lots of people. And if we're just clear about there is a \(T_{\mu\nu}\), in two dimensions we can write it down. You can argue, and Stephen Fulling was taking me to task about how to represent the terms in that, you know, where they most appropriately belong and so forth, and that's possibly an issue. But it's only an issue of semantics really because it's there for all to see. We can predict from those calculations exactly what any given observer moving on any trajectory is going to see and everything has got to be self consistent because it's a covariant object, the expectation value of \(T_{\mu\nu}\). 

\ford I had a comment on those models. An important part of that is that what you've discussed is only one side of the mirror. Those particular trajectories are ones that accelerate away from an observer and emit a thermal spectrum of particles, let's say to the right, but of course there's also correlated radiation going off to the other side of the mirror. As long as you take those in to account there is always a pure state. And including if you wrap around the universe if you include what goes off on both sides of the mirror then definitely everything stays in a pure state. It's just that in this model the thing that goes off to, let's say, the left side of the mirror, or away from the mirror, is analogous to what goes down the horizon. Except here it's not lost, it's available to be caught later.

\park Can I just make a comment? When I was working on my PhD thesis in the 1960's I worried about something like an information loss. Namely, pairs of particles are created by the expanding universe and one pair goes off in to the distance and the other one is near us, say, so we can't observe the other member of the pair but it's correlated. If you look at the Einstein-Rosen-Podolsky paradox, I was wondering is there any kind of correlation where if we make a measurement, or rather if somebody in this galaxy far away that we don't see, makes a measurement of one of those photons in that pair, it's correlated with the photon sitting here in our galaxy or coming toward us and we should then have our photon somehow correlated. There's no obvious way to get the correlation between the two because of relativity and so that brought me to the question of what's going on, how do you end up destroying such correlations between the particles in the pair created? At that time that was sort of like an information paradox. It reminds me of what goes on inside the horizon of a black hole and outside, this question which has come up about the correlations between the members of the pair. It seems to be no problem in the expanding universe, I guess, ultimately. I had a solution, you can solve the problem in another way in the expanding universe. But I guess for the black hole it doesn't work because of the total evaporation that's supposed to take place. Anyways, it's interesting.

\full Since the Penrose diagram of star collapsing to a black hole is on the board I want to make a quick comment following up on Paul's comment in response to Laura's question. I put in a collapsing body. I just want to say that I believe very strongly in this picture that I drew with the blue arrows, where there is something going in to the black hole out here. It is not originating in the body. I believe that, because in the two dimensional model I can calculate it exactly. No approximations. No uncertainty in what the formulas are. It's very simple formulas and they describe this. Positive flux going out here towards \(\mathcal{I}^+\), negative flux going here over the horizon, in to the black hole. This positive flux does not originate down here near the collapsing body. Let me remove that. I understand, Laura, why you're concerned about the fact that the curvature that is acting as the source of this. And that was another big contribution of Bill Unruh's, to notice that in two dimensions, specifically, the conservation law for the stress tensor is a set of first order differential equations that tells you that the \(T_{vv}\) and \(T_{uu}\), the independent components of the stress tensor that are not determined by the curvature, are determined by first order equations that have the curvature scalar itself as the source. Your question, as I understand it from your question of Leonard several days ago: how is it possible that this phenomenon is not invariant under time reversal? Why does the radiation go in to the future and not in to the past? The point is that this is a very special situation which is similar to classical radiation from a uniformly accelerated charge and also analogous to the quantum field theory phenomena called the Klein paradox or the Schwinger effect where there is a degeneracy of the vacuum state. For thermodynamic reasons we usually assume that you start with nothing and radiation comes out. But there are other solutions in which the radiation is coming in and disappearing and other solutions in which you have as much coming out as you have coming in. That's a very close analogy to the distinction between the Hartle Hawking state and the Unruh state. %I have a number of references I'd like to give you but I don't want to carry on.

\mers So are you saying that the collapse is not crucial to obtaining Hawking radiation?

\full The role of the body, the collapsing body, is to produce a Schwarzschild metric that extends all the way down to \(r=2M\). Once you get that vacuum solution, in two dimensions it's not a vacuum solution but it is essentially a reduction of the four dimensional vacuum solution to the \(t, r\) plane, then you can do quantum field theory in that background and you find this kind of Hawking effect. But you don't need the body to produce the particles directly, it's a two step process.

\mers When we do the calculation for the quantum field we define the state at past infinity. Suppose you choose the Unruh vacuum, then you discover what happens near the surface of the collapsing star.

\full In the Unruh vacuum you don't even need a collapsing star.

\mers Yes, that's my question. Then you sit at future infinity with a detector that clicks when a Hawking radiation particle arrives there. Are you claiming that you can actually get the Hawking flux without the collapse phase? Is that what you are saying?

\full Yes, if you take the Unruh vacuum in the full spacetime without any collapsing body there at all then you get that result. You also get it if you start with a static star back here and then allow it to collapse. The point of the Unruh vacuum was that it mocked up that result without actually having to consider the collapsing body explicitly.

\mers So a static object, or nearly static object such as black hole, would give you the Hawking flux?

\full It's not really static.

\mers Well it evaporates, a solar mass black hole takes \(10^{75}\) years to evaporate. That's pretty static.

\full If you start with, say, a neutron star or something, and then it collapses, then it's not a static situation.

\mers No no, we're going in circles now, I said forget about collapse because you claim the collapse is not important. We start with a static star and then have this nearly stationary black hole.

\full I'm saying ignore the star entirely. Don't worry about it.

\mers Yes, so you still get the Hawking flux.

\full The point is, you have a horizon. You have something that looks like the Schwarzschild metric including the place where the coordinate system changes from static to\ldots

\mers Yes, that's exactly the situation I'm looking at. You have a static star and then there is a black hole, no collapse. The black hole takes such a long time to evaporate you can think of it as nearly stationary. Are you claiming that you start with nothing and suddenly there is a Hawking flux for this nearly stationary black hole without collapse?

\full I'm claiming, and I'll probably get the formulas wrong, but the basic formulas are \( \frac{\partial}{\partial v} T_{vv} = R \) plus metric coefficients and so on that I don't remember.

\mot The point is the equations are time-reversal symmetric but it's not the initial that you're imposing, you have to argue that those are the right conditions to impose. You could argue about that. But the initial conditions that are being imposed are time asymmetric.

\full Thank you, that's precisely my point. Thank you for making it more clearly.

\bard Right, so if you really have a static star then there is no Hawking radiation. There's also no horizon. But the existence of a future horizon means that the spacetime is not static because there is that region inside the future horizon that is definitely not static.

\ford I think there is no plausible way to generate the Unruh vacuum without collapse.

\mers Exactly.

\ford Formally I agree with what Steve's saying but I don't think in terms of physics without collapse you have any plausible way to get an Unruh vacuum state.

\mers That was the question, can you get it without collapse?

\ford In that sense the collapse is absolutely crucial.

\full I agree with Larry but that doesn't mean that the outward radiation flux is originating inside the body.

\mers I never made a statement about it originating inside the body, I was just interested in the time when that is produced.

\ford I think there is one thing, of course the original Hawking derivation from the problem of trans-Planckian modes. That is, the modes that enter the body have to be extremely high and then be redshifted. Of course, as Jim pointed out in his talk, it's not very well defined where the particles are created. They're apparently created somewhere in this region outside the collapsing body.

\mers No, that problem it's easily addressed by speaking in terms of the stress energy tensor. The average wavelength of those particles is bigger than the size of the star so of course it doesn't make sense to say they are produced inside or outside or somewhere \(3M\) away. 

\ford There is a difference between the stress energy and particles, we have to be careful about that. I think there is one thing that I don't think has been mentioned in this meeting and should be brought up, that's the Corley and Jacobson idea to get rid of the trans-Planckian modes by having a nonlinear dispersion relation. Of course that has its price, it means you're giving up local Lorentz invariance. But if you do that, they've shown that you can in fact get Hawking radiation without trans-Planckian modes through a mode regeneration mechanism that creates the needed modes just when they're needed to be populated. Of course they still need the collapse in order for their process to work but it's happening outside the star and you don't need these modes at very high frequency skimming close to the horizon in their model.

\mers Just one comment, there is still confusion about that: how important the collapse is to actually producing the flux, to the point that you'll hear people bring up the analogy with the Schwinger pair creation. That argument goes: well you can get a Schwinger pair created under a constant electric field. Again the confusion there is the electric field might be a constant but the potential associated with it is not. That's exactly what is producing the Schwinger pair. I just wanted to make the point.

\mot No, I disagree with that. The gauge invariant quantity \textit{is} the electric field. Once again the boundary conditions, the initial conditions, that are asymmetric.

\mers The electric potential is not though.

\mot You can do it in a static gauge, that's a gauge dependent statement. The electric field is time-reversal invariant. A constant, uniform electric field is a time-reversal invariant state of the field and there exist states in which particles are coming in from infinity and being destroyed, but you don't use those states because you think that's unphysical and you put a different initial condition in which the particles are created. I think the situation is analogous here. I think it's less clear, to me, so I'll use this opportunity to say I would like to add one more thing on that list, which I'm sure from your point of view is the most extreme. That is that the Unruh state is not necessarily the unique initial condition and there are nearby states in which there are large effects near the horizon and something may happen at the horizon at formation. Now the only reason that's considered extreme in this community is because physics is upside down. You're willing to accept singularities, you're willing to accept information loss which usually in other areas in physics would be considered a problem. But if you take that point of view then having something happen at the horizon looks extreme. On the other hand, trying to explain things by what I would call `normal physics,' namely counting states and ensembles we have about 150 years of experience with, that is the way normal physics works. Not information being lost and singularities. But that's my pitch, I'll end.

\stod I have a question that maybe somebody here can answer, but before I get to my question I want to say that I agree with Larry about what he said a few minutes ago about burning the piece of paper. That's what the point I was trying to make with my example a couple of days ago about a uranium anti-uranium nucleus producing a pion gas. It's very easy to start with a pure state and get something that looks thermal. There's no big mystery about the whole thing and think that is what Larry was trying to say. Anyway, here is my question: in atomic or nuclear physics when we emit a photon we say the photon is emitted in the s-wave or the p-wave. It has angular momentum zero or one or two. Can somebody tell me what is the angular momentum of the photons from Hawking radiation? If you naively multiply the energy times the size of the black hole you get a rather low angular momentum which would mean that there are correlations in some sense in different directions.

\ford It's emitted in all \(l\) modes but the angular momentum barrier highly suppresses the very high modes.

\stod So I made a mistake? The angular momentum is not low?

\ford It's intermediate, but it's a range of angular momenta.

\stod Okay so then you've got correlations, some kind of quantum correlations in different directions. This question, originally I was asking Jim at breakfast to explain what Hawking said and there was something happening on two-spheres at infinity which of course you would get if you have spherical harmonics with low angular momentum. Where is the peak of the angular momentum distribution you're talking about?

\ford I think it depends a little bit on the spin of the field you're dealing with but it's on the order of a few.

\stod Let's take spin zero fields.

\ford Jim do you remember?

\bard I'm not sure.

\dav We can just Google it.

\stod The angular momentum is a dimensionless number so it must be some number you concoct up from existing quantities.

\bard If you're talking about electromagnetic radiation there is no s-wave, for instance. 

\ford Yes. The angular momentum barrier is proportional to \(l^2\) so it's going to grow very greatly as \(l\) gets large compared to 1. If you want a number: 2, 3, 4. Something like that.

\stod Yes, that's what I sort of got. Does that mean then that if you have low angular momentum, of course there's some kind of correlations in different directions.

\bard In the case of Hawking radiation it's very difficult for high angular momentum particles with the Hawking energy to get through the potential barrier. It's a very large centrifugal barrier centered around \(r = 3M\) which prevents modes from going from inside to outside. Most of the Hawking radiation you expect to be in the lowest possible multipoles. 

\stod Yes, so then you've got correlations in different directions. There's some kind of coherence.

\mot There's nothing unusual about that.

\stod No, but it means that there's something coherent in the problem, right, with respect to different directions.

\bard Sure, Hawking radiation does not come out in one specific direction. The quantum state, as it evolves, you have the Hawking radiation going out in all directions. There are correlations between different directions but you can't say that a given Hawking quantum comes out in a particular direction. It comes out in all directions.

\stod Yes but when you have an atom decaying in the s-wave of course it's equal in all directions but there's a certain coherence in which you don't have a totally random statistical emission.

\bard Sure. In connection with what Emil was talking about, the fact that you have the Unruh state which is well behaved on the horizon and then neighboring states which aren't. There's an interesting paper by Lenny Susskind which came out in July called `The typical state paradox, diagnosing horizons with complexity.' As a typical Lenny Susskind, paper there's a lot of conjecture, but I think it is an interesting paper and addresses this issue of how most quantum states in the vicinity of the horizon will involve things like firewalls and very messy physics. If you're talking about something evolving from nice initial conditions, as in the case of the Unruh vacuum, then while there may be some tendency to develop in to these much more complex states, the timescale for that, he argues, is much longer than the evaporation time of the black hole so it's not really relevant to the evaporation problem. If you're interested, read the paper.

\mers We have about three or four minutes left for this session so I'm wondering if we could go quickly through points (iii) and (iv) in what Stephen wrote. I can say something about point (iii), we had three people talking on Monday about it. I suppose the problem is very well defined, there is a collapsing star or a black hole and there is a Hawking flux associated with it, the question is how does that flux modify the evolution of the star, of the collapsing star or the black hole. In terms of procedure and formalism, all we have to do is solve a set of Einstein equations and the stress energy tensor conservation, and we are done. Whatever comes out of it that is the answer. We saw three different approaches to the problem of the backreaction, by Gerard 't Hooft, Jim Bardeen, and myself. They look different, I am curious I think there is quite a bit of overlap between the three approaches. For example Jim and I were both talking about negative energy fluxes going in to the hole and doing something to the interior. That will be an interesting problem, to see how the three approaches overlap. When it comes to point (iv), quantum gravity, I think you may have noticed, that word always comes up when we are stuck with a kind of problem that we either don't understand or where our theories break down, such as Einstein's theory of gravity and quantum field theory. That's almost like a code word for saying we don't understand singularities and we don't understand what happens to these trans-Planckian modes being blue-shifted to infinite energies near the horizon. I suggest we take a vote on singularities. We all know that mathematically we do end up with singularities at the center of a black hole, that's been known since the 1930's, but mathematically and physically are two different things, and we have learned that not all mathematical objects do necessarily exist. I'd like to take a vote on how many people think there are really such physical objects in the universe, singularities.

\bard In terms of backreaction I would just say that given, as I believe along with Stephen Fulling, that Hawking is not generated in a collapsing star, and you can calculate the backreaction using the semi-classical energy momentum tensor. It does not have any singularity near the horizon, the fractional effects are of order \(m_{\text{Planck}}/M^2\) in a freely falling frame, and you can calculate the evolution of this assuming that since the semi-classical energy momentum tensor for a Schwarzschild black hole is spherically symmetric, you can still have a spherically symmetric spacetime, to first order anyway in the backreaction. You can calculate the change in the metric as a result of the backreaction using the Einstein equations, again just to first order outside, what I claim the semi-classical approximation allows you to do. But it's a straightforward calculation and you find that nothing strange happens. The geometry stays Schwarzschild, with small corrections.

\ford If you're asking about mathematical singularities in classical relativity I think, like most people, I would say that quantum gravity probably has to do something to it. Probably close to the Planck scale. But what, I have no idea. And whether that would solve an information problem is also a lot less clear. I think I would have to turn in to something else but what is not clear at all to me.

\duf I don’t know the answer to the information paradox any more than the rest of you but one question I would ask is: in looking for a theory that might give us the right answer, does it first give us the right answer for the microscopic Bekenstein-Hawking entropy? If it doesn’t do that, its chances of answering this more difficult question are not very high. Now string and M-theory have an answer to the first question, maybe there are other theories that do. But I’d like someone to address the question of how they think they’re going to solve the information paradox without first solving the entropy problem. 

\ford Let me just comment quickly on Mike's question. Certainly string and M-theory have a calculation of entropy for extremal or near-extremal black holes, so it's a partial answer, but it appears that it's a state counting argument. It would not surprise me to expect that there are a lot of theories that, when they're worked out, will be able to do similar state counting arguments. I agree that that's something that is a necessary requirement but I think it's far from sufficient. You need a lot more than counting states to be able to say what's going to happen near a singularity.

\end{dialogue} 

		\pagebreak
		\subsection{Status Report \\ \textit{\small Stephen Hawking} }
				%auto-ignore

It has been a pleasure for me to participate in the Hawking radiation conference and spend a week of intense discussions with old friends and colleagues on fundamental problems in black hole physics. It is many years since we experienced similar excitement on this topic in the forty years since Hawking radiation was discovered. As is often the case, a breakthrough in an important field creates a set of new and more difficult puzzles. In this case, the discovery of Hawking radiation leads to the mystery of the information loss paradox. 

Soon after I discovered that black hole's evaporate I realized, since we cannot measure the quantum state of what fell in the the black hole, then we could not recover this information when the black hole had completely evaporated. I raised this concern in my paper `Breakdown of predictability in gravitational collapse' in 1976. There is a whole body of literature dedicated to the information loss paradox ever since. Various attempts to solve the problem have been tried and have, so far, failed. The problem was recently restated in the more dramatic description of firewalls around a black hole. In Jan 2014 I suggested that black holes may not exist at all. Such a possibility requires a precise understanding of the origin of Hawking radiation and implies that singularities do not exist. The singularity theorem can be evaded by quantum effects. We heard three talks on Monday by 'T Hooft, Mersini-Houghton, and Bardeen exploring such a possibility through backreaction. The overlap in the approaches in the three proposals remains to be further explored. 

Hawking radiation observed at future infinity is produced by vacuum pair creation near the surface due to the changing gravitational surface near the collapsing star, as I described in 1973. The precise location in the exterior where pair creation takes place, cannot be given accurately since particles lose meaning in curved spacetime. But the stress energy tensor is well defined locally. That the process is asymmetric in the collapse phase is crucial to particle creation and the breaking of symmetry. When the star has collapsed into a black hole, the horizon temperature is given by the temperature of the thermal bath of Hawking radiation. The average wavelength of the produced particles for a star with mass \( M \) is about order \( M \), and the time interval during which the star collapses is order of \( M^{-1} \), and thus the uncertainty principle is satisfied. 

Black holes are not stationary. They evaporate, albeit very slowly. A solar mass black hole takes about \(10^{75}\) years to evaporate. Since black holes evaporate, the process of pair creation continues although at an incredibly low rate. 

Indirect evidence suggests that black holes exist. We will soon know the answer to this question observationally when data from the Event Horizon Finder experiment becomes available. Even if black holes exist, information about the particles that fell in them is not lost. I described my proposal for solving the information loss paradox on Tuesday. 

The solution to the information paradox lies in the supertranslation invariance of the gravitational scattering matrix that connects the ingoing and outgoing particles. I realized that the BMS symmetry, with supertranslations as its subgroup, can be defined on the horizon of a stationary black hole. Every time an ingoing particle falls through the horizon, the horizon shifts by a certain amount. A supertranslation moves each point of \(\mathcal{I}^+\) a distance \(\alpha\) to the future along the null geodesic generator of the horizon on \(\mathcal{I}^+\). This is equivalent to the advanced time \(v\) being shifted by an amount \(\alpha\) where \(\alpha\) is a function on the two sphere. The information about the horizon shift induced by the incoming particles falling though the horizon is stored in the supertranslation. This corresponds to a delay of the emission of the wavepacket registered as part of Hawking radiation at \(\mathcal{I}^+\). The information about ingoing particles, registered in the form of emission delays of the outgoing particles, is scrambled. Although scrambled, this information is fully recovered since the scattering matrix is supertranslation invariant and thus preserves unitarity. A cutoff of high energy modes near the horizon is not needed. I conclude that the solution to information loss paradox is supertranslation. 

The solution of the information paradox can be applied to a black hole on any background, including those that are not necessarily asymptotically flat.

		\pagebreak
		\subsection{Concluding Remarks \\ \textit{\small Paul Davies} }
				%auto-ignore

\begin{dialogue}

\dav Ladies and gentlemen let me first start by thanking the organizers of this meeting. It has been quite an exhausting week, five and a half days, I think I've probably heard as much about evaporating black holes in the last five and a half days as I need for the next forty years. But we've certainly done the subject proud. Of course we wouldn't be here if it were not for Stephen's work forty years ago. The worst part of every conference, I'm sure you'll all agree, is the group photo and you'll remember this being done a couple of days ago. I dug this one out, this is another group photo from 1970, summer of 1970, and here is Stephen in the front row with Martin Rees and there's a very youthful looking Paul Davies standing in the back. It's a pleasure to be here all these years on and to be with so many old friends and colleagues, Stephen in particular. Thank you Stephen for summing up so well in the last ten or fifteen minutes what your position as regards the whole black hole information issue. 

\vspace{0.1in}

I want to take a somewhat broader sweep and what I have prepared here is a mixture of things we've been talking about over the last few days and one or two things we have only just touched upon, so these are really what I would regard as un-resolved issues. I think this subject is not closed, there's still plenty of work for young people to do which is good. 

\vspace{0.1in}

Let me begin first with the observational aspects. A lot of people forget that Stephen's first paper on this was called black hole explosions and there the emphasis was less on things like information, more on whether it might be possible to observe the end-state of black holes as an astronomical phenomenon. Now the point here is that if microscopic black holes were to form in the very early universe then over the age of the universe the runaway evaporation process would become quite explosive and they'd disappear in a puff of charged particles. If this takes place in an astronomical background magnetic field then there would be a pulse of electromagnetic radiation which could be detected and observations were carried out to see if this phenomenon was taking place. It would be truly wonderful if such pulses of radiation were detected. Well that whole subject sort of faded away, partly because I think people exhausted the observational capabilities although that could be easily resurrected with something like the square kilometer array, but I think also because of a general sense that microscopic black holes were unlikely to exist. These would be primordial black holes made in the very dense phase just after the big bang. Current thinking is that they're probably not there. But I wonder if we're being a little too quick in dismissing the possibility of black holes of subatomic dimension. So maybe there is still room for resurrecting this observational program. 

\vspace{0.1in}

I don't want to dwell anymore on that because mostly the people here are theoreticians and we have all sorts of discussions and I'm not going to summarize the back and forth of the last several days about the information paradox, breakdown of unitarity, existence of firewalls, the stress energy momentum tensor and the backreaction, and a lot of issues that have swirled around depend on the consistency of the semi-classical approximation and where it might break down. Whether it breaks down at Planck scales or something above that. These are really unresolved issues. I always remember Mike Duff years and years ago drawing caution to the use of the semi-classical approximation in which we couple the expectation value of \(T_{\mu\nu}\) to a classical geometry, and the circumstances under which that may or may not apply. I think there are many unresolved issues around that. 

\vspace{0.1in}

The last one Stephen Fulling put in at point 1.5 or point 0.5 or something, he quite rightly says that we really weren't worried about information loss back in the good old days because there was a singularity inside the black hole, the information could go into it and disappear. It's true that quantum gravity may change the singularity, and we don't know in to what, but it's entirely likely that it changes it in to something with a more complex or complicated topology. These things come in to fashion and go out of fashion but there have certainly been phases when people have thought of black holes as portals to other asymptotically flat regions of spacetime. Other universes if you like, connected by wormholes or bridges or spacetime foam. There's all this sort of terminology, a lot of it going back to Wheeler in the 1950's. We certainly cannot rule out that even if the singularity is removed, that quantum gravity replaces it by a more elaborate topology. Under those circumstances unitarity is meaningful, of course, only if we take a god's eye view on the entire spacetime and not just in our restricted region of spacetime. There seems to me lots of escape clauses that could in the future be used to get around the information loss problem. For those people that have a problem with it. Everyone here is working under the assumption that quantum mechanics has got to be the correct and the ultimate theory of the universe as a whole and it doesn't break down anywhere on any scale or size or complexity or any other parameter. If you go to the great wide world outside and you talk to physicists very few of them would be so bold. I think they would all entertain the possibility that we cannot extrapolate unitary evolution to the universe as a whole. I'm sort of agnostic on that point but many theoretical physicists would think it's simply going too far.

\vspace{0.1in}

I want to turn now to some of the things that only got touched upon. In the discussion we just had the question of trans-Planckian modes was raised. This is Stephen's original construction where we have to trace the modes from \(\mathcal{I}^+\) back through the collapsing star to \(\mathcal{I}^-\) and perform a Bogolubov transformation between the in-states and the out-states. This famously leads to these exponential factors which is the same as the exponentially escalating red-shift and so what this means is that Hawking radiation with, say everyday photon wavelengths out here, would lead to stupendously high frequencies down here. The question is what should we do about this? Should we simply say well this is just a formal mathematical construction, it doesn't really matter, it is just an intermediate part of a calculation which gives us a result which we believe and so we'll ignore this. And it has been largely ignored but Ted Jacobson has recently resurrected it. It was pointed out way back in the mid 70's that this is an issue which perhaps should be addressed. What we would really like is to be able to derive the Hawking radiation by some sort of alternative calculation that doesn't get tangled up with this. The point being that if you put a cutoff in here, say at the Planck length or a trillion Planck lengths, then this whole calculation gets messed up. So that's one issue I don't propose any more about.

\vspace{0.1in}

The question of gravitational entropy: so Fay gave us a lovely talk yesterday in which she introduced this set of ideas that go all the way back to Tolman, really, in the 1930's, but was particularly championed by Roger Penrose in the late 70's. The issue is very easily explained, when we look at the history of the universe the state of the universe just after the big bang was one of smoothness and uniformity, the ripples in the CMB, remember, are only one part in \(10^5\) or so. The universe starts out with matter and energy distributed very smoothly and then over time the material is clumped together. This type of gravitational aggregation or clumping is what drives most of the irreversible processes we see in daily life, gives us the so-called arrow of time, and the natural end-state of the clumping is, of course, a black hole. We can think of the black hole as the completion of the process of aggregation, and for the black hole we have a precise formula for the entropy. Many people, and particularly, as I've mentioned, Penrose, feel that somehow one ought to be able to capture the concept of entropy and inexorable rise of entropy, in this case the inexorable rise of clumpiness, by some sort of measure of the degree of inhomogeneity of the gravitational field. The square of the Weyl tensor was one idea that was mentioned briefly in passing. This issue is unresolved. One of the big problems about trying to generalize gravitational entropy from the black hole case to this more general gravitational field is that the black hole formula has Planck's constant in it and this is a collection of stars. There's no obvious connection here with quantum effects. 

\vspace{0.1in}

Yesterday Fay raised the issue about how far the generalized second law can be generalized. The generalized second law, let me remind you, is where we include the entropy of black hole horizons with the overall thermodynamics of the system so that there can be a tradeoff, an exchange of energy and entropy between a black hole and its environment. I think we all believe that if you take the sum of the black holes and the entropy of the environment that the generalized second law will be respected. That is, the total entropy is not going to go down. The question is: is the horizon area always a good measure of entropy? In particular when we extend horizons to cosmology and the classic cases, deSitter space, when we do that is it always the case that, say one square centimeter of black hole horizon, is worth the same as one square centimeter of cosmological horizon. If we wish to make a strong statement that the generalized second law should apply in all cases to all horizons and all matter in the universe, can we use that as a filter to rule out theories where that goes wrong? And, again, we had some discussion about that. I want to just pick up a couple of points. 

\vspace{0.1in}

Let's take deSitter space as the case that is much studied. It is certainly true that a particle detector sitting at the center of coordinates in deSitter space will click click click and register a thermal spectrum with a temperature that depends on the radius of deSitter space, or on the cosmological constant, that seems like a very close analogy to black hole horizon entropy. It's been taken as such, I think, by most people since. But when you dig a little bit more deeply in to this there are some differences, it's not so completely clear. The first thing is that the stress tensor that you calculate for the vacuum state that gives rise to the thermal response does not correspond to thermal radiation. When you calculate the stress tensor of the Hawking effect out at infinity it looks just like a flux of thermal radiation, or in the Hartle-Hawking case a bath of thermal radiation. You might expect deSitter space to have a \(T_{\mu\nu}\) corresponding to a bath of thermal radiation, it doesn't. That would be that, that's what you actually get. It just renormalizes the cosmological constant. Is there an information paradox for deSitter space? Well I'm not sure what that even means. You can allow a galaxy to go across the deSitter horizon, it's gone, it's disappeared from your patch of the universe. How much information is hidden behind it? Well, an infinite amount. So unlike in the case of the black hole where, as Bekenstein told us many years ago, the implosion of \(N\) particles in to a black hole means we've lost roughly \(N\) bits of information, in the case of deSitter space it seems to be infinite. deSitter space doesn't evaporate, it doesn't give us back entropy in the form of thermal radiation that we might have previously attributed to the horizon. So there are these differences and the question is do we have any right, given the somewhat suspect nature of the deSitter event horizon, do we have any right to suppose that a generalized second law will apply there as well? 

\vspace{0.1in}

Well this is a question that actually, and Larry Ford mentioned this a couple of days ago, that he and I worked on quite some years ago because we looked at the case where there's a black hole at the center of the universe, so to speak, and here's the deSitter horizon. So we have two horizons, this one is hot and that one is cooler, there's a flow of heat from the black hole to the deSitter horizon and eventually the black hole evaporates and you're just left with deSitter space. If you want to be really clever you can put a shell of matter around the black hole, a static shell of matter, and that depresses the effective temperature outside of the system so you can then get a back-flow of radiation, the deSitter horizon can actually be hotter than the black hole horizon. You can play those sorts of games and you can ask if there's a tradeoff of horizon area, if it's going from here to here or the other way, if the black hole is giving some of its horizon area to deSitter space or vice versa, will the books always balance? 

\vspace{0.1in}

I don't know if Larry Ford has the same recollection as I do, but I think we were working in Newcastle and we had all the equations up on the board and it turned out that Don Page was coming. This is a story about Don Page which you may appreciate. Don Page was coming to give a seminar and he walked in and said `What are you doing?' and we explained this, he took up the chalk and he started writing on the board and then he wrote more and more and we sort of sat back. He covered the entire board with chalk and then at the bottom `greater than or equal to zero.' That's it. Proved. It's a typical Don Page story that one. That's the paper that came out of it.

\vspace{0.1in}

Anyway, the upshot is that if you try to look at the interplay between black hole horizon and deSitter horizon everything comes out in the wash. Some years later with Tamara Davis, who incidentally was at Nordita when it was in its Copenhagen phase, and Charley Lineweaver, we looked at a gas of black holes in a universe with a cosmological constant and then just evolved that with the Friedmann equations. Thus allowing the gas of black holes to disappear across the deSitter horizon and taking their entropy with it. Then the question is, does the deSitter horizon increase in area enough to pay for the black holes that you've lost touch with? The answer is: yes it does. 

\vspace{0.1in}

So all that seems good. But now, okay we can believe that it may work with deSitter horizons because they're sort of more or less static, but what about more general cosmological models where you have an event horizon but it's time dependent? That would be a situation, for example, if the dark energy were not to be precisely constant. You can still have a horizon but it becomes a time dependent horizon. What happens then? That's something I took a look at some years ago. You can define an instantaneous horizon, incidentally the question about the teleological nature of horizons came up at this meeting. I think it is really important to remember that a horizon is the boundary of the past of \(\mathcal{I}^+\) and it depends on all the things that are going to happen in the future. A lot of the difficulty, I think, in wrapping our heads around what's going on with black holes and so forth is that somehow whatever is happening when we're talking about a horizon has no local significance, only globally it's defined. You might imagine that in the case where the horizon is time dependent, as it would be in this case, that's the horizon radius and this is the conformal coordinate defined in the familiar way. You can define a horizon like that. You might expect under those circumstances there would be maybe a violation of the generalized second law but it turns out that there isn't. So long as this energy condition, \(\rho + p \geq 0\) is satisfied then the horizon area is non-decreasing with time. This is like a cosmological version of the Hawking area theorem. I can give the reference to that paper to anybody who wants it.

\vspace{0.1in}

You can solve some models exactly, this is a radiation filled model with a \(\Lambda\) term, that's its behavior, and then you can look at a horizon volume with the radiation entropy. So you've got an entropy to start of with, you've got just radiation entropy plus horizon entropy, and then you can evolve it forward and at the end all the radiation is gone and it's just deSitter space. And again you check that the horizon area just keeps going up, and it does. This is even more remarkable because there's no well defined temperature in the case of time dependent horizons. For deSitter space I've showed you that there is, told you that there is, but in the case of time dependent horizons a particle detector would not even have a steady state response let alone a thermal spectrum. Yet it seems that the GSL (generalized second law) is doing its best to be satisfied even in these sorts of cases. 

\vspace{0.1in}

You can ask yourself well, what sort of violence do you have to do to physics and cosmology in order to get a decrease in the generalized entropy. You can do that. There's certainly cases where you can do it, one of these has come up already. When we have higher order terms in the gravitational Lagrangian, there can be circumstances there in which you can get a violation of the generalized second law. The case that is perhaps of most contemporary interest is if we have a violation of the energy condition, that is if the dark energy is increasing with strength over time, \(\rho + p < 0\), or the \(w\) parameter is less than minus one. Under those circumstances we live in an approximation to deSitter space but the deSitter parameter gets bigger and bigger with time, or smaller and smaller the thing accelerates faster and faster. That leads to the famous big rip demise of the universe. If you take the trouble to calculate the total entropy on the approach to the big rip, which is to say that you take any sort of radiation that might be around plus the horizon entropy, then this certainly shrinks down to zero as you approach the big rip. 

\vspace{0.1in}

You \textit{can} get violations of the generalized second law but that leads me to the question: so what? Do we simply say, well, the generalized second law can be generalized beyond what we originally thought but not to everything; or should we, as Fay said, cling to it with Eddington's quote, and say that if we discover scenarios where the generalized entropy decreases we should reject those scenarios? There are really two attitudes you can take about violations of something as fundamental as the second law of thermodynamics. One is to reject any theory that possesses solutions corresponding to a decrease of total entropy, the other is that you just reject those solutions. The problem about just rejecting the solutions is that we can always imagine a super civilization that can manipulate spacetime geometry and matter such as to contrive one of the solutions that would violate the generalized second law and thus, in effect, create a \textit{perpetuum mobile}. I don't propose to dwell any more on the philosophical aspects of that, but if we really are going to take, as Fay I think pointed out so well, Stephen's great contribution has been to draw the connection between thermodynamics, gravitation, quantum field theory. There seems to be something deep still to be explicated about that linkage, that if we're to do that we're going to regard that as sort of a foundational principle for doing physics and cosmology. Anything we place at that foundation we had better take seriously and so we might well want to eliminate theories, and there are going to be many many of them, your favorite theory, it might be string theory or it might be something else, if it's found to be in violation of this generalized second law. Then we might want to reject it on those grounds. That's sort of ending on a dramatic note.

\vspace{0.1in}

The last thing I want to talk about is another trip down memory lane and some of you know we've had some discussions on this in the last few days. Back in the 1970's when we were struggling to make sense of calculating the stress-energy-momentum tensor to put in the semi-classical equations there was a sort of dispute going on, known as the in-in or in-out dispute, between two groups of people. I think we all agree that in the semi-classical approximation we have some sort of geometrical terms on the left and something on the right representing the quantum field, the question is what is this quantity here in the wedge around the \(T_{\mu\nu}\)? Bryce DeWitt, who was very influential in getting this whole subject off the ground, drew inspiration from Schwinger's effective action as applied to QED. When you do that you're led very naturally from an action principle to consider the quantity in-out. That is, take the \(T_{\mu\nu}\) operator and sandwich it between an in-state, doesn't have to be a vacuum state but it's simplest to think of that, an in-vacuum and an out-vacuum, and this is a normalization factor. That quantity, for those people who came under DeWitt's influence, was the thing that tended to be put on the right hand side of these semi-classical gravitational equations. Is that so? Sure, here we have Dowker and Critchley and there it is. Here we have Hartle and Hu, there it is, the in-out quantity. And here, I'm pleased to say, we have Gibbons and Perry. So, again, that's the in-out. This was considered, you know, maybe the right thing to do. But a number of others of us insisted that what you should put on the right hand side is a true expectation value and not the in-out. They're related through the Bogolubov transformation in a complicated way. The key thing is all the divergences, which were the main purpose of studying this (how to get rid of the infinite terms in the right hand side quantity), all these divergences are the same in the in-in or the in-out. It didn't really matter for what we regarded as the main job, which was to get rid of those divergences. There the matter may have rested because, I think, by about 1980 everyone was persuaded that this expectation value was the appropriate quantity. 

\vspace{0.1in}

Everything may have been left there were it not for this gentleman, Yakir Aharonov, one of the founders of modern quantum mechanics. I've been collaborating with him for about the last ten years on a sector of quantum mechanics which has always been there but it's been overlooked in most of the history of the subject. An extraordinary thing to think all these years on that ordinary, basic, non-relativistic quantum mechanics can still contain a sector with predictive power that has not been tested until recently, but it does. This has to do with what are known as weak measurements. I'm not going to go in to this in any detail but I just want to give you a broad-brush idea. When we make a measurement in quantum mechanics, a normal strong or projective or von Neumann type of measurement, in an idealized system you couple the quantum system to, say, von Neumann measuring apparatus and then, to use the politically unfavorable incorrect language, you collapse the wave function. That's a strong projective collapse. Now in a weak measurement you turn down the interaction between the measured the system and the measuring apparatus to allow arbitrarily small coupling. Why would you do that? Because then that means that the results of your measurement, the noise smears them all out and it's no use to you. And it's true that you don't collapse the wavefunction but you don't get any information out, however you can compensate for that by having a very large ensemble of identically prepared systems. An identical initial state. Then you can take a statistical average over that large ensemble. That's something you can do, you can do a lab experiment to actually test this. The statistical average is called the weak value, and there it is, that's Aharonov's weak value. You take the initial state, this is the operator corresponding to the observable of interest, you take the final state, and you divide by that quantity there. In classical mechanics you can't do this because in classical mechanics the initial state suffices to fix the states for all time. In quantum mechanics if you have a large ensemble of identically prepared systems there will always be a sub-ensemble that will satisfy a final boundary condition as well as an initial boundary condition. You can take that sub-ensemble and consider weak measurements in between the preparation phase and the final phase. That's what this weak value represents. Of course that's just what DeWitt was telling us about. 

\vspace{0.1in}

This sort of long running, not dispute, but long running confusion as to what to put on the right hand side of the semi-classical equations, I think, is at last resolved. We now have a physical interpretation of DeWitt's version and that physical interpretation is that this quantity is the weak value of the stress-energy-momentum tensor. It's the value you would get if you make repeated weak measurements in a system where, if you have identically prepared systems, if you imagine having an ensemble of them, then if you take a sub-ensemble that satisfies a final boundary condition where that out-state is not just the unitary evolution of the in-state, if you do that then that's the value you get. The question is, is that relevant at all to the real universe? Aharanov has conjectured that the effect of a special natural final boundary condition would seem to us as a new fundamental law of nature. If we live in a universe where there's not only a natural initial state, maybe the Hartle-Hawking state, but also a natural final state then there may well be consequences. I'm not going to dwell to much on this. Aharanov calls this quantum miracles. You might notice something very strange. One case that's just come out, and we've been discussing it, is if you were to take the initial state of a collapsing state to be the Unruh vacuum, you could take the final state to be the Boulware vacuum, and then you would certainly have a divergence. This in-out quantity, this weak measurement, this weak value, would certainly diverge on the horizon. There would be a firewall. Interpreted in that way you could certainly have firewalls. The question is, what is the final boundary condition on the universe? Can we say anything at all about the final state of the universe? We all have this prejudice that somehow the universe should be born in a simple or natural quantum state but we're largely agnostic as to how it's going to end up in the future. You could take the attitude that we shouldn't be prejudiced. That the universe may have been born simple and it may die simple. If it's the case, as Charlie Misner was saying yesterday, that we're living at an atypical phase between two deSitter spaces, the inflationary one and the one in the far future of the universe dominated by dark energy, we're just in this transition phase, it may be rather natural to pick a vacuum in the final phase and vacuum in the initial phase. Because the initial vacuum will not propagate unitarily to form the final vacuum, the weak values that you would get from cosmological measurements would depart from the expectation values which is the thing that we have known about. 

\vspace{0.1in}

As it seems fitting at a conference like this where deliberations have been really rather conservative, I thought I might end on a speculative note. The subject of cosmology is normally regarded as purely observational or theoretical, the question is could it be experimental? So just for a bit of fun I thought I would say well, yes, there has been at least one cosmology experiment. This was by Bruce Partridge back in the 1970's. It was to test the Wheeler-Feynman absorber theory of radiation by taking a microwave antenna to the top of a mountain and beaming it out in to the universe and in to an absorbing material and comparing the two. In the Wheeler-Feynman theory, because the existence of retarded radiation depends on a future absorber, a future boundary condition of the universe, it would show up in the power drain of the antenna. Well needless to say, he didn't find anything but it's interesting that this was actually a cosmological experiment. So, in that spirit, I'd like to finish by proposing this experiment: that we have a natural initial condition for the universe, which would be some deSitter space, something corresponding to inflation, we pick one in the end of the universe, the dark-energy dominated phase, then the question is how would we know? Of course if it is open, you know, who is the great selector at the end of the universe, but let's leave that aside, how would we know? Well here's an experiment that may be able to tell you. If you shine a laser beam in to the sky out of the plane of the Milky Way then a simple calculation shows that most of the photons are going to escape to infinity, they won't be absorbed. And if it's the case that the far future of the universe is the vacuum state of the electromagnetic field then those photons had better not be emitted in the first place because otherwise it can't be a vacuum. And so imagine that we have this absorbing material and there's a gap here, this is a photo-detector, and here we have the decay of a particle, and this is an EPR type setup, and so we can detect a photon here and deduce that there's one going the other way there, and then we rotate the whole thing like that, and what you should find is that if this is pointing up to the sky in the appropriate direction then you don't detect anything here. 

\stod You then what?

\dav You would not detect a photon here under those circumstances \textit{if} it is the case that the far future of the universe is a vacuum state. And so with that somewhat tongue-in-cheek proposal to convey this entire field in to the experimental domain I think I would just like to thank once again the organizers and it's great to see old friends, I particularly thank Laura for her tireless work in putting this together and in arranging so much food. I think maybe a week or two of reduced intake is probably called for for everybody. %(36:57)

\vspace{0.1in}

I wasn't terribly clear exactly whether I am supposed to sort of round it up and say `goodbye, go home, drive safely,' or are we going to have some Q\&A and, in particular, is Stephen going to come back and answer some questions? I'm going to hand it over to you because you're in charge.

\end{dialogue} 		
	
		\pagebreak
		\subsection{A Tribute to Stephen Hawking \\ \textit{\small Michael Duff} }
				%auto-ignore

\begin{dialogue}

\duf Laura, may I take two minutes to make a personal comment to Stephen? Stephen, you may recall that I spoke at the 1974 conference where you made your famous announcement. In those forty years I've come to value, as has everyone else, your contributions to theoretical physics. But some of us are grateful to you in ways you may not be aware of and I'd be kicking myself if I missed this golden opportunity to tell you why. In 2008 I was diagnosed with Parkinson's disease, a degenerative disorder of the central nervous system, as you know. My first reaction was to feel sorry for myself, why me? But then I said to myself `look at Stephen Hawking, he's lots worse off than you but he didn't sit around feeling sorry for himself, he grasped life by the scruff of its neck, lead life to the full, and so can you.' 

\vspace{0.1in}

Now, while I've got everyone's attention, I want to point out that my own college, Imperial College, happens to be the home of the Parkinson's UK brain bank and they're looking for donations. If anybody wants to contribute to this area of research they're looking for healthy brains as well as Parkinson's ones. Personally I think it's a no-brainer. Although there is no cure for Parkinson's it responds well to medication and I feel very fortunate that seven years down the line the medication has so-far kept in check the worst symptoms and I don't think it's affected my research. 

\vspace{0.1in}

That brings me, Stephen, to the reason why I'm telling you all this. Whatever the future may bring, and I think I speak for others in situations similar to mine, we'd like you to know that if we can muster just a fraction of the courage that you have mustered, if we can summon just a fraction of that mulish stubbornness in the face of adversity that you have summoned, and if, above all, we can retain the same sense of humor that you've retained, then we'll be very lucky people. I wanted to say it and to say it out loud: thank you Stephen, you're an inspiration to us all.

\end{dialogue}

\pagebreak
\section{Closing Remarks}% \\ \textit{\small Laura Mersini-Houghton} }
	
			%auto-ignore

\begin{dialogue}

\mers It's lunch time but I'd like to thank a few people before we close and head for lunch. Thank you Stephen [Hawking], thank you Paul [Davies], and thank you Mike [Duff] for your remarks. That was beautiful, thank you! I would like to invite a few people up here for two minutes. They have really been the driving force behind everything that was arranged locally for this conference: Katrine, Elizabeth, Pouya, Yen Chin. Please come up here. And Malcolm, who forgot he was on the organizing committee himself, when he thanked the organizers. These are the people that made everything possible and it's been more than hard work. What I have enjoyed and loved is that they made us all feel part of a family. Besides the hard work, they've also extended all their warmth, and hospitality, to all of us. It's been a roller coaster at times. Poor Yen Chin had to survive three women shouting at him from all sides. But everybody has been absolutely wonderful and helpful, and dedicated to making this happen. Thanks to you Malcolm (Perry) and Yen Chin (Ong) for being on the organizing committee. Thanks to you Katrin (Morck), you arranged everything I asked from you, from the public lecture venue to this beautiful room for our conference. And you, Elizabeth (Yang), who every time the phone rang (which was daily) everything was fixed by that afternoon – thank you. And Pouya for doing everything we asked you to help us with.

Thank you all of you for coming and participating in the conference. I don't know what you were thinking when I wrote to you and invited you to this conference only 2 months ago, in June, but I am delighted that you are here! People have been wondering and have asked me: `how did you end up gathering this amazing group of scientists in one room, and why did you think of doing it?' In my case it was very simple. We need to solve the black hole information loss problem. So, I was thinking: `who are the people whose work I deeply respect and admire?' and that was the list of people that I sent an invitation to. And here we are. Thank you so much for coming and contributing to the conference. Thank you. And many thanks to Nordita for being very supportive and for being the first to sponsor this conference, after which, everyone else followed. I am very grateful to UNC-Chapel Hill for putting their full support from day one behind my initiative to hold this conference. The conference is now closed.

\end{dialogue}	
	
\pagebreak
\section{References}
	
			%auto-ignore


\begin{thebibliography}{10}


%--------------------------------------------
% 't Hooft references

\bibitem{tHooft:85}
	G. 't Hooft,
  %``On the Quantum Structure of a Black Hole,''
  Nucl. Phys. B {\bf 256}, 727 (1985).
  DOI:10.1016/0550-3213(85)90418-3.
	
\bibitem{tHooft:96}
	G. 't Hooft,
  %``The Scattering matrix approach for the quantum black hole: An Overview,''
  Int. J. Mod. Phys. A {\bf 11}, 4623 (1996).
  DOI:10.1142/S0217751X96002145
  [arxiv:9607022 [gr-qc]].

\bibitem{tHooft:15}
  G. 't Hooft,
  %``Diagonalizing the Black Hole Information Retrieval Process,''
  arXiv:1509.01695 [gr-qc].
  %%CITATION = ARXIV:1509.01695;
	

%--------------------------------------------
% Mersini-Houghton references

\bibitem{Mersini:14} 
  L. Mersini-Houghton,
  %``Backreaction of Hawking Radiation on a Gravitationally Collapsing Star I: Black Holes?,''
  Phys. Lett. B {\bf 738}, 61 (2014).
  DOI:10.1016/j.physletb.2014.09.018
  [arXiv:1406.1525 [hep-th]];
	
\bibitem{Mersini_Pfeiffer:14}
	L. Mersini-Houghton and H. P. Pfeiffer,
  %``Back-reaction of the Hawking radiation flux on a gravitationally collapsing star II,''
  arXiv:1409.1837 [hep-th].
  %%CITATION = ARXIV:1409.1837.


%--------------------------------------------
% Bardeen references

\bibitem{Bardeen:14}
	J. M. Bardeen,
	%``Black hole evaporation without an event horizon,''
	arXiv:1406.4098 [gr-qc].
	

%--------------------------------------------
% Hawking references

\bibitem{Hawking:15} 
  S. W. Hawking,
  %``The Information Paradox for Black Holes,''
  arXiv:1509.01147 [hep-th].
  %%CITATION = ARXIV:1509.01147
 

%--------------------------------------------
% Hawking-Perry references

\bibitem{Hawk_P_S:16} 
  S. W. Hawking, M. J. Perry, and A. Strominger,
  %``Soft Hair on Black Holes,''
  Phys. Rev. Lett. {\bf 116}, no. 23, 231301 (2016).
  doi:10.1103/PhysRevLett.116.231301
  [arXiv:1601.00921 [hep-th]].


%--------------------------------------------
% Rovelli references

\bibitem{Rovelli:14A} 
  A. Barrau and C. Rovelli,
  %``Planck star phenomenology,''
  arXiv:1404.5821 [gr-qc].

\bibitem{Rovelli:14B} 
  C. Rovelli and H. M. Haggard,
  %``Black hole fireworks: quantum-gravity effects outside the horizon spark black to white hole tunneling,''
  arXiv:1407.0989 [gr-qc].
	
	
%--------------------------------------------
% Rovelli + Vidotto references

\bibitem{Rovelli_Vidotto:14A} 
  C. Rovelli and F. Vidotto,
  %``Planck stars,''
	arXiv:1401.6562 [gr-qc].
	
\bibitem{Rovelli_Vidotto:14B} 
  A. Barrau, C. Rovelli, and F. Vidotto,
  %``Fast Radio Bursts and White Hole Signals,''
	arXiv:1409.4031 [gr-qc].
	
	
%--------------------------------------------
% Vidotto references

\bibitem{Vidotto:15} 
  A. Barrau, B. Bolliet, F. Vidotto, and C. Weimer,
  %``Phenomenology of bouncing black holes in quantum gravity: a closer look,''
	arXiv:1507.05424 [gr-qc].
	
	
%--------------------------------------------
% Kiefer references

\bibitem{Kiefer:95} 
  J. G. Demers and C. Kiefer,
  %``Decoherence of Black Holes by Hawking Radiation,''
	Phys. Rev. D {\bf 53}, 7050 (1996).
  DOI: 10.1103/PhysRevD.53.7050
	[arXiv:9511147 [hep-th]].


%--------------------------------------------
% Kiefer + Moniz references

\bibitem{Kiefer_Moniz:08} 
	C. Kiefer, J. Marto, and P. V. Moniz,
	Annalen der Physik 18, 722 (2009).
	DOI:	10.1002/andp.200910366
	[arXiv:0812.2848 [gr-qc]].


%--------------------------------------------
% Parker references

\bibitem{Parker:15} 
	L. Parker,
	%``Creation of quantized particles, gravitons, and scalar perturbations by the expanding universe''
	J. Phys. Conf. Ser. {\bf 600}, 012001 (2015) .
	DOI: 10.1088/1742-6596/600/1/012001 
	[arXiv:1205.5616 [astro-ph.CO]].
	
\bibitem{Parker:12}
	L. Parker,
	%``Particle creation and particle number in an expanding universe''
	J. Phys. A	{\bf 45}, 374023  (2012).
	DOI: 10.1088/1751-8113/45/37/374023
	[arXiv:1205.5616 [astro-ph.CO]].


%--------------------------------------------
% Mottola references

\bibitem{Mottola:15} 
	P. O. Mazura and E. Mottola,
	%``Surface Tension and Negative Pressure Interior of a Non-Singular `Black Hole' ''
	Class. Quant. Grav. {\bf 32}, 215024 (2015).
	DOI: 10.1088/0264-9381/32/21/215024
	[arXiv:1501.03806 [gr-qc]]


%--------------------------------------------
% Louko references

\bibitem{Louko:14A}	
	J. J. Louko,
	%``Unruh-DeWitt detector response across a Rindler firewall is finite''
	JHEP {\bf 2014}, 142 (2014).
	DOI: 10.1007/JHEP09(2014)142 
	[arXiv:1407.6299 [hep-th]].

\bibitem{Louko:15}	
	E. G. Brown and J. J. Louko,
	%``Smooth and sharp creation of a Dirichlet wall in 1+1 quantum field theory: how singular is the sharp creation limit?''
	JHEP {\bf 2015 }, 061 (2015).
	DOI: 10.1007/JHEP08(2015)061 
	[arXiv:1504.05269 [hep-th]].

\bibitem{Louko:14B}	
	E. Martin-Martinez and J. J. Louko,
	%``Particle detectors and the zero mode of a quantum field''
	Phys. Rev. D {\bf 90}, 024015 (2014).
	DOI: 10.1103/PhysRevD.90.024015 
	[arXiv:1404.5621 [quant-ph]].
	


%--------------------------------------------
% Duff references

\bibitem{Duff:15}	
	L. Borsten and M. J. Duff,
	%``Gravity as the square of Yang–Mills?''
	Phys. Scripta {\bf 90}, 108012 (2015). 
	DOI: 10.1088/0031-8949/90/10/108012 
	[arXiv:1602.08267 [hep-th]].


%--------------------------------------------
% Stelle references

\bibitem{Stelle:15A}
	H. L\"u, A. Perkins, C. N. Pope, and K. S. Stelle,
	%``Spherically Symmetric Solutions in Higher-Derivative Gravity''
	Phys. Rev. D {\bf 92}, 124019 (2015).
	DOI: 10.1103/PhysRevD.92.124019 
	[arXiv:1508.00010 [hep-th]].
	
\bibitem{Stelle:15B}	
	H. L\"u, A. Perkins, C. N. Pope, and K. S. Stelle
	%``Black Holes in Higher Derivative Gravity''
	Phys. Rev. Lett. {\bf 114}, 171601 (2015).
	DOI: 10.1103/PhysRevLett.114.171601

\end{thebibliography}
\end{document}